\documentclass{article}

\usepackage{arxiv}
\usepackage{times}
\usepackage{newtxtext,newtxmath}
\usepackage{graphicx}
\usepackage{amsmath}
\usepackage{booktabs}
\usepackage{url}
\usepackage{hyperref}
\usepackage{xcolor}
\usepackage[table]{xcolor}
\usepackage{tabularx}
\usepackage{caption}
\usepackage{siunitx}
\usepackage[normalem]{ulem} 
\usepackage{comment}
\usepackage{multirow}
\usepackage{multicol}
\usepackage{float}  
\usepackage{booktabs, array, xcolor, colortbl, siunitx, adjustbox}
\usepackage{subcaption}
\usepackage[
  datamodel=acmdatamodel,
  style=acmauthoryear,
  backend=biber,
  giveninits=true,
  uniquename=init
  ]{biblatex}
\definecolor{lightgray}{gray}{0.93}
\newcounter{suppnum}
\newcommand{\suppsection}[2]{%
  \refstepcounter{suppnum}\label{#2}%
  \section*{Supplementary S\thesuppnum: #1}%
}
\newcommand{\supp}[1]{S\ref{#1}}

\definecolor{lightblue}{RGB}{200,220,255}   
\definecolor{lightred}{RGB}{255,200,200}    

\definecolor{ForestGreen}{RGB}{34,139,34}

\definecolor{fig1_blue}{RGB}{64, 98, 187}
\definecolor{fig1_pink}{RGB}{255, 102, 196}

\definecolor{lightblue}{RGB}{200,220,255}   
\definecolor{lightred}{RGB}{255,200,200}    

\AtBeginDocument{%
  }

\title{Happy Young Women, Grumpy Old Men? Emotion-Driven Demographic Biases in Synthetic Face Generation}

\author{ Mengting Wei \\
        University of Oulu \\
        Oulu, Finland \\
        \texttt{mengting.wei@oulu.fi} \\
        \And
        Aditya Gulati\\
	    ELLIS Alicante\\
        Alicante, Spain \\
	    \texttt{aditya@ellisalicante.org} \\
        \And
        Guoying Zhao \\
        University of Oulu \\
        Oulu, Finland \\
        \texttt{guoying.zhao@oulu.fi}
        \And
        Nuria Oliver\\
	    ELLIS Alicante\\
        Alicante, Spain \\
	    \texttt{nuria@ellisalicante.org} \\
}

\renewcommand{\shorttitle}{Happy Young Women, Grumpy Old Men?}

\begin{document}

\maketitle

\begin{abstract}
{\bf Background:} 
    Synthetic faces from text-to-image (T2I) models pervade digital media, yet their demographic biases under emotionally conditioned prompts remain poorly understood.
    
    {\bf Objectives:}
    We aim to systematically audit how emotionally conditioned prompts affect demographic and perceived-attractiveness biases in synthetic faces generated by T2I models, with particular attention to intersectional patterns and cross-ecosystem differences across model families.
    
    {\bf Methods:}
    We audited eight (four Western and four Chinese) T2I models. We generated 56,000 faces under seven prompt conditions: a neutral baseline and six emotion conditions. We quantified biases in gender, race, age, and perceived attractiveness using information-theoretic divergence metrics. We further conducted intersectional analyses across combined demographic attributes and compared patterns between the Western and Chinese model groups to assess cross-ecosystem consistency and divergence in bias behavior.
    
    {\bf Results:}
    All models show strong overrepresentation of young faces, and most also overrepresent White-coded individuals. Intersectional analysis reveals compound underrepresentation or near-erasure of specific demographic combinations, such as young~$\times$~female~$\times$~Black faces, which are largely absent across models and are not captured by single-attribute audits. Western and Chinese models exhibit broadly similar demographic bias patterns despite being developed for different target markets. Emotion prompts act as additional demographic selectors: negatively valenced emotions (including sadness and fear) consistently shift outputs toward White, middle-aged, male-coded faces. This produces a valence-driven mapping that is also associated with lower perceived attractiveness in generated faces.
    
    {\bf Conclusions:}
   These findings indicate that demographic bias in T2I face generation is both pervasive and shaped by emotional conditioning. They underscore the need for intersectional, emotion-conditioned, and multilingual demographic audits as part of standard pre-deployment evaluation practices.
\end{abstract}


\section{Introduction}

\begin{figure}
    \centering
    \includegraphics[width=\linewidth]{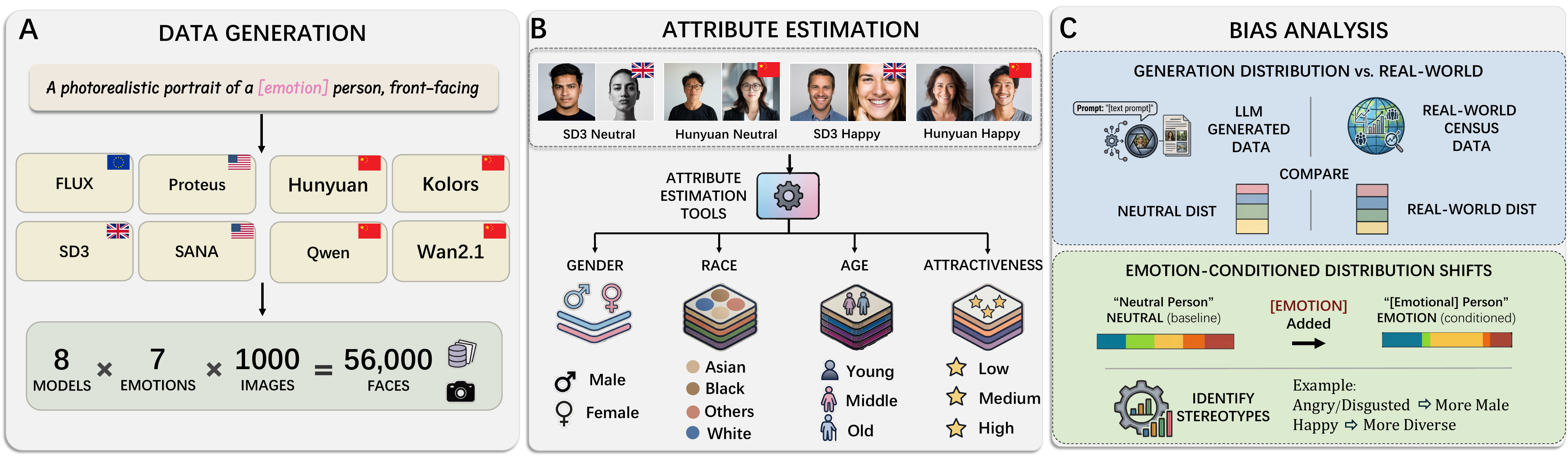}
    \caption{Study pipeline. \textbf{(A)} A standardized prompt is submitted to eight T2I models (four Western, four Chinese) across seven emotion conditions (\textit{happy}, \textit{sad}, \textit{surprised}, \textit{fearful}, \textit{disgusted}, \textit{angry} and a \textit{neutral} baseline condition with no emotion specified), producing 56,000 faces in total. \textbf{(B)} Gender, race, age, and attractiveness are estimated automatically for each face. \textbf{(C)} Two analyses are applied: demographic distributions under the neutral condition are compared against global census data and between regional model groups, and emotion-conditioned prompts are assessed for shifts in the generated demographic profiles.}
    \label{fig:overview}
\end{figure}

Synthetic faces are now generated at industrial scale. Text-to-image (T2I) models (a term we use broadly to include multimodal large language models capable of image generation) power creative tools, advertising pipelines, and social media applications reaching millions of users globally, producing photorealistic human faces from short textual prompts. The demographic makeup of those faces, however, is not a neutral reflection of human diversity. It is shaped by biases that were latent in the models' training data, objectives, and architectures \cite{bianchi2023accessible}. As synthetic imagery becomes woven into the visual fabric of public life---appearing in stock photography, AI-generated social content, training data for downstream systems, and interactive agents---skewed generative outputs risk systematically normalizing distorted representations of who exists, who is trusted, and whose appearance is presented as normative \cite{raji2020audit,doh2026facct}.

Prior work has found that T2I models overrepresent light-skinned, young, and conventionally attractive faces, underrepresent women of color and older individuals, and exhibit \emph{algorithmic lookism}, \emph{i.e.}, the systematic association of positive attributes such as intelligence, competence, happiness, and trustworthiness with physical attractiveness \cite{luccioni2023stable,howard2025uncovering,gulati2024lookism,Yang_2025_RacialBiasAIGeneratedImages,Wu_2025_GenderBiasStableDiffusion,doh2026facct}. Computer vision systems more broadly show systematic performance gaps along demographic lines \cite{buolamwini2018gender,schwemmer2020diagnosing}, and generative models can amplify biases present in the training data or introduce entirely new ones through their internal representations, even when training corpora are carefully curated \cite{huang2025generative,huber2024bias}. Yet three critical dimensions remain underexplored.

First, existing audits predominantly examine demographic attributes in isolation: gender or race, rarely both, almost never jointly with age. However, social inequality is inherently intersectional: discrimination and erasure emerge most strongly at the intersections of gender, race, age, and other identity axes \cite{crenshaw2013,hanna2020,crenshaw1991mapping}. Generative models are particularly prone to erasing intersectional subgroups such as older women of color, whose low prevalence in the training data makes them susceptible to stereotyping or  omission \cite{huber2024bias,AlDahoul_2024_AIgeneratedFacesStereotypes,Wu_2025_GenderBiasStableDiffusion,doh2026facct}. Representational bias therefore cannot be adequately characterized by marginal statistics alone: it requires examining who is over- or underrepresented in the \emph{joint} demographic space.

Second, prior audits have focused primarily on neutral-expression faces. However, facial expressions are central to social cognition, eliciting trait inferences about competence, warmth, and status, and are closely tied to cultural norms and identity \cite{adolphs2002,ekman1992argument,todorov2008}. The \emph{attractiveness halo effect}---whereby faces perceived as more attractive are also judged as happier, more competent, and more trustworthy \cite{nisbett1977halo,gulati2024beautiful}---provides a relevant link: if generative models associate positive emotions with conventionally attractive and demographically narrow faces, they may reproduce well-documented human cognitive biases in synthetic outputs at scale \cite{bjornsdottir2024stereotypes,gulati2025beauty,reis1980physical}. Despite this, emotion-conditioned demographic bias in T2I models has received limited systematic investigation, although prior work suggests that generated outputs can shift meaningfully with emotional context \cite{mehta2024emotional}.

Third, most bias research has focused on Western-origin models, even as China has become a major producer of T2I systems deployed globally \cite{khanal2025}. Cultural psychology has shown that beauty standards, facial aesthetics, and emotion display rules are culturally constructed \cite{matsumoto1990,zhan2021,rhee2018}, and that language-based models can encode systematic cultural patterns reflecting their training data contexts \cite{lu2025}. 
This raises the question of whether such cultural encoding leads to meaningful differences in bias patterns across development ecosystems---potentially requiring culture-specific mitigation strategies---or whether similar biases emerge regardless of model origin. 
A related and previously underexplored issue is whether bias patterns observed under English-language prompts persist when models are evaluated in their primary deployment languages. This distinction has direct implications for the validity of cross-lingual and cross-regional evaluations of T2I systems.

Addressing these three gaps requires an audit that is intersectional, emotion-conditioned, and cross-cultural, a combination that remains uncommon in the existing literature. We present an audit of T2I-generated faces that jointly examines all three dimensions at scale: 56,000 faces generated from eight models (four Western and four Chinese, grouped by region of development) across seven prompt conditions (a neutral baseline and six canonical emotions).  Demographic attributes (gender, race, and age) were estimated using FairFace \cite{karkkainenfairface} and attractiveness was estimated using an ensemble of computer vision models pre-trained on three facial attractiveness datasets (AHEAD \cite{gulati2024beautiful}, CelebA \cite{liu2015faceattributes}, and SCUT-FBP5500 \cite{liang2018scutfbp5500}). All models were extensively validated on real and synthetically generated images (Supplementary~\supp{supp:attr}).

Bias is quantified using information-theoretic divergence metrics relative to both global population distributions and model-specific neutral-prompt baselines.

We find that the eight models substantially misrepresent global demographics: the 20--39 age group accounts for over 80\% of generated faces in six models, and the young\,$\times$\,female\,$\times$\,Black intersection is produced at a mean probability of only 1.3\%, corresponding to a suppression factor of $0.25\times$ relative to an unbiased generator. Emotion prompts further amplify these disparities. The four negative emotions consistently increase male-coded representation and reduce perceived attractiveness (46\% of faces rated low-attractiveness under Disgusted vs.\ 3.4\% under neutral conditions), yielding a valence-driven pattern that differs from typical human emotion stereotypes. This pattern constitutes what we term a \emph{social grammar}: a learned system in which emotional cues are associated with characteristic demographic profiles rather than simply modifying facial expression. Western and Chinese models exhibit broadly similar bias patterns despite differing cultural and development contexts, suggesting that this behavior may reflect a structural property of current T2I systems rather than a region-specific effect. 

These findings motivate the need for intersectional, emotion-conditioned, and multilingual bias audits as part of standard pre-deployment evaluation for generative image models. In the next section, we provide full details of the experimental design, model specifications, attribute estimation methods, and bias metrics. 

\section{Materials and Methods}\label{sec:methods}

\subsection{Evaluated Models}
We evaluate eight popular diffusion-based T2I models, summarized in Table \ref{tab:model_comparison}: four developed by Western groups (FLUX-Schnell \cite{labs2025flux1kontextflowmatching,flux2024}, Proteus V0.3 \cite{proteusv03modelcard}, Stable Diffusion 3.5 medium \cite{stabilityai2024stable35} and SANA 1.5 \cite{xie2025sana1_5}) and four developed by Chinese groups (Hunyuan-DiT \cite{li2024hunyuan-dit}, Qwen-Image \cite{wu2025qwenimage} with LoRA for facial detail, Kolors \cite{kolors2024report} and Wan2.1 \cite{wan2025wan}). For brevity, we refer to the models as follows: FLUX, Proteus, SD3, SANA, Hunyuan, Qwen, Kolors and Wan2.1.

The eight models span both U-Net-based \cite{ronneberger2015u} and Transformer-based \cite{vaswani2017attention} denoising architectures. All models were released between 2023 and 2025 and are publicly available on HuggingFace or equivalent repositories, selected based on popularity and availability at the time of data collection. The selection deliberately spans the capability and style spectrum: all eight are publicly deployed systems used in production, not research prototypes. Additionally, including models across a range of output quality provides a comprehensive audit of what users actually encounter. Experiments are conducted on the LUMI supercomputer \footnote{https://lumi-supercomputer.eu/} (LUMI-G partition), hosted at CSC, Kajaani, Finland. Each compute node is equipped with 4× AMD Instinct MI250X GPUs (128 GB HBM2e memory per GPU, 3.2 TB/s memory bandwidth), a single 64-core AMD EPYC 7A53 ``Trento'' CPU, and 512 GB DDR4 RAM. Nodes are interconnected via HPE Cray Slingshot-11 at 800 Gbit/s per node. Job scheduling is managed by SLURM. Experiments used Python 3.12, PyTorch 2.5.1, and ROCm 6.2.3, run inside a Singularity container. All experiments use a fixed random seed of 1024 for reproducibility. Full dependency versions are provided in the accompanying code repository.

\subsection{Emotion Prompts}
We generated synthetic faces using the prompt template:
\texttt{``A photorealistic portrait of a \textit{[emotion]} person, front-facing.''}
where \texttt{[emotion]} was replaced by one of six canonical emotions: Happy, Sad, Angry, Surprised, Disgusted, and Fearful. A neutral baseline was created by omitting the emotion slot. We selected this minimal template after iterative piloting. Holding all prompt elements constant except the emotion token helps isolate demographic differences across conditions, reducing confounding from syntactic variation or co-occurring lexical content. 

We assess robustness to prompt wording using four complementary analyses: (1) the same valence-driven pattern appears across the eight architecturally diverse models spanning two development ecosystems; (2) near-synonym prompts (e.g., ``sad'' vs.\ ``unhappy'') yield similar demographic distributions (Supplementary~\supp{supp:unhappy}); (3) Chinese-language prompts reproduce the same bias patterns in Chinese models (Supplementary~\supp{supp:chinese}); and (4) Western and Chinese models converge on a similar valence-based mapping (Section~\ref{subsec:rq2}). These results provide convergent evidence that the observed patterns are not driven by idiosyncratic prompt wording but persist across prompt variants, languages, and model families.

The six emotions are grounded in the basic-emotion framework \cite{ekman1992argument,ekman1993facial} and in the circumplex model of affect \cite{russell1980circumplex}, which organizes emotional states along valence and arousal dimensions. Regarding valence, happiness is positive; sadness, anger, fear, and disgust are negative; and surprise is ambiguous, depending on context. The six emotions display recognizable facial configurations across cultures and are consistently annotated in major emotion recognition datasets (CK+ \cite{kanade2000comprehensive}, RAF-DB \cite{li2017reliable}, FER+ \cite{barsoum2016training}, AffectNet \cite{mollahosseini2017affectnet}), providing a standardized and widely comparable basis for evaluation.

Note that our study measures \emph{which demographic groups} are generated in response to emotion prompts, not whether the resulting faces would be accurately or consistently perceived as expressing those emotions across cultures. Our analysis is therefore independent of debates on the universality of facial affect.

\subsection{Image Generation}
\label{sec:methods.imageGeneration}
For every model and emotion condition, we generated 1,000 images, yielding 56,000 faces in total (8~models $\times$ 7~conditions $\times$ 1,000). Bootstrap 95\% confidence intervals on almost all reported TVD and JS values have widths $<$0.03 (Supplementary~\supp{supp:divergence}), confirming that this sample size is sufficient to reliably detect the bias magnitudes we report. All models were run with default inference hyperparameters. No seed was fixed so as to reflect typical deployment diversity.

Face detection was performed using the dlib-based detector \cite{king2009dlib} bundled with the FairFace model \cite{karkkainenfairface}. Images in which no face was detected were discarded and replaced with newly sampled generations, so that each mode-condition cell retains exactly 1{,}000 valid face images for analysis. The per-model, per-condition rejection rates are reported in Table~\ref{tab:rejection_rate}. Across all models, Fear consistently yields the highest rejection rates, because several models frequently generate images of a person covering their face with both hands, resulting in heavy occlusion that prevents reliable face detection. Cartoon-style or otherwise non-photorealistic outputs are implicitly excluded by this pipeline, as they fail the dlib face-detection step and are therefore never included in the analyzed sample.

\subsection{Attribute Estimation}\label{subsec:attr}
Demographic attributes (gender, race, and age) were extracted from each generated face using FairFace \cite{karkkainenfairface}. We use FairFace for demographic annotation because it is widely adopted in prior work on face attribute analysis and has been shown to perform robustly on diverse sets of both real and synthetic faces. Reported performance includes 97\% accuracy for gender, 90\% for race, and 73\% average accuracy for age (composite across three age datasets; see below), consistent with evaluations on synthetic faces from the eight audited models (Fig.~\ref{fig:syn_cm}). 
In addition, recent comparative analyses of face-attribute classifiers used in T2I auditing suggest that FairFace exhibits low demographic disparities across groups \cite{doh2026facct}, reducing the risk that classifier bias could confound measured generative bias. 
Facial attractiveness was assessed using using an ensemble of four computer vision models trained on three different facial attractiveness datasets (AHEAD \cite{gulati2024beautiful}, CelebA \cite{liu2015faceattributes}, and SCUT-FBP5500 \cite{liang2018scutfbp5500}). The individual models in the ensemble exhibited comparable performance on real images. However, the ensemble outperformed each of its constituent models on synthetic images. We therefore estimate the attractiveness of the synthetically generated faces by means of the ensemble. Details regarding model selection and evaluation are provided in Supplementary~\supp{supp:attr}. 

\paragraph{Gender, Race and Age Estimation}
We evaluated FairFace on the CFD dataset \cite{ma2015chicago} for gender and race classification, and on FACES \cite{ebner2010faces}, APPA-REAL \cite{agustsson2017appareal}, and FGNET \cite{ranking2016PAMI} for age estimation (full confusion matrices in Supplementary~\supp{supp:attr}). FairFace achieves 97.0\% accuracy for gender and 90.0\% for race; age accuracy is 86\% on the FACES dataset and lower on APPA-REAL and FGNET, which include children and adolescents not present in FACES (composite across the three datasets: 73.0\%), with misclassifications bounded to adjacent age groups.

\paragraph{Synthetic-face validation}
To validate FairFace directly on the images it is applied to, we generated a controlled set using the eight audited models: 100 images per combination of gender, race, and age group using explicit attribute prompts (\emph{e.g.}, \texttt{``a photorealistic portrait of a young Asian female''}). Age was specified with the keywords \textit{girl/boy} (0--9), \textit{teenager} (10--19), \textit{young} (20--39), \textit{middle-aged} (40--59), and \textit{old} (60+). FairFace predictions were compared against the prompt-specified ground truth.

The resulting confusion matrices (Figure~\ref{fig:syn_cm}) show that age estimation errors are largely confined to adjacent categories and tend to average out at the aggregate level. The 60+ group, which is central to our age-related findings, is classified with comparatively high accuracy. Asian and White faces are reliably distinguished across all eight models, and gender classification is near-perfect.

Emotion-conditioned validation across 4{,}000 faces per emotion condition (Figure~\ref{fig:syn_emotion}) confirms that these properties are stable across all six emotions: gender recall remains high ($\geq 0.97$), Asian--White confusions remain low ($\leq 0.01$, except $0.02$ under Happiness), and cross-group age confusions (young, middle-aged, old) remain small and consistent across conditions.

Overall, we do not observe systematic classifier error patterns that align with or could plausibly account for the demographic shifts reported in our main results. Full validation results are provided in Supplementary~\supp{supp:attr}.

\paragraph{Binary gender classification}

FairFace produces a binary gender classification (male/female), consistent with prior algorithmic audits of T2I systems \cite{buolamwini2018gender,Wu_2025_GenderBiasStableDiffusion,AlDahoul_2024_AIgeneratedFacesStereotypes}. Because the images under analysis are synthetically generated and do not correspond to real individuals, this measurement captures how gender is \emph{visually encoded in model outputs}, \emph{i.e.}, the proportion of male-coded versus female-coded faces produced under a given prompt, rather than the gender identity of any person depicted. Systematic asymmetries in these proportions are interpreted as representational bias in the model’s outputs.

\paragraph{Four-class race taxonomy}

Initially, we considered five race categories: White, Black, Asian (East and Southeast Asian combined), Indian, and Latino/Hispanic. However, the confusion matrices on the CFD dataset (Supplementary~\supp{supp:attr}, Figure~\ref{fig:cm}) revealed substantially lower classification performance for the Latino/Hispanic and Indian categories compared to the other groups. Retaining these categories would therefore risk propagating higher classifier uncertainty into downstream bias estimates.

We thus merged categories into four groups: White, Black, Asian, and Others (combining Indian, Latino/Hispanic, and Middle Eastern faces). This aggregation improved classification stability across groups and is consistent with prior work adopting coarser-grained racial groupings in fairness analyses \cite{buolamwini2018gender,mehrabi2021survey}. Furthermore, the main conclusions regarding racial disparities revealed in our analyses remain unchanged under this grouping. 

We conducted two follow-up studies to evaluate the distinguishability of the race categories grouped into ``Others'': a human-annotation agreement study measuring inter-rater agreement on faces from these groups relative to a White/Black/Asian control sample, and an embedding-space separability study for these groups. The human-rater study ($N=19$) found that raters correctly identified White, Black, and Asian faces with accuracies ranging from 89\% to 97\%, but frequently confused Latino, Indian, and Middle Eastern faces with one another (Figure~\ref{fig:humanRaceStudy_supp}). The embedding-space study found that Indian, Latino, and Middle Eastern face embeddings form more overlapping, less separable clusters than White, Black, and Asian embeddings in the FairFace representation space (Figure~\ref{fig:fairface_embeddings}). Both studies are described in detail in Supplementary~\supp{supp:attr.humanStudy}.

These results indicate that the four-class collapse reflects a structural property of visual-only demographic inference for these categories, rather than a deficiency specific to the classifier we used. 
Nonetheless, we acknowledge that grouping these four categories limits the representation of groups that are already subject to persistent discrimination. 
Following Khan \& Fu \cite{khan2021one} and Doh et al. \cite{dohposition}, we recognize that racial taxonomies are socially constructed, historically contingent, and limited in global applicability. We therefore use this four-category scheme as a measurement device to capture broad representational patterns in synthetic outputs, rather than as a claim about intrinsic or universal human categories. 

Adopting a continuous skin-tone or ethnicity representation (\emph{e.g.}, the Fitzpatrick scale) or colorism-oriented analyses, in place of discrete categories, remains a complementary direction for future work (see Section \ref{sec:discussion}).

\paragraph{Attractiveness estimation}
Perceived facial attractiveness is not a purely aesthetic property: via the halo effect, faces judged as more attractive also tend to receive more positive attributions of competence, trustworthiness, and social desirability \cite{nisbett1977halo,todorov2008,gulati2024beautiful}. 
In the context of generative AI---where synthetic faces are widely used in advertising, stock imagery, and journalism---systematic associations between emotion prompts and perceived attractiveness may influence which demographic groups are represented as more or less socially desirable. This makes attractiveness an important dimension of bias with potential downstream social implications \cite{gulati2024lookism}.

Few datasets provide attractiveness ratings. CelebA \cite{liu2015faceattributes} and SCUT-FBP5500 \cite{liang2018scutfbp5500} are the most widely used, but both have low demographic diversity. Additionally, SCUT-FBP5500's ratings come from a small pool of 60 human ratings, each providing feedback of over 5,000 faces, which limits rater diversity and increases the likelihood of rater fatigue. AHEAD \cite{gulati2025thesis} is, to our knowledge, the only face attractiveness dataset that jointly spans age, gender, and race diversity. To assess attractiveness, we evaluated four different classifiers. First, a ResNet-50 trained on the training split of the AHEAD dataset. Second, three state-of-the-art models that had been pretrained on CelebA and SCUT-FBP5500: a ViT (CelebA) \cite{celeba_attr_model}, a SEResNeXt50 with ComboLoss (SCUT-FBP5500) \cite{comboloss_attrmodel}, and a ResNet-18 regression model (SCUT-FBP5500) \cite{scut_r18_attrmodel}. We finetuned these three alternatives on the training split of the AHEAD dataset, replacing each output layer with a new three-class (low/medium/high) head that matches AHEAD's rating scheme. As a result, the four classifiers were trained/finetuned and evaluated on the same demographically diverse data. Finally, we built an ensemble with the four models and compared its performance to that of each individual classifier.

On a held-out AHEAD validation set, the ensemble reached 82.1\% accuracy (macro-F1 0.815), within 1.2 percentage points of the best-performing single classifier (ViT, 83.3\%). Its performance was comparable to that of SEResNeXt50 (82.1\%), and exceeded that of ResNet-50 (80.8\%) and ResNet-18 (76.9\%). Additionally, we evaluated the four individual classifiers and their ensemble on 800 synthetic faces generated by the eight audited models, using prompts containing either ``attractive'' or ``plain'' to assess their performance on the target data. The ensemble achieved the best performance across faces generated with the eight models. We therefore selected the ensemble for all subsequent analyses of facial attractiveness. The details about the fine-tuning and evaluation of the models can be found in  Supplementary~\supp{supp:attr}, Table~\ref{tab:altattr_validation} and Table~\ref{tab:scorer_comparison}.

We further assessed potential demographic variation in the performance of the attractiveness ensemble classifier using the AHEAD dataset. Results are reported in Supplementary~\supp{supp:attr}. We observe no significant differences in classification performance across racial groups. In contrast, performance varies significantly across age groups ($\chi^2, p < 0.01$), with the highest error rate observed for middle-aged faces (24\%), followed by older (19\%) and younger (8\%) faces. We also find a significant difference across gender groups ($\chi^2$, $p < 0.001$): the attractiveness ensemble misclassifies 22.7\% of female faces compared with 9.3\% of male faces, suggesting a higher error rate on female faces. Note that we quantify attractiveness shifts using total variation distance (TVD) between paired distributions, with each condition compared against the neutral baseline within the same demographic group. This approach makes the analysis robust to monotonic calibration shifts in the classifier output. We therefore expect the observed differences in classifier error rates across age and gender groups to have minimal impact on the reported attractiveness shifts. Further details on the attractiveness classifier and its validation are provided in Supplementary~\supp{supp:attr}.

\subsection{Reference Distributions}

Bias was assessed relative to two complementary reference distributions.

\paragraph{Global population baseline}
Model outputs were compared against 2024 global population statistics from the United Nations World Population Prospects database (\url{https://population.un.org/wpp/}). Gender and age distributions were derived directly from UN tables. For race, no direct global statistics exist, so we followed the country-to-race assignment strategy used in prior demographic fairness work \cite{mehrabi2021survey,raji2019actionable,buolamwini2018gender}: each country's population was assigned to one racial group and aggregated. The racial categories were derived from UN regional groupings and then mapped to the four-class taxonomy used in the study: \textit{White} (Europe and Northern America); \textit{Black} (Sub-Saharan Africa); \textit{Asian} (Eastern Asia and South-Eastern Asia); and \textit{Others} (India, Latin America and the Caribbean, and remaining groups). Each grouping corresponds to an explicit column in the UN table. For multi-racial countries such as the United States and Canada, official national census statistics \cite{tian2025countingrace, hou2023changing} were incorporated to approximate the racial breakdown, which substantially improves accuracy over single-group assignment.

\paragraph{Neutral-prompt baseline}
The second reference is the model's own output under the neutral (no-emotion) prompt. Comparing emotion-conditioned distributions against this baseline isolates the demographic effect attributable specifically to the emotion prompt, factoring out each model's inherent generative prior that we measure with the neutral expression prompt. 

\subsection{Bias Metrics}
Standard classification-fairness criteria (demographic parity, equalized odds) do not apply to generative systems. We therefore use information-theoretic divergence measures --- Kullback-Leibler (KL) divergence, Jensen-Shannon (JS) divergence, and Total Variation Distance (TVD) --- which directly quantify deviations between probability distributions and handle continuous attributes such as age \cite{cascone2025framework}. We assess bias through three layers: \emph{baseline bias} (neutral-prompt output vs.\ global population statistics), \emph{intersectional bias} (over- and underrepresented attribute combinations), and \emph{emotion-conditioned bias} (distributional shift induced by each emotion prompt relative to the neutral baseline).

\paragraph{Individual demographic attributes}

Let $P_{\mathrm{world}}(d)$ denote the real-world distribution of a demographic attribute $d$ (gender, race, or age), and let $P_{\mathrm{T2I}}(d)$ denote the distribution of synthetic faces generated under neutral prompts. Bias is quantified using the Kullback--Leibler (KL) divergence:

\begin{equation}
D_{\mathrm{KL}}\!\left(P_{\mathrm{world}}(d) \parallel P_{\mathrm{T2I}}(d)\right)
=
\sum_{i \in \mathcal{C}_d}
P_{\mathrm{world}}(d_i)
\log
\frac{P_{\mathrm{world}}(d_i)}{P_{\mathrm{T2I}}(d_i)},
\end{equation}

\noindent where $\mathcal{C}_d$ denotes the set of categories of attribute $d$, computed independently for gender, race, and age. As KL divergence is unbounded, to provide bounded, symmetric complements we also compute the Jensen--Shannon (JS) divergence $D_{\mathrm{JS}} \in [0,1]$ and the total variation distance (TVD) $\in [0,1]$:

\begin{equation}
D_{\mathrm{JS}}\!\left(P_{\mathrm{world}}(d), P_{\mathrm{T2I}}(d)\right)
=
\frac{1}{2}
D_{\mathrm{KL}}\!\left(P_{\mathrm{world}}(d) \parallel M_d\right)
+
\frac{1}{2}
D_{\mathrm{KL}}\!\left(P_{\mathrm{T2I}}(d) \parallel M_d\right),
\end{equation}

\noindent where $M_d = \tfrac{1}{2}\!\left(P_{\mathrm{world}}(d) + P_{\mathrm{T2I}}(d)\right)$, and

\begin{equation}
\mathrm{TVD}\!\left(P_{\mathrm{world}}(d), P_{\mathrm{T2I}}(d)\right)
=
\frac{1}{2}
\sum_{i \in \mathcal{C}_d}
\left|
P_{\mathrm{world}}(d_i) - P_{\mathrm{T2I}}(d_i)
\right|.
\end{equation}

\paragraph{Intersectional shift}

The intersectional analysis examines the joint distribution over $k=3$ demographic attributes (gender, race, age). Let $\mathbf{d} = (d_1, d_2, d_3)$ represent a specific intersectional configuration (\emph{e.g.}, female $\times$ Asian $\times$ 60+) drawn from the Cartesian product $\mathcal{C}_{d_1} \times \mathcal{C}_{d_2} \times \mathcal{C}_{d_3}$ of all category sets.

For emotion-conditioned analysis, no global joint statistics exist across all three attributes simultaneously, so we compare the joint distribution under emotion $e$, $P_{\mathrm{T2I}}(\mathbf{d} \mid e)$, against the neutral-prompt baseline $P_{\mathrm{T2I}}(\mathbf{d} \mid e_0)$. The emotion-induced intersectional shift is:

\begin{equation}
    \label{eq:kl_intersectional}
D_{\mathrm{KL}}^{\mathrm{intersectional}}(e \,\|\, e_0)
=
\sum_{\mathbf{d} \in \mathcal{C}_{d_1} \times \cdots \times \mathcal{C}_{d_k}}
P_{\mathrm{T2I}}(\mathbf{d} \mid e)
\log
\frac{P_{\mathrm{T2I}}(\mathbf{d} \mid e)}
     {P_{\mathrm{T2I}}(\mathbf{d} \mid e_0)} .
\end{equation}

\noindent This divergence quantifies how an emotion prompt redistributes demographic mass across intersectional configurations, capturing compounded shifts invisible in marginal attribute analyses. We report the symmetric JS analogue alongside the KL:

\begin{equation}
D_{\mathrm{JS}}^{\mathrm{intersectional}}(e, e_0)
=
\frac{1}{2}
D_{\mathrm{KL}}\!\left(
P_{\mathrm{T2I}}(\mathbf{d} \mid e)
\,\middle\|\,
M(\mathbf{d})
\right)
+
\frac{1}{2}
D_{\mathrm{KL}}\!\left(
P_{\mathrm{T2I}}(\mathbf{d} \mid e_0)
\,\middle\|\,
M(\mathbf{d})
\right),
\end{equation}

\noindent where $M(\mathbf{d}) = \tfrac{1}{2}\!\left(P_{\mathrm{T2I}}(\mathbf{d} \mid e) + P_{\mathrm{T2I}}(\mathbf{d} \mid e_0)\right)$.

\paragraph{Emotion-conditioned bias}

To identify which demographic categories drive an emotion-induced shift, we define the per-category KL contribution. Let $d$ be a single attribute with categories $\mathcal{C}_d$, and let $P_{\mathrm{T2I}}(c \mid e)$ and $P_{\mathrm{T2I}}(c \mid e_0)$ denote the probability of category $c$ under emotion $e$ and the neutral prompt $e_0$, respectively. The per-category KL contribution is:

\begin{equation}
\mathrm{KL}_{d}(c \mid e, e_0)
=
P_{\mathrm{T2I}}(c \mid e)\,
\log
\frac{P_{\mathrm{T2I}}(c \mid e)}
     {P_{\mathrm{T2I}}(c \mid e_0)} .
\end{equation}

\noindent Summing over all $c \in \mathcal{C}_d$ recovers the standard KL divergence between the emotion-conditioned and neutral distributions. A larger per-category value indicates a stronger emotion-induced shift for that group, enabling direct comparison across demographic categories and emotions.

\section{Results}
\label{sec:results}

\subsection{All T2I Models Misrepresent Global Demographics}
\label{sec:results.baseline}

\begin{figure*}[ht]
    \centering
    \includegraphics[width=1\linewidth]{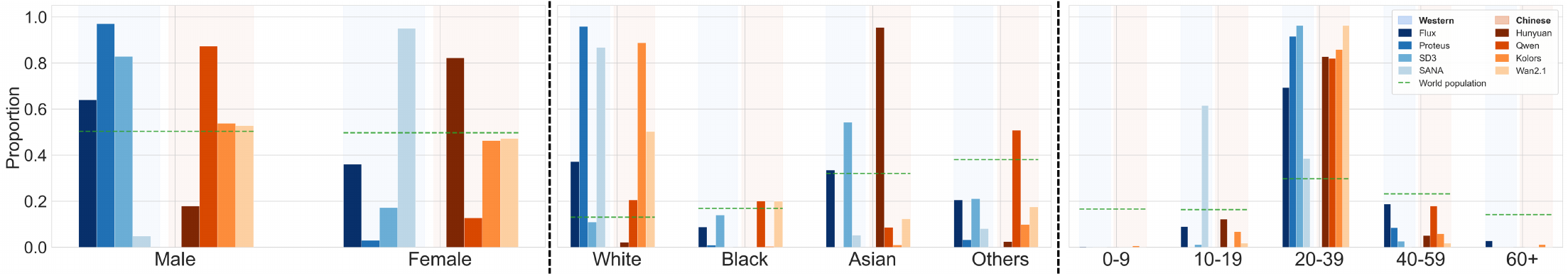}
    \caption{Individual demographic distributions of neutral faces generated by the audited models (Western models in blue and Chinese models in red). From left to right, the graphs show the distributions of gender, race, and age, compared against the corresponding real-world population estimates, marked by green dashed horizontal lines.} 
    \label{fig:netural_dist}
\end{figure*}

\paragraph{Individual demographic attributes} 
Figure~\ref{fig:netural_dist} summarizes the distributions of gender, race and age in the generated faces by the eight T2I models compared to real-world population estimates (green dashed lines). 
All eight models exhibit clear and systematic deviations from real-world population statistics. In terms of gender (left graph), the real-world baseline is close to parity, yet most models generate faces with a pronounced gender imbalance. Specifically, four models (FLUX, Proteus, SD3 and Qwen) substantially overrepresent male faces, with male proportions exceeding 60\%–90\% of the generated samples, while two models (SANA and Hunyuan) predominantly generate female faces, accounting for over 80\% of the samples. Only two models (Kolors and Wan2.1) generate a relatively balanced gender distribution.

Regarding race (middle graph), the disparities with the reference distributions are even larger. Three models (Proteus, SANA and Kolors) overwhelmingly generate White faces, with percentages above 80–90\%, despite the real-world proportion being around 11\%. Asian and Black populations are consistently underrepresented by nearly all models, whereas the “Others’’ category appears only minimally except in the case of Qwen. These results highlight a substantial racial imbalance, with all evaluated models producing distributions far from global demographic baselines. Even excluding Proteus (the model with the largest bias) the remaining seven models show race TVD values ranging from 0.23 to 0.74 (Supplementary~\supp{supp:divergence}), all in the substantial-to-extreme bias range. Thus, the core finding does not depend on any single model.

A similar pattern emerges in the age distribution of the generated faces (right graph). All models heavily favor the 20–39 age group, exceeding 80\% of all generated faces in 6 models, while severely underrepresenting children (0–9), adolescents (10–19), middle-aged adults (40–59), and especially older adults (60+). This pronounced skew toward young adults is consistent across all  models. A notable case is SANA, which has a disproportionately large probability ($>60\%$) of generating faces of very young (10-19) White females.

All but one (Wan2.1 for gender) of the model--attribute combinations significantly deviate from global population distributions ($\chi^2$ goodness-of-fit, $p < 0.05$; Supplementary~\supp{supp:chisquare}), with all TVD values falling in the moderate-to-extreme bias range except gender for Kolors and Wan2.1. Figure~\ref{fig:divergence_metrics} shows KL, JS, and TVD with 95\% bootstrap confidence intervals per model and attribute. Exact values are tabulated in Supplementary~\supp{supp:divergence}.

\subsubsection{Intersectional Analysis}
\label{sec:results.baseline.intersectional}
\begin{figure*}[ht]
    \centering
\includegraphics[width=1\linewidth]{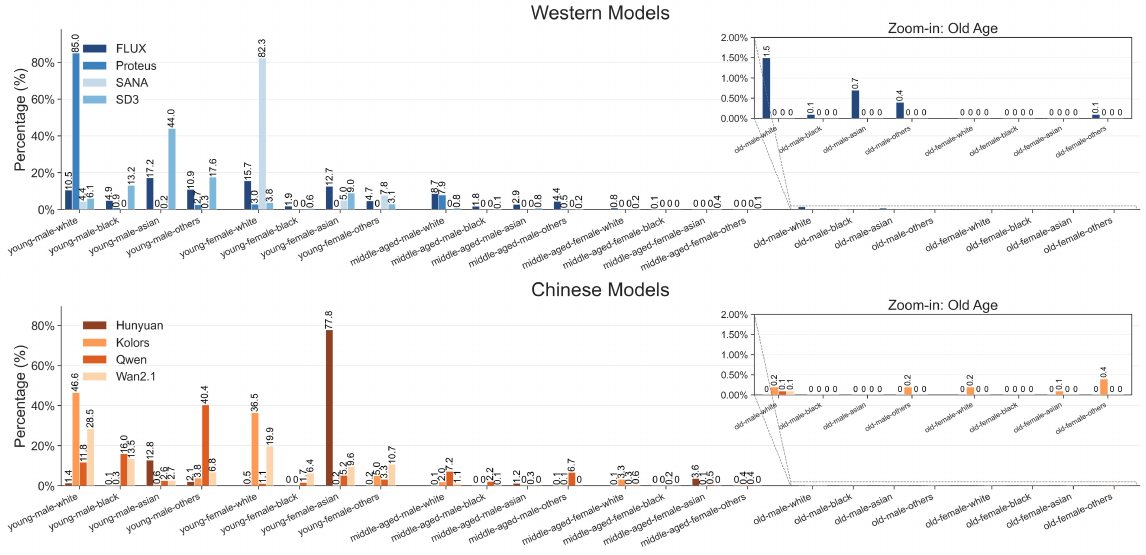}
    \caption{Intersectional Age–Gender–Race distributions of the faces generated by Western (top) and Chinese (bottom) models.}
    \label{fig:inter}
\end{figure*}

Next, we analyze the joint distribution of all attributes to study intersectional biases in the synthetically generated faces.
Figure~\ref{fig:inter} depicts the joint intersectional distributions over age, gender and race of the faces generated by each model using the emotion-neutral prompt. To improve interpretability and reduce sparsity in age-related statistics, the age categories of 0–9, 10–19, and 20–39 are aggregated into a single \emph{young} category, ages 40–59 into a \emph{middle-aged} category, and ages 60 and above into an \emph{old} category. The gender (2 groups) and race (4 groups) categories are unchanged. With the collapsed 3-level age taxonomy, there are $2 \times 4 \times 3 = 24$ intersectional cells per model.

The generated outputs show a strong concentration of probability mass on a small subset of demographic combinations: on average, only 7 of 24 intersectional cells (29\%) have non-negligible representation. We summarize two main patterns.

\textbf{(1) High proportion of near-zero intersectional cells.}
For each model, we define near-zero representation as $P(\mathbf{d}) < 0.01$ (\emph{i.e.}, fewer than 10 out of 1,000 samples in a given cell). Across models, 16.5 $\pm$ 3.3 of 24 cells (69\%) fall into this category, ranging from 11 (FLUX) to 20 (Proteus and SANA) (Supplementary~\supp{supp:intersectional}). Cells involving middle-aged and older individuals are consistently underrepresented across models, though not uniformly across gender and race(Table~\ref{tab:least_represented}). Among the eight most severely underrepresented cells (mean $P(\mathbf{d}) < 0.001$), seven involve the \emph{old} age group, with a majority of females (5/8) and a mixed distribution of racial groups (3 Black, 2 Asian, 2 Others and 1 White). Age, rather than race, seems to be the dominant axis of erasure at the extreme tail of the intersectional distribution.

\textbf{(2) Compound underrepresentation of young $\times$ female $\times$ Black faces.}
The young $\times$ female $\times$ Black (YFB) intersection is consistently rare, despite the fact that ``young" and ``female" are well represented (see Table~\ref{tab:top5_regional}), indicating limitations of marginal-only analyses. To quantify this pattern, we compare observed frequencies to an independence-based baseline: $
P_{\text{exp}} = P_{\text{world}}(\text{young}) \cdot P_{\text{world}}(\text{female}) \cdot P_{\text{world}}(\text{Black}) \approx 0.0527 (5.27\%)$.

Four of eight models assign 0\% probability to the YFB  intersection, and no model reaches the world-population baseline. The mean observed rate across models is 1.3\% (95\% bootstrap CI: [0.21\%, 2.93\%]), corresponding to a suppression factor of 0.25$\times$, with the entire interval below the baseline. As an illustration, FLUX produces 36\% female faces and 8.8\% Black faces (values that would not be flagged in marginal audits), but only 1.9\% YFB. Table~\ref{tab:yfb} in Supplementary~\supp{supp:intersectional} reports model-level results.

We highlight YFB specifically because it illustrates erasure that is invisible to marginal-attribute audits, yet it is not necessarily the numerically smallest intersectional cell. Several cells combining rarer age and race categories (\emph{e.g.}, old $\times$ female $\times$ Black, old $\times$ female $\times$ Asian; Table~\ref{tab:least_represented}) show equal or more
extreme near-zero representation, but their scarcity is unsurprising given that each of their constituent marginals is already rare in the generated outputs. YFB is diagnostic precisely because its marginals (young, female) are individually well represented (Table~\ref{tab:top5_regional}), so single attribute audits would not flag it, yet the joint cell has almost zero representation.
Moreover, the underrepresentation of YFB is an illustration of historically documented compound erasure that echoes a documented failure mode in face-analysis systems: Buolamwini and Gebru's Gender Shades audit \cite{buolamwini2018gender} found that commercial face-analysis systems perform worst on Black women, the same population that \emph{DeGraffenreid v.\ General Motors} had rendered legally invisible to single-axis discrimination doctrine and that motivated the intersectionality framework that we base our work on \cite{crenshaw2013}. Thus, YFB's near-erasure from generative outputs suggests that this specific, previously documented pattern of exclusion recurs in generative AI models. 

Table~\ref{tab:yfb-bias} reveals two different patterns of failure:

\textbf{(1) Collapse driven by marginal race deficits.}
In the first cluster---SANA ($P_m(\text{Black}) = 0\%$), Hunyuan ($0.1\%$), Kolors ($0.3\%$), and Proteus ($0.9\%$)---YFB is effectively absent because Black faces are rarely generated in these models overall. In these cases, the absence of the intersectional cell is consistent with marginal race distributions and would be detectable using single-attribute race audits.

\textbf{(2) Intersectional suppression beyond marginals.}
In the second cluster---FLUX, SD3, and Qwen---$P_m(\text{Black})$ is non-zero ($8.8\%$, $13.9\%$, and $20.1\%$ respectively), yet observed YFB rates ($1.9\%$, $0.6\%$, and $1.7\%$) remain below the model-independence baselines ($2.49\%$, $2.33\%$, and $2.10\%$). This indicates a deviation from the independence expectation that is not captured by single-attribute analyses and emerges from the intersectional analysis.

Wan2.1 is the only model to exceed the world-population baseline ($6.40\%$ vs.\ $5.27\%$). This outcome coincides with relatively high marginal representation of Black faces ($20.0\%$) and a near-balanced gender distribution ($47.2\%$ female), which together yield an independence expectation close to the global reference rate.

In the case of models with non-zero YFB rates, the observed suppression (mean 1.3\% vs.\ 5.27\% expected) is substantially below the world baseline. For cases with zero observed YFB, the result reflects complete absence of samples in the corresponding intersectional cell rather than measurement noise. For non-zero cases, the magnitude of the deviation is larger than what would be expected from known marginal classifier error rates alone (see Supplementary~\supp{supp:attr}).

\begin{table}[ht]
\centering
\caption{Intersectional bias analysis: Percentage of young $\times$ female $\times$ Black (YFB) faces across image generation models. Almost all models generate fewer YFB faces than the world-population baseline ($P_{\text{world}} = 0.627 \times 0.497 \times 0.169 = 5.27\%$).}
\label{tab:yfb-bias}
\begin{adjustbox}{max width=\textwidth}
\renewcommand{\arraystretch}{1.2}
\begin{tabular}{
  l
  S[table-format=3.2, table-space-text-post=\%]
  S[table-format=3.2, table-space-text-post=\%]
  S[table-format=2.2, table-space-text-post=\%]
  S[table-format=2.2, table-space-text-post=\%]
  S[table-format=2.2, table-space-text-post=\%]
  S[table-format=2.2, table-space-text-post=\%]
}
\toprule
\textbf{Model}
  & \textbf{{$P_m(\text{young})$}}
  & \textbf{{$P_m(\text{female})$}}
  & \textbf{{$P_m(\text{Black})$}}
  & \textbf{{YF base}}
  & \textbf{{YFB base}}
  & \textbf{{Obs.\ YFB}} \\
\midrule
\cellcolor{lightblue}
Flux    & {78.50\%} & {36.00\%} & {8.80\%}  & {28.26\%} & {2.49\%} & {1.90\%} \\
\cellcolor{lightblue}
Proteus & {91.60\%} & {3.00\%}  & {0.90\%}  & {2.75\%}  & {0.02\%} & {0.00\%} \\
\cellcolor{lightblue}
SD3     & {97.40\%} & {17.20\%} & {13.90\%} & {16.75\%} & {2.33\%} & {0.60\%} \\
\cellcolor{lightblue}
SANA    & {100.00\%}& {95.10\%} & {0.00\%}  & {95.10\%} & {0.00\%} & {0.00\%} \\
\midrule
\cellcolor{red!10}
Hunyuan & {94.90\%} & {82.20\%} & {0.10\%}  & {78.01\%} & {0.08\%} & {0.00\%} \\
\cellcolor{red!10}
Qwen    & {82.10\%} & {12.70\%} & {20.10\%} & {10.43\%} & {2.10\%} & {1.70\%} \\
\cellcolor{red!10}
Kolors  & {93.00\%} & {46.20\%} & {0.30\%}  & {42.97\%} & {0.13\%} & {0.00\%} \\
\cellcolor{red!10}
Wan2.1  & {98.10\%} & {47.20\%} & {20.00\%} & {46.30\%} & {9.26\%} & {6.40\%} \\
\bottomrule
\end{tabular}
\end{adjustbox}
\par\smallskip
{\footnotesize \textit{Note:} YF base $= P_m(\text{young}) \times P_m(\text{female})$;
YFB base $= P_m(\text{young}) \times P_m(\text{female}) \times P_m(\text{Black})$,
both under independence assumption.}
\end{table}

\subsection{Western and Chinese Models Converge on Similar Biases}\label{subsec:rq2}

Figure~\ref{fig:netural_dist} shows no clear or systematic differences in the output distributions of Western vs.\ Chinese models: in both groups, deviations from global statistics are model-specific rather than region-dependent (see Figure~\ref{fig:divergence_metrics} and Table~\ref{tab:divergence_compact}). We formally quantify this homogenization with two metrics:

\begin{table}[ht]
\centering
\caption{Jensen--Shannon (JS) divergence between the aggregate distributions of Western (W) and Chinese (C) models and within-group mean pairwise JS, under the neutral condition. The last row reports the same quantities at the intersectional (joint gender--race--age) level.}
\label{tab:js_between_regions}
\begin{tabular}{lc|cc}
\toprule
\textbf{Attribute}  & \textbf{JS (W vs C)} & \textbf{Within W} & \textbf{Within C}  \\
\midrule
Gender  & 0.0063 & 0.3025 & 0.1334  \\
Race    & 0.0229 & 0.3059 & 0.4640  \\
Age     & 0.0309 & 0.2194 & 0.0487  \\
\hline
Intersectional (joint) & 0.0634 & 0.3887 & 0.3681  \\
\bottomrule
\end{tabular}
\end{table}

\textbf{(1) Between-group JS divergence.} We compute the aggregate distribution for each region by averaging the output distributions of the four models within that group. We denote the resulting mean distribution for the Western models as $\bar{P}_W$ and for the Chinese models as $\bar{P}_C$. Table~\ref{tab:js_between_regions} reports $D_{\mathrm{JS}}(\bar{P}_W, \bar{P}_C)$ for each marginal attribute, together with the within-group mean pairwise JS across all $\binom{4}{2}=6$ pairs within each region. The between-group JS is $0.006$ (gender), $0.023$ (race), and $0.031$ (age)---in the three cases an order of magnitude smaller than the within-group mean pairwise JS for Western models  and Chinese models. Within these eight models, regional origin therefore accounts for far less of the observed variation in outputs than individual model differences within each group.

\textbf{(2) Between-group JS divergence (intersectional).} The marginal comparisons above treat gender, race, and age separately. At the intersectional level---the joint distribution over all three attributes simultaneously---the same conclusion holds.

The last row of Table~\ref{tab:js_between_regions} shows that the JS divergence between the aggregate Western and Chinese intersectional distributions ($D_{\mathrm{JS}} = 0.0634$) is substantially smaller than the within-group mean pairwise JS for Western models ($0.3887$) and Chinese models ($0.3681$), confirming that, across the eight audited models, regional origin accounts for less of the observed variation in intersectional outputs than individual model differences within each group. Full pairwise results are provided in Table~\ref{tab:js_within}, Supplementary~\supp{supp:intersectional}.

Beyond aggregate metrics, the race distributions show limited systematic regional differentiation: despite China’s predominantly East Asian population, most Chinese models still overrepresent White faces, broadly similar to Western models. Hunyuan is an exception, generating a high proportion of Asian faces (approximately 90\%) relative to other groups.
To examine whether this pattern is driven by prompt language, we additionally evaluate the Chinese models using Chinese-language prompts (Supplementary~\supp{supp:chinese}). The observed biases persist across both English and Chinese inputs. For Hunyuan, Qwen, and Wan2.1, the proportion of generated White faces \emph{increases} under Chinese prompts than under English prompts.

These results suggest that prompt language alone is unlikely to account for the observed racial distributions, and that the underlying generative prior may play a stronger role in shaping outputs.

Despite this overall homogenization under English prompts, one model reveals a language-specific deviation. Kolors generates only 0.30\% female faces under Chinese-language prompts (95\% bootstrap CI: [0.00\%, 0.70\%]), compared to 46.20\% under English prompts. This effect is not observed in Hunyuan, Qwen, or Wan2.1 under the same Chinese prompts (Supplementary~\supp{supp:chinese}). The Chinese prompts used in this evaluation are gender-neutral, containing no explicit gender markers and only a generic reference to a person. The effect is therefore specific to a single model rather than a general property of Chinese-language prompting.
The finding suggests a model-specific interaction between language encoding and demographic generation in Kolors, leading to a substantial shift in gender representation under Chinese prompts that is not apparent under English evaluation.

\subsubsection{Region-Specific Demographic Comparison}

\begin{figure*}[t]
    \centering
    \includegraphics[width=1\linewidth]{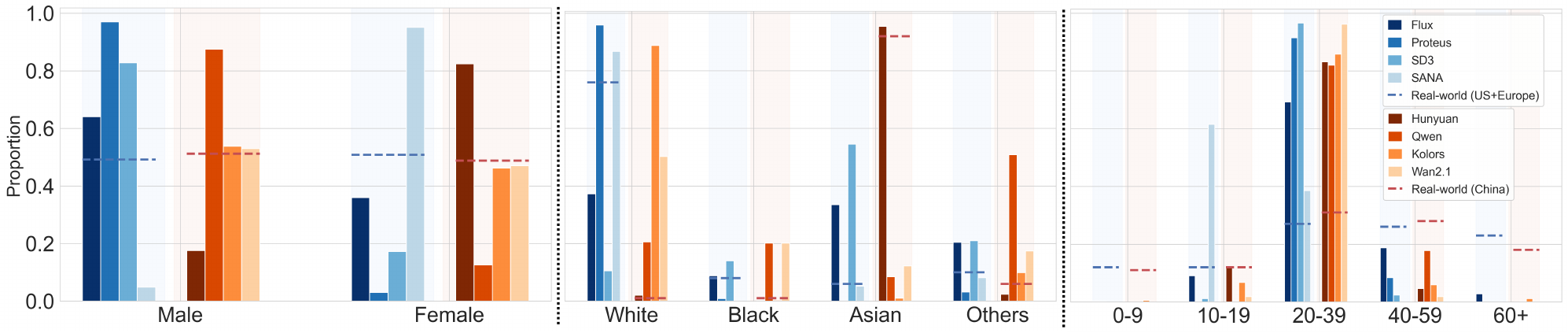}
    \caption{Demographic distributions of the generated faces compared against region-specific real-world baselines. Western models are evaluated against U.S. and European statistics; Chinese models against Chinese population statistics. Dashed lines indicate the regional reference.} 
    \label{fig:separate}
\end{figure*}

The analysis above compares all models against global demographic baselines. In this section, we study whether models are biased relative to the demographics of their own intended user base: U.S. and European populations for Western models, and Chinese population statistics for Chinese models (Figure~\ref{fig:separate}).

For gender, the pattern is unchanged. Six models exhibit clear gender bias: four models have a male bias and two models a female bias, regardless of the reference population. 

For race, the pattern differs across model groups. Western models show \emph{reduced} overrepresentation of White faces when evaluated against Western regional baselines, with models such as FLUX and SANA producing outputs that are closer to U.S. and European population distributions than when evaluated against global baselines. This suggests that part of the apparent White overrepresentation under global comparisons may reflect differences between training data composition and evaluation reference distributions. Chinese models show a different pattern. Despite China’s predominantly East Asian population (approximately 91\% Han Chinese), most Chinese models generate Asian faces at rates below the corresponding regional baseline. Hunyuan is an exception, producing substantially higher proportions of Asian faces.
This discrepancy indicates that demographic alignment is not consistently preserved even when evaluated against region-specific references, highlighting differences in how models reflect local population structure across development contexts.

Age bias is consistent across both model groups and both reference distributions. All models heavily overrepresent the 20--39 age group and systematically erase older adults, a pattern that holds whether the reference is the global population or a regional one. Across all eight audited models, age bias is therefore robust to reference distribution choice and cannot be explained as an artifact of using global rather than local baselines. 

\subsection{Emotions Function as a Demographic Filter}
\label{sec:results.emotions}

Emotion prompts systematically shift the demographic composition of generated faces: $93.8\%$ of model--emotion--attribute combinations show a statistically significant distributional shift after Bonferroni correction (Supplementary~\supp{supp:emotion}). Table~\ref{tab:claim_c} ranks these shifts by mean TVD averaged across the eight models and three demographic attributes (gender, race, and age). Anger produces the largest overall shift (mean TVD\,=\,0.48) and Happiness the smallest (0.17). Negatively-valenced, high-arousal emotions consistently induce larger demographic shifts than positive ones.

\begin{table}[h]
\centering
\caption{Mean TVD per emotion averaged across all models and attributes.}
\label{tab:claim_c}
\begin{tabular}{clc}
\toprule
\textbf{Rank} & \textbf{Emotion} & \textbf{Mean TVD} \\
\midrule
1 & Happy     & 0.1688 \\
2 & Sad       & 0.2355 \\
3 & Surprised & 0.2700 \\
4 & Fearful   & 0.3149 \\
5 & Disgusted & 0.3760 \\
6 & Angry     & \textbf{0.4825} \\
\bottomrule
\end{tabular}
\end{table}

\subsubsection{Demographic shifts per emotion}
\begin{figure*}[ht]
    \centering
    \includegraphics[width=1\linewidth]{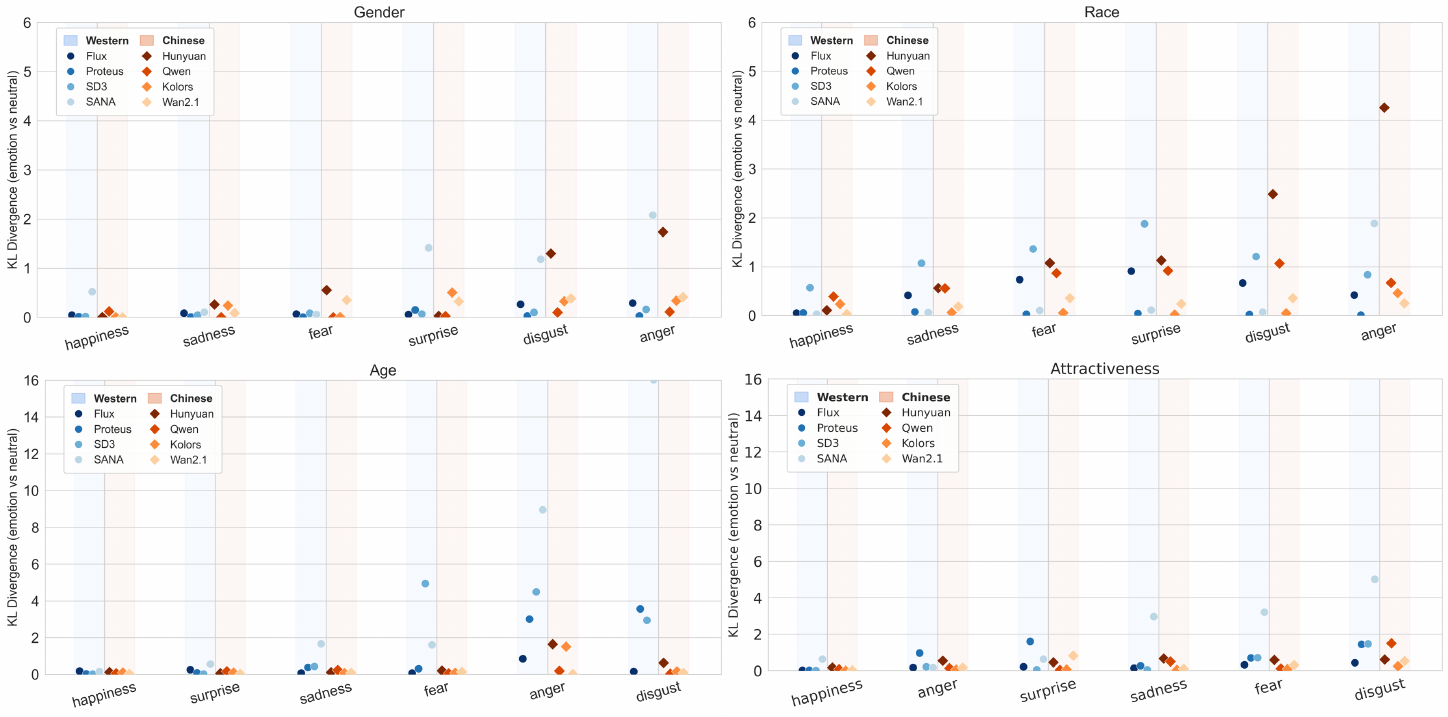}
    \caption{Emotion-induced demographic distribution shifts across models. KL divergence between emotion-conditioned and neutral demographic distributions is shown for different attributes. Emotion categories are ordered by the overall KL divergence averaged across models. Blue/red background corresponds to Western/Chinese models, respectively.}
    \label{fig:emotion_kl}
\end{figure*}

The magnitude of emotion-induced shifts varies substantially across emotions and attributes. Age exhibits the largest KL divergence overall: for several models, age-related KL values increase sharply under high-arousal, negatively valenced emotions (Fear, Anger and Disgust), exceeding shifts in gender, race, and attractiveness. Gender divergences are moderate but consistent across emotions. Figure~\ref{fig:emotion_kl} shows the per-model breakdown, and significance fractions per emotion--attribute combination are in Supplementary~\supp{supp:emotion}, Figure~\ref{fig:attr_chisq}.

\begin{figure}[htbp]
    \centering
    \includegraphics[width=1\linewidth]{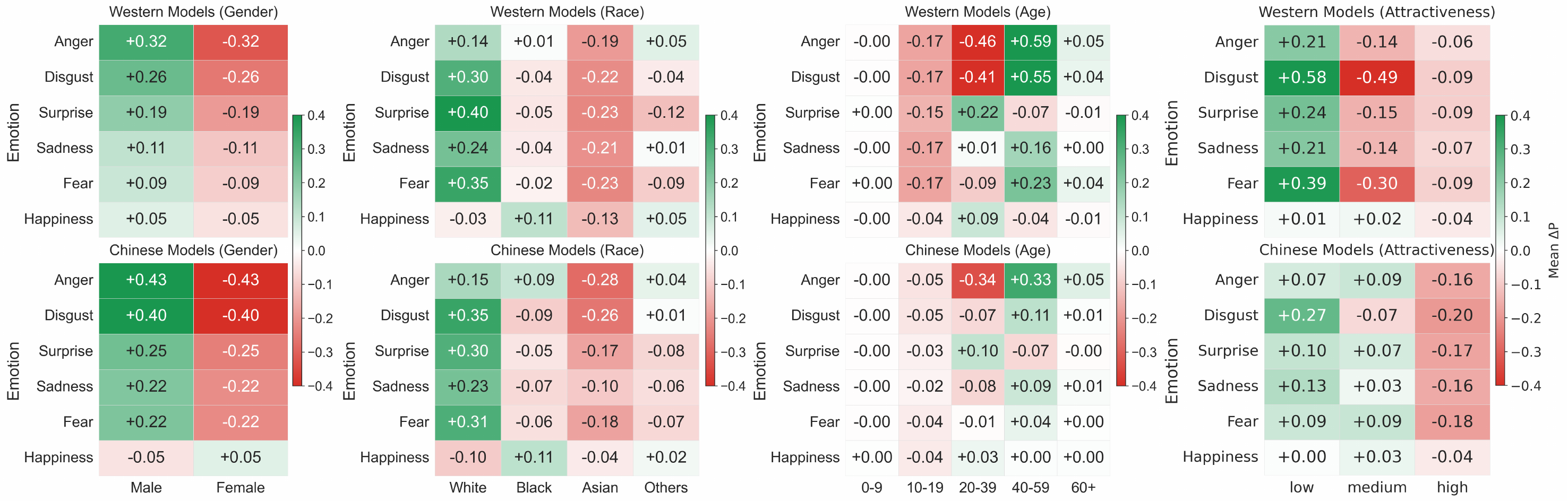}
    \caption{Heatmaps of the mean distributional shifts ($\Delta P$) of facial attributes induced by emotional prompts, relative to the neutral-prompt baseline, across Western (top) and Chinese (bottom) models.  Positive values indicate an increase in that demographic category's representation; negative values indicate a decrease.} 
    \label{fig:emotion}
\end{figure}

\paragraph{Directional shifts per emotion.}
Figure~\ref{fig:emotion} shows the mean $\Delta P$ (shift in proportion relative to neutral) for each demographic category, emotion, and model group. 
It reveals consistent directional patterns across models: under negatively valenced emotions, the models generate a larger proportion of male, White, and older faces, reducing female, Asian, and young representation.
Happiness produces the smallest and most uniformly distributed shifts. The direction of shifts is consistent between the Western and Chinese model groups. The Others category (which pools Indian, Latino, and Middle Eastern faces) shows no comparably consistent directional shift: across the six emotions, $\Delta$Others ranges from $-9.93$pp under Surprise to $+4.67$pp under Anger as presented in Table~\ref{tab:race_emotion}, in contrast to the uniform increase in White-coded and decrease in Asian-coded representation. We do not interpret this instability, since the category aggregates three demographically and socially different groups (see Section \ref{sec:discussion}), and resolving it would require a finer-grained taxonomy capable of distinguishing which of the pooled sub-groups is being selectively suppressed or amplified by a given emotion.

Anger produces the largest gender shift and Disgust yields the largest proportion of significant distributional shifts, directionally consistent across the eight models (per-model Bonferroni-corrected tests: Supplementary~\supp{supp:emotion}, Figure~\ref{fig:attr_chisq}). Under Disgust, the proportion of White faces rises from 49.0\% to 81.4\% while Asian faces decrease from 26.3\% to 2.1\% (Supplementary~\supp{supp:emotion}, Table~\ref{tab:race_emotion}). Anger and Disgust also produce a dramatic age shift across all models, largely replacing young faces with middle-aged ones (Supplementary~\supp{supp:emotion}, Table~\ref{tab:age_emotion2}). Under Anger the 20--39 age group drops from 80.3\% to 40.4\% while the 40--59 group rises from 7.5\% to 53.8\%. Disgust produces a similar but less pronounced shift.

\paragraph{A valence-primary demographic mapping.}
Across all models, gender representation shifts consistently for the four negatively valenced emotions: Anger, Disgust, Fear, and Sadness each increase male-coded representation relative to the neutral baseline (Figure~\ref{fig:emotion}; Supplementary~\supp{supp:emotion}, Table~\ref{tab:gender_emotion}).  This pattern differs from documented human emotion stereotypes, in which anger and disgust are more often associated with masculinity, while sadness and fear are more often associated with femininity \cite{plant2000gender,bjornsdottir2024stereotypes}, and from text-only language models, which tend to preserve these distinctions \cite{plazadelarco2024angrymensadwomen}. 
Instead, T2I models exhibit a coarser valence-driven mapping in which negatively valenced emotions are associated with increased male-coded, White-coded, and older faces, reducing emotion-specific variation along demographic dimensions. 

This pattern is consistent across architecturally diverse models from two development ecosystems, suggesting a shared property of current T2I training pipelines rather than model-specific effects. The role of Surprise is discussed in Section~\ref{sec:discussion}. 

\subsubsection{Intersectional Shifts}

We further analyze how different emotional prompts induce shifts in the intersectional joint distribution of gender, race and age relative to the emotion-neutral baseline. The results are summarized in Table~\ref{tab:intersectional_kl_js}. Each cell reports
KL\,/\,JS for each model--emotion pair, comparing the emotion-conditioned joint distribution to that model's own neutral-prompt baseline. Because KL is unbounded and can be inflated by near-zero-probability cells (Eq.\,4), JS (bounded in $[0,1]$) is the more directly comparable quantity both across models and across emotions, and an asterisk marks divergences exceeding the practical-significance threshold $D_{\mathrm{JS}} > 0.05$ used throughout the
paper. The cell with the highest value in the table corresponds to Hunyuan under Anger ($D_{\mathrm{JS}} = 0.90$), 0.16 above the next-largest cell (Proteus under Disgust, $D_{\mathrm{JS}} = 0.74$), indicating an unusually severe reorganization of the joint gender--race--age distribution for this specific model--emotion pairing. Only 3 of the 48 model--emotion cells fall below the practical-significance threshold (Proteus and Wan2.1 under Happiness, Kolors
under Fear), and occur under Happiness or Fear, the emotions under
which some models leave the joint demographic distribution effectively
unchanged. The largest cell corresponds to a Chinese model (Hunyuan) and the second-largest to a Western model (Proteus). The near-zero cells are
likewise split across both groups, indicating that the magnitude of
intersectional reshaping is model-specific rather than a property of the
Western/Chinese grouping. Across models, Happiness consistently induces the smallest intersectional shift, as reflected by the low KL and JS divergences for most models, which is consistent with our earlier findings on marginal distributions. In contrast, high-arousal, negatively valenced emotions (Anger, Fear and Disgust) tend to lead to substantially larger shifts, indicating that emotional conditioning reshapes demographic co-occurrence patterns. 

\begin{table}[h]
\centering
\caption{KL/JS divergences of the joint distributions of gender, race and age combined between the emotion-specific vs neutral prompts. $^*$ indicates practically significant differences ($JS > 0.05$) following prior work. Lower values indicate smaller emotion-induced biases. Largest value is marked bold and the smallest is underlined.}
\label{tab:intersectional_kl_js}
\small
\setlength{\tabcolsep}{3pt}
\begin{tabularx}{\linewidth}{l *{6}{>{\centering\arraybackslash}X}}
\toprule
\textbf{Model} & \textbf{Hap.--Neu} & \textbf{Sad.--Neu} & \textbf{Anger--Neu} & \textbf{Fear--Neu} & \textbf{Sur.--Neu}  & \textbf{Dis.--Neu} \\
\midrule

\cellcolor{lightblue} Flux &
0.27 / 0.09$^*$ &
0.64 / 0.24$^*$ &
1.27 / 0.43$^*$ &
1.15 / 0.37$^*$ &
1.54 / 0.53$^*$ &
1.08 / 0.40$^*$ \\

\cellcolor{lightblue} Proteus &
\underline{0.13} / \underline{0.03} &
0.47 / 0.12$^*$ &
3.12 / 0.69$^*$ &
0.45 / 0.11$^*$ &
\underline{0.26} / \underline{0.09}$^*$ &
3.55 / \textbf{0.74}$^*$ \\

\cellcolor{lightblue} SD3 &
0.63 / \textbf{0.23}$^*$ &
\textbf{1.50} / \textbf{0.41}$^*$ &
5.05 / 0.69$^*$ &
\textbf{5.64} / \textbf{0.70}$^*$ &
\textbf{2.19} / \textbf{0.64}$^*$ &
3.72 / 0.62$^*$ \\

\cellcolor{lightblue} SANA &
0.56 / 0.14$^*$ &
0.89 / \underline{0.07}$^*$ &
10.20 / 0.64$^*$ &
0.85 / 0.08$^*$ &
1.50 / 0.37$^*$ &
\textbf{15.57} / 0.54$^*$ \\

\midrule

\cellcolor{red!10} Hunyuan &
0.20 / 0.06$^*$ &
0.81 / 0.17$^*$ &
\textbf{12.77} / \textbf{0.90}$^*$ &
1.47 / 0.30$^*$ &
1.62 / 0.31$^*$ &
3.19 / 0.69$^*$ \\

\cellcolor{red!10} Qwen &
0.57 / 0.19$^*$ &
0.74 / 0.25$^*$ &
0.95 / 0.30$^*$ &
1.23 / 0.35$^*$ &
1.32 / 0.42$^*$ &
1.14 / 0.40$^*$ \\

\cellcolor{red!10} Kolors &
\textbf{0.92} / 0.12$^*$ &
\underline{0.39} / 0.14$^*$ &
4.31 / 0.49$^*$ &
\underline{0.14} / \underline{0.04} &
0.57 / 0.24$^*$ &
\underline{0.58} / \underline{0.18}$^*$ \\

\cellcolor{red!10} Wan2.1 &
0.28 / 0.04 &
0.70 / 0.14$^*$ &
\underline{0.73} / \underline{0.24}$^*$ &
0.98 / 0.28$^*$ &
0.52 / 0.20$^*$ &
1.02 / 0.28$^*$ \\
\bottomrule
\end{tabularx}
\end{table}

\subsubsection{Attractiveness shifts per emotion}
Beyond demographic composition, we examine perceived attractiveness, motivated by the well-documented attractiveness halo effect linking facial appearance to attributions of competence, trustworthiness, and social desirability \cite{nisbett1977halo}.

Across emotion conditions, the four negatively valenced emotions substantially increase the low-attractiveness share relative to neutral (3.4\%): Disgust by 42.6 percentage points, Fear by 23.8, Sadness by 17.2 and Anger by 13.9 (Table~\ref{tab:attractiveness}). Surprise yields an increase of the low-attractiveness faces in 17.0 percentage points whereas Happiness produces essentially no change (+0.9 points). The six emotions reduce the high-attractiveness share from the neutral baseline of 16.0\%: by 11.3 (anger) to 14.3 (disgust) percentage points for the negative emotions, 13.2 points for Surprise and 3.8 points for Happiness, which retains the largest high-attractiveness share of 12.2\% (Table~\ref{tab:attractiveness_emotion}).
We report emotion- and demographic-conditioned changes in low-attractiveness rates averaged across all eight models as $\Delta P_{\text{low-att}}(g, e) = P(\text{low-att} \mid g, e) - P(\text{low-att} \mid g, e_0)$ in Supplementary~\supp{supp:emotion}, Table~\ref{tab:delta_lowatt}.

The observed shift towards low-attractiveness under Anger is observed across all demographic groups with similar magnitude: +4.8 percentage points for male and +6.6 for female faces, +14.7 for White and +14.9 for non-White faces, and +8.5 for young and +2.6 for middle-aged faces (Supplementary~\supp{supp:emotion}, Table~\ref{tab:delta_lowatt}). The gender estimates for Anger are based on four models only, because too few female faces are generated under Anger for the remaining models. The 60+ group is omitted for the same reason.
Neutral low-attractiveness shares are 2.8\% for both White and non-White faces, and 2.4\% for young and 18.1\% for middle-aged faces.

The demographic comparisons above should be read with some caution. The attractiveness classifier's error rate differs significantly across gender (22.7\% for female faces vs.\ 9.3\% for male faces) and across age groups (24\% for middle-aged vs.\ 8\% for young faces), though not across race groups (Section~\ref{sec:methods}). Because the classifier does not perform equally well across these groups, direct comparisons of attractiveness shifts between male and female faces, or between age groups, are less straightforward than comparisons within a single group relative to its own neutral baseline. The direction of the gender error, however, works against rather than toward the pattern we report: in the confusion matrix (Supplementary~\supp{supp:attr}, Table~\ref{tab:confusion_matrices_gender}), actual-medium-attractiveness female faces are far more often misclassified as high-attractiveness than as low-attractiveness (56 of 257 vs.\ 35 of 257), whereas for male faces this asymmetry is much smaller (4 of 304 vs.\ 27 of 304). The classifier therefore tends to assign female faces higher, not lower, attractiveness scores than their ground-truth label warrants, the opposite of what would be needed to explain the shifts reported here as a classifier artifact. We nonetheless report the gender and age comparisons above as indicative rather than definitive.

\section{Discussion}\label{sec:discussion}

The faces generated by state-of-the-art text-to-image models are not demographically neutral. In the absence of demographic specification, all models in our study overrepresent young faces (under 40), and most overrepresent individuals estimated as White. 
We further find that emotional prompts do not only modify facial expression, but they also shift the demographic composition of generated outputs, changing which groups are more likely to be depicted as angry, sad, or fearful in a manner that is broadly consistent across models. 
These results suggest a \emph{social grammar}: a learned association structure in which emotional cues are linked to characteristic demographic profiles rather than solely to facial expressions, shaping demographic representation in emotion-conditioned generation.

\paragraph{The myth of the blank slate.}
The White-and-young default observed across several models is consistent with the documented overrepresentation of WEIRD (Western, Educated, Industrialized, Rich, Democratic) populations in Internet-based training corpora \cite{henrich2010weirdest,birhane2021multimodal}. T2I models trained on such data not only reproduce these distributions but tend to amplify them, as generation is drawn toward dominant modes in the learned visual prior \cite{bianchi2023accessible}. 
Our results extend this observation in a cross-cultural setting. Chinese models exhibit similar White-centric and age-homogeneous defaults as Western models, despite their different intended deployment contexts. At the same time, there is substantial variation between individual models within each regional group, often exceeding differences between groups, suggesting that model-specific design and training choices play a larger role than regional origin alone.
One model, Hunyuan, diverges from this pattern, generating approximately 90\% Asian faces under neutral prompts, which illustrates that demographic alignment with a target population is technically achievable. In contrast, other Chinese models do not reflect local population structure even when evaluated in Chinese (Supplementary~\supp{supp:chinese}), suggesting that demographic priors are not solely determined by input language and may be embedded at the level of the generative prior. Age distributions show a similar pattern of concentration. Across models, youth is consistently overrepresented, in alignment with social desirability in a commercially dominant visual culture \cite{an2025,sufian2025}. In some cases this concentration is extreme. For example, SANA generates over 60\% very young White female faces under neutral prompts, indicating a strong collapse toward a narrow socially valorized visual profile.

\paragraph{Intersectional erasure is not reducible to marginal bias.}
This observation has direct implications for auditing practice. Marginal distributions, which form the basis of most demographic fairness evaluations, cannot capture compound forms of representational loss.
Even under neutral prompts, we observe substantial intersectional sparsity: on average 69\% of intersectional cells (16.5 of 24) contain near-zero probability mass across models, and the young$\times$female$\times$Black intersection appears at a mean probability of only 1.3\% (a suppression factor of 0.25$\times$ relative to a population-based expectation). 
Emotion conditioning further redistributes probability mass across the joint demographic space, with intersectional divergence measures (Table~\ref{tab:intersectional_kl_js}) indicating shifts that are not visible in any single attribute marginal.
This pattern is consistent with intersectionality theory, which predicts that multiple axes of identity can interact to produce non-additive forms of disadvantage \cite{crenshaw2013}. Our results are consistent with this expectation already at the neutral-prompt baseline. 

The specific intersectional cell that we highlight, young $\times$ female $\times$ Black carries specific historical weight along several converging lines. First, \cite{crenshaw2013} is grounded in \emph{DeGraffenreid v.\ General Motors} (1976), in which Black women plaintiffs were denied legal standing because antidiscrimination doctrine could only recognize race discrimination (measured against Black men) or sex discrimination (measured against white women). Black women's compounded experience was structurally invisible to a framework that only evaluates one axis at a time, similarly as YFB is invisible to marginal-only demographic audits. Second, Buolamwini and Gebru's Gender Shades audit \cite{buolamwini2018gender} established that darker-skinned women are the demographic group most poorly served by commercial face-analysis systems. Our results show a structurally analogous erasure recurring in generative AI models. Third, the erasure of Black women from ordinary, aspirational visual representation (as opposed to a narrow set of stereotyped tropes) has a documented history in media and visual-culture scholarship \cite{collins2000black,hooks2014black}, and imaging \emph{technology} itself has a parallel history of encoding racial defaults, from Kodak's Shirley-card color calibration standards that systematically misrepresented darker skin tones for decades \cite{roth2009looking} to the generative priors we document here. Read against these histories, a mean representation rate of 1.3\% against a 5.27\% population-based expectation is not simply a data-sparsity artifact but a continuation, in a new technological setting, of previously documented patterns of exclusion, patterns that, we emphasize, are neither unique to nor exhausted by this single intersectional cell.

These findings suggest that single-attribute fairness metrics, which remain widely used in both the algorithmic fairness literature and emerging governance frameworks, may not capture compound forms of representational imbalance in generative models.

\paragraph{Emotion functions as a demographic coding system.}
A central finding is that emotion words in T2I prompts act as a mapping from affect space to demographic space: rather than only modifying facial expression, emotion tokens are associated with systematic shifts in the demographic composition of the generated faces. Across the eight models, negatively valenced emotions are associated with higher proportions of older, male-coded, and White-coded faces. This pattern indicates that the effect is not limited to the rendering of a particular facial expression: emotional conditioning changes \emph{who} is generated. The fact that the direction and magnitude of the shift vary with emotional valence further suggests that gender representation is coupled with the model's learned associations between emotion and facial appearance. 

The consistency of this pattern across architecturally diverse models and two development ecosystems suggests that it is not easily attributable to architecture-specific effects alone. A plausible contributing factor is the structure of multimodal training data. Prior work has shown that both text corpora and image-text datasets encode demographic co-occurrence patterns between affective language and visual attributes \cite{caliskan2017semantics,garg2018word,kiritchenko2018examining}. Analyses of LAION-5B further show systematic associations between emotion-related captions and demographic attributes, including gendered and racialized patterns \cite{unmaskinglaion5b2026}, and that a substantial fraction of downstream bias in diffusion models can be predicted from such co-occurrences in training data \cite{girrbach2026laion400m}. These results are consistent with the interpretation that diffusion models learn emotion-conditioned demographic priors from training data, such that emotion tokens influence not only expression but also the demographic distribution of generated faces. For the Chinese models in our study, this interpretation is indirect: the observed cross-ecosystem consistency in output patterns suggests a similar mechanism may be present, but their training data have not been publicly audited at comparable depth.

The specific form of this learned mapping diverges from documented human stereotypic knowledge in a theoretically informative way. Human perception research has reported emotion-specific gender stereotypes, in which anger and disgust are more often associated with masculinity, while sadness and fear are more often associated with femininity \cite{plant2000gender,bjornsdottir2024stereotypes}. 
In contrast, the models we study do not preserve this differentiation: negatively valenced emotions (and particularly Anger and Disgust) are associated with increased male-coded representation, as discussed below. This suggests a coarser representation organized primarily along a valence axis, rather than emotion-specific demographic templates. Moreover, this pattern differs from that observed in text-only language models, which tend to reproduce more differentiated stereotype structures consistent with human judgment studies \cite{plazadelarco2024angrymensadwomen,schwemmer2020diagnosing}. One possible explanation is that image-text training introduces stronger marginal visual co-occurrence signals, which may dominate more fine-grained emotion-specific associations present in text-only corpora.
This interpretation is consistent with prior analyses of multimodal training data \cite{unmaskinglaion5b2026,girrbach2026laion400m} and with the cross-architecture consistency of our findings, though direct causal attribution remains an open question.

Surprise helps refine this interpretation. Although valence-ambiguous in the circumplex model of affect \cite{russell1980circumplex}, it produces substantial demographic shifts, with 79.5\% male-coded representation (Table~\ref{tab:gender_emotion}) and a relatively high TVD rank (Table~\ref{tab:claim_c}), exceeding both Sadness (73.9\%) and Fear (73.3\%). 
This suggests that valence alone is insufficient to fully account for the observed ordering. In our analyses, Surprise groups with the high-arousal negative emotions (Anger, Disgust, Fear) rather than with other valence-ambiguous or positive conditions. Across the six emotions, the observed TVD ordering is broadly consistent with a joint interpretation in which both valence and arousal may contribute to demographic displacement, with higher-arousal emotions tending to produce larger shifts. Sadness, which is typically lower arousal in the circumplex model, shows more moderate displacement (TVD = 0.23) than high-arousal negative emotions (Fear: 0.31, Disgust: 0.38, Anger: 0.48). Happiness produces the smallest displacement (TVD = 0.17).
This interpretation is observational and does not directly measure arousal in the training data, as arousal is inferred from the circumplex model rather than empirically estimated. Direct validation would require training-data analyses of arousal-conditioned demographic co-occurrences, which are beyond the scope of this study.

Attractiveness reinforces the structure observed in emotion-conditioned generation. Negative emotions reduce the high-attractiveness share from the neutral baseline of 16.0\% (Table~\ref{tab:attractiveness}) to between 1.7\% (Disgust) and 4.8\% (Anger), whereas Happiness leads to the smallest reduction in high-attractiveness faces among the six emotions.
These patterns are consistent with the co-occurrence between affect and perceived attractiveness described in the attractiveness halo effect, in which positive expressions are more often associated with higher attractiveness and negative expressions with lower attractiveness in human perception \cite{nisbett1977halo,gulati2024beautiful}, and with the distributional analysis of \textcite{doh2026facct}, who measure the distance between trait-conditioned and attractiveness-conditioned distributions of generated faces.

These findings extend prior work on emotional bias in generative image systems \cite{mehta2024emotional} by showing that similar directional patterns appear across multiple model families and development contexts.

\paragraph{Masculinization of negative emotion outputs compounds intersectional penalties.}
Across models, negative emotional conditioning systematically shifts the gender distribution of generated faces toward male-coded representations. The distributional shift is strongest for Anger and Disgust and particularly pronounced among Chinese models, whereas Happiness (the only positively valenced emotion examined) produces little change in the gender distribution. 

This pattern is consistent with longstanding evidence that gender and emotion and intertwined in the perception and interpretation of faces. Gender stereotypes systematically associate some emotions with men and others with women. \citet{plant2000gender}, for example, found that anger and pride were stereotypically associated more strongly with men, whereas sadness and fear were more strongly associated with women. Furthermore, these stereotypes also affected the interpretation of facial expressions: identical or ambiguous expressions were interpreted differently depending on the perceived gender of the face. Thus, gender can influence the interpretation of emotion, while emotional appearance can simultaneously provide cues about gender. 

The physical structure of facial expressions provides an additional mechanism through which these associations can arise. \citet{adams2015intersection} show that gender-related facial appearance and emotional expression share overlapping physical and social cues, particularly those associated with facial maturity and dominance. Anger, in particular, tends to be associated with mature and dominant facial characteristics, whereas fear is associated with more babyish facial characteristics \cite{marsh2005fear}. These overlaps make emotion and gender especially susceptible to being jointly represented: a model conditioned to generate a particular emotional appearance may consequently alter other facial characteristics that are statistically associated with that emotion in its training distribution.

Our results extend this literature from the perception of existing faces to the \emph{generation} of faces. The masculinization under Anger is compatible with the established association between anger, dominance, and masculine-coded facial appearance. However, the shift is not confined to Anger. Disgust, Fear, and Sadness also produce measurable movement toward male-coded representations. This broader pattern is important because it suggests that the observed effect cannot be reduced to the reproduction of a single anger--masculinity stereotype. Instead, negative emotional conditioning appears to activate a wider region of the model's learned representation in which negative affect is associated with male-coded facial characteristics. The stronger effects observed in Chinese models further indicate that the magnitude of this association is not universal and may depend on training-data composition, model design, or the cultural and visual distributions from which emotion--gender associations are learned.

More importantly, the gender shift interacts with the attractiveness shift in a way that produces an asymmetric outcome. Female-coded faces experience larger increases in low-attractiveness levels than male-coded faces for three of the four negative emotions examined: Anger, Disgust, and Fear. The disparity is largest for Disgust, where the increase in low-attractiveness representation is 9.1 percentage points greater for female-coded than male-coded faces. Although the attractiveness classifier exhibited a higher error rate for female faces, it nevertheless tended to assign them higher attractiveness scores, not lower. These findings run counter to the pattern that would be expected based solely on differential classifier error. Thus, the conditions that shift the generated population toward male-coded faces also tend, in most cases, to impose a stronger attractiveness penalty on the female-coded faces that remain.

Disgust additionally produces one of the strongest masculinizing shifts in the generated-face distributions (+0.26 in Western models and +0.40 in Chinese models) while simultaneously producing the largest gender disparity in the increase in low-attractiveness faces. This suggests that Disgust may constitute a particularly strong trigger for gendered representational associations. The prompt appears to reorganize the demographic and appearance characteristics of the generated population. This finding matters because facial attractiveness is not socially neutral. Research on the attractiveness halo effect shows that attractive faces tend to receive more favorable evaluations on other characteristics, including intelligence and trustworthiness \cite{gulati2024beautiful}. Consequently, a systematic increase in low-attractiveness female representations may have implications beyond appearance if the generated faces are subsequently used as representations of people or social groups.

The resulting disparity is therefore not adequately described as either gender bias or attractiveness bias in isolation. It instead reflects a conditional interaction between emotion, gender, and appearance, with the magnitude of that interaction varying across emotions and model origins. Emotion-conditioned generation therefore does not only modify facial expression: it can systematically reorganize gender representation and, within that representation, the distribution of appearance-related attributes.

\paragraph{Implications for downstream representation}

The societal stakes of this coupling extend beyond the statistics. T2I models are now integrated into advertising, journalism, social media, and stock imagery, where generated faces may reach large audiences without consistent disclosure of their synthetic origin. This exposure is not demographically neutral: generative-AI tools are disproportionately adopted and used by young people. EU survey data show that 63.8\% of 16-24-year-olds had used generative AI tools in 2025, nearly twice the general-population rate of 32.7\% (aged 16-74),\footnote{Eurostat, ``64\% of 16-24-year-olds used AI in 2025'', 2026,
https://ec.europa.eu/eurostat/web/products-eurostat-news/w/edn-20260210-1} and US survey data show a parallel pattern, with adoption of AI tools and weekly-use rates highest among adults under 30.\footnote{AP-NORC, ``Young adults are leading the way in AI adoption,'' https://apnorc.org/projects/young-adults-leading-the-way-in-ai-adoption/} The audience most exposed to these systems' outputs therefore substantially overlaps with the age group whose own representation these systems
severely distort under negative-emotion prompts (Section \ref{sec:results.emotions}), and are among the groups most susceptible to the effects of such representations: adolescence and young adulthood are developmental periods of active identity formation, during which self-concept is particularly responsive to media exposure \cite{avci2024socialmedia}.

Repeated exposure to images that systematically associate negative emotions with particular demographic profiles is consistent with concerns from cultivation theory regarding the long-term effects of skewed media environments on social perception \cite{morgan2010state}. 
Initial experimental evidence supports the plausibility of such effects: a preregistered survey experiment found that exposure to non-inclusive AI-generated faces increases participants’ racial and gender biases, while exposure to more inclusive faces can reduce them \cite{AlDahoul_2024_AIgeneratedFacesStereotypes}. Whether similar effects arise specifically from the emotion-conditioned demographic structure documented here, and whether they persist over time in the manner observed in traditional media contexts \cite{dasgupta2001malleability}, remains an open empirical question. 

Nevertheless, the scale and speed of T2I generation make this question increasingly consequential. Generative systems can produce large volumes of synthetic imagery at speeds and costs that exceed those of media production systems \cite{bianchi2023accessible}. Consequently, even distributional biases that appear modest at the level of individual generations may become socially consequential when reproduced across large quantities of imagery. The findings here therefore motivate further work not only on demographic bias in generated faces but on how affective conditioning interacts with demographic and appearance attributes to shape the representational environments produced by generative AI. 

\paragraph{Kolors reveals a cross-lingual alignment gap.}
The female-erasure result observed in Kolors (Supplementary~\supp{supp:chinese}) provides a diagnostic contrast within a set of Chinese-language models developed within broadly similar contexts, yet exhibiting different behavior along a single linguistic axis.

We consider two non-exclusive explanations for this discrepancy. First, a training-data hypothesis, suggesting that Chinese-language image-text pairs may associate gender-neutral person references with male-skewed visual representations, leading the model to reproduce these associations regardless of the alignment achieved in other language interfaces. Second, an alignment-scope hypothesis, suggesting that demographic balancing may be more effective in the English prompt space, while the Chinese-language encoder remains insufficiently aligned, resulting in language-dependent demographic outputs.

These mechanisms may operate independently or jointly. Importantly, Kolors generates an approximately balanced gender distribution under English prompts (approximately 50\% female) while producing near-complete female underrepresentation under Chinese prompts. This pronounced discrepancy suggests that demographic outputs are strongly conditioned on prompt language, even within the same model.

This form of cross-lingual divergence may remain undetected by conventional evaluations conducted in a single language. More generally, these findings highlight a limitation of monolingual fairness assessments: evaluating a multilingual system exclusively through English prompts may obscure biases that emerge in other linguistic contexts. Multilingual auditing is therefore essential for identifying language-dependent disparities and assessing fairness under cross-cultural model deployment conditions.

\paragraph{Governance implications.}
Different actors bear responsibility for addressing the biases identified in this paper. First, \emph{model developers} should disclose the demographic composition and geographic provenance of training data. The persistence of White overrepresentation in Chinese models prompted in Chinese demonstrates that demographic bias is embedded in the generative prior itself, making dataset characterization a necessary requirement for deployment transparency \cite{gebru2021datasheets}. Second, \emph{platform operators} integrating T2I models into content pipelines should provide more granular demographic controls and monitoring mechanisms. Emotion prompts produced extremely large shifts in intersectional demographic distributions (up to JS\,=\,0.90 for Hunyuan under Anger; Table~\ref{tab:intersectional_kl_js}), on top of baseline race biases already reaching TVD\,=\,0.81 under neutral prompting (Supplementary~\supp{supp:divergence}). Current systems often require users to explicitly specify demographic attributes in prompts in order to avoid unintended demographic defaults when expressing affective intent, yet many users may be unaware that such specification is necessary to obtain demographically balanced outputs. Third, \emph{regulators and standards bodies} should require  multilingual and intersectional bias audits prior to deployment. The female-erasure effect observed in Kolors was not detectable under an English-only evaluation, demonstrating that monolingual auditing underestimates representational harms in globally deployed systems. Emerging AI governance frameworks that rely on single-modality, single-language evaluations are therefore unlikely to capture the full scope of demographic bias. More broadly, the Wan2.1 result demonstrates that substantially more globally balanced intersectional generation is achievable within current architectures, even without specialized intersectional mitigation techniques, when marginal race and gender distributions are calibrated toward global demographic distributions. These findings suggest that demographic calibration benchmarks could become part of future pre-deployment evaluation and certification frameworks. Fourth, \emph{affected communities and end-users} should be engaged directly in the audit and mitigation process rather than treated only as the subjects of measurement. Concrete mechanisms include participatory design workshops with the demographic groups most affected by the compound erasure we document (e.g.,
the young $\times$ female $\times$ Black intersection, Section \ref{sec:results}), structured channels for reporting representational harms in deployed systems,
and training resources that help non-technical stakeholders interpret and contest biased outputs. Treating demographic bias as a technical property to be measured by researchers and disclosed by developers risks
overlooking the perspectives of the communities the biases most affect. Participatory mechanisms are a necessary complement to the disclosure,
control, and audit measures discussed above.

\paragraph{Limitations.}
Our audit spans eight T2I models covering four architectural families (U-Net and Transformer variants), two development ecosystems, and a three-year release window (2023--2025). Closed-source systems (\emph{e.g.}, DALL-E 3, MidJourney, Imagen) and future fine-tuned model versions remain untested, and findings may not generalise to those architectures. 

All prompts were in English. While Chinese-language evaluation (Supplementary~\supp{supp:chinese}) confirms that biases persist across languages, English prompts may themselves encode Western visual norms before reaching the generative prior. Results characterise statistical tendencies across 1,000 sampled outputs per condition and are not deterministic predictions for individual generations. 

The analysis covers six canonical emotions plus a neutral baseline. Compound emotions, continuous affect dimensions, and culturally specific emotional concepts remain unexamined. The convergent evidence described in Section \ref{sec:methods} indicates that the findings are not specific to the tested prompt wording. However, systematic paraphrase variation across the full semantic neighbourhood of each emotion word remains a priority for future work. 

Gender is treated as binary, for the epistemic reasons discussed in the Methods \cite{scheuerman2019computers}. Non-binary-inclusive auditing requires methodologically distinct approaches, such as participatory audits with non-binary communities, rather than improved classifiers. Similarly, the four-class race taxonomy is coarse and U.S.-centric \cite{khan2021one,dohposition}. Since Indian, Latino, and Middle Eastern faces are pooled into a single \emph{Others} category, our method cannot detect over- or under-representation \emph{within} that category, even though these groups together correspond to a substantial share of the world's population and may be subject to their own patterns of erasure. This grouping is not desirable, but our  validation studies (Section \ref{subsec:attr}) indicate that it reflects a structural necessity given current estimation tools rather than an arbitrary methodological choice. As with gender, addressing this limitation would require a different methodological  approach rather than an improved classifier. Continuous representations, such as skin-tone metrics or colorism-oriented analyses, could offer complementary granularity and are left for future work.

Several audited models incorporate RLHF-based alignment or demographic-balance filters that may attenuate the generative prior's demographic skew. The bias magnitudes reported here may therefore underestimate some aspects of the underlying prior's demographic biases. Models with stronger alignment may exhibit smaller absolute TVD values while retaining the same directional patterns. 

The study documents demographic bias patterns but does not establish the precise causal mechanism linking training decisions to outputs. The convergence of Western and Chinese models---presumably trained on different corpora in different languages---on the same valence-primary mapping points toward training data as the primary driver, consistent with evidence that large-scale web-scraped corpora encode systematic demographic co-occurrences with emotional content \cite{girrbach2026laion400m}. Mechanistic attribution, identifying which specific training data subsets or alignment steps produce the observed patterns, requires access to proprietary training details unavailable for most audited models and is a priority for future work. 

Finally, this framework is diagnostic rather than interventionist: it quantifies demographic biases but do not evaluate debiasing interventions or downstream effects on human social cognition. Establishing those links is the essential next step for translating these audit findings into causal understanding and actionable remediation.

The generation of human faces is among the most socially consequential capabilities of contemporary AI systems. Our findings show that, across eight state-of-the-art models spanning Western and Chinese development ecosystems, prompts with emotions activate a consistent social grammar: a learned demographic ordering in which emotional cues consistently select who appears. Negative affect is disproportionately associated with older, male, White-coded, and less attractive faces, while Happiness, the only positive-valence condition in this study, is the single emotion that departs from this pattern on every axis: it is the only condition that shifts generation toward younger faces and away from the White-coded default, and the only one that does not impose the sharp attractiveness penalty seen under every negative emotion. This social grammar is not an incidental artifact of imperfect technology, but a structural feature of models trained on visual cultures that already encode unequal social representations. Addressing it requires not only better models, but governance mechanisms commensurate with the scale, realism and speed with which these generative models now shape the visual environments through which human social cognition operates.

\section*{Acknowledgments}

\textbf{Funding:} M.W. is supported by the China Scholarship Council and the ELSA Mobility Fund. A.G. and N.O. are supported by a nominal grant received at the ELLIS Unit Alicante Foundation from the Regional Government of Valencia in Spain (Resolución de la Conselleria de Industria, Turismo, Innovación y Comercio, Dirección General de Innovación), along with grants from the European Union’s Horizon Europe research and innovation programme: ELIAS (grant agreement 101120237), and ELLIOT (grant agreement 101214398) and a grant from Intel. A.G. is additionally supported by a grant from the Banc Sabadell Foundation. G.Z. is supported by the Research Council of Finland (former Academy of Finland) Academy Professor project EmotionAI (grants 336116, 359894), HPC project FaceCanvas (grant number 364905), the University of Oulu \& Research Council of Finland Profi 7 (grant 352788), and EU HORIZON-MSCA-SE-2022 project ACMod (grant 101130271). Views and opinions expressed are those of the author(s) only and do not necessarily reflect those of the European Union, the European Commission or the European Health and Digital Executive Agency (HaDEA). Neither the European Union nor the European Commission can be held responsible for them.
\textbf{Author contributions:} M.W., A.G. and N.O. designed the experiments. M.W. and A.G. collected the data, and performed the analyses. M.W., A.G., G.Z., and N.O. jointly conceived the study. N.O. and G.Z. supervised the research. M.W., A.G., N.O. wrote the manuscript and all authors approved it.
\textbf{Competing interests:} The authors declare no competing interests.
\textbf{Data and materials availability:} Data is available at \href{https://huggingface.co/datasets/mengtingwei/emotion\_bias}{\url{https://huggingface.co/datasets/mengtingwei/emotion\_bias}}. Code is available at \href{https://github.com/weimengting/EmotionBias-SyntheticFaces}{\url{https://github.com/weimengting/EmotionBias-SyntheticFaces}}.
\textbf{Generative AI Usage Statement:} In addition to the generative AI tools used in the experiments, the authors used ChatGPT 5.2 and Claude during the preparation of this manuscript to assist with the linguistic quality, grammatical correctness, and the formatting of tables in the text. The authors reviewed and modified all the generated content as needed and take full responsibility for the final integrity and accuracy of the publication.

\newpage

\printbibliography

\newpage

\renewcommand{\thefigure}{S\arabic{figure}}
\renewcommand{\thetable}{ST\arabic{table}}
\renewcommand{\theequation}{S\arabic{equation}}
\renewcommand{\thepage}{S\arabic{page}}
\setcounter{figure}{0}
\setcounter{table}{0}
\setcounter{equation}{0}
\setcounter{page}{1}

\begin{center}
\section*{Supplementary Materials for ``Happy Young Women, Grumpy Old Men? Emotion-Driven Demographic Biases in Synthetic Face Generation''}

Mengting Wei, 
	Aditya Gulati, 
	Guoying Zhao,
    Nuria Oliver\\

\end{center}

\suppsection{Audited T2I Models}{supp:models}

\begin{table}[H]
\centering
\caption{The eight T2I models audited in this study. Rows shaded blue are Western-origin models; rows shaded red are Chinese-origin models. DiT~=~Diffusion Transformer; EN~=~English; ZH~=~Chinese.}
\label{tab:model_comparison}
\small
\setlength{\tabcolsep}{4pt}
\begin{tabular}{l l l r l l r l}
\toprule
\textbf{Model} & \textbf{Developer} & \textbf{Architecture} & \textbf{Params} & \textbf{Lang.} & \textbf{Resolution} & \textbf{Steps} & \textbf{License} \\
\midrule
\multicolumn{8}{l}{\textit{Western models}} \\[2pt]
\rowcolor{lightblue}
FLUX    & Black Forest Labs  & DiT         & $\approx$12B & EN        & 1024$\times$1024 & 4  & Apache-2.0 \\
\rowcolor{lightblue}
Proteus & HuggingFace comm.  & UNet        & 3.5B         & EN        & 1024$\times$1024 & 20 & GPL-3.0 \\
\rowcolor{lightblue}
SD3     & Stability AI       & DiT         & $\sim$3--4B  & EN        & 1024$\times$1024 & 40 & SA Community \\
\rowcolor{lightblue}
SANA    & NVIDIA             & DiT         & 1.6B         & EN        & 1024$\times$1024 & 20 & NSCL v2 \\[4pt]
\multicolumn{8}{l}{\textit{Chinese models}} \\[2pt]
\rowcolor{red!10}
Hunyuan & Tencent            & DiT         & 17B          & ZH \& EN  & 1024$\times$1024 & 50 & Tencent Comm. \\
\rowcolor{red!10}
Qwen    & Alibaba            & DiT + LoRA  & 20B          & ZH \& EN  & 1024$\times$1024 & 50 & Apache-2.0 \\
\rowcolor{red!10}
Kolors  & Kuaishou           & UNet        & 8.9B         & ZH \& EN  & 1024$\times$1024 & 50 & Apache-2.0 \\
\rowcolor{red!10}
Wan2.1  & Wan-AI             & Video DiT   & 1.3B         & ZH \& EN  & 832$\times$480   & 50 & Apache-2.0 \\
\bottomrule
\end{tabular}
\end{table}

\suppsection{Image Regeneration Rates Across Models}{supp:regenRates}

\begin{figure}[H]
    \centering
    \includegraphics[width=0.8\linewidth]{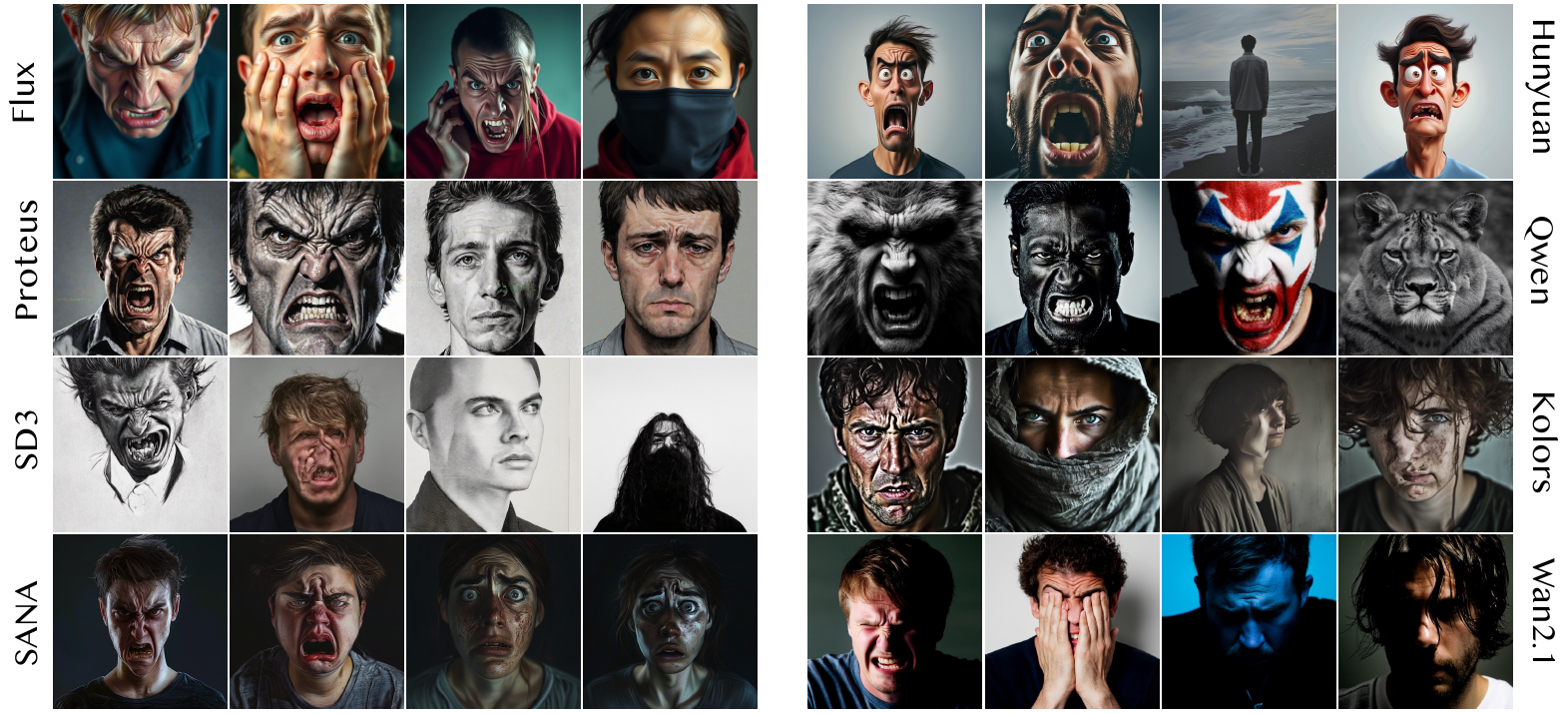}
    \caption{Examples of invalid face generations across models, including occluded, unrealistic, and distorted faces.}
    \label{fig:occlusion}
\end{figure}

As described in Section \ref{sec:methods.imageGeneration}, images produced by certain models were regenerated due to the presence of occlusions in some outputs (Figure \ref{fig:occlusion}). Table \ref{tab:rejection_rate} reports the proportion of occluded images across models and emotion categories. All images containing occlusions were excluded and subsequently regenerated to ensure a total of 1,000 images per model for each emotion.

\begin{table}[htbp]
\centering
\caption{Per-model face-detection rejection rates (\%) across all emotion conditions. Rejection rate = 1 $-$ (detected faces / 1000 generated images).}
\label{tab:rejection_rate}
\begin{tabular}{lcccccccc}
\toprule
\textbf{Model} & Angry & Disgusted & Fearful & Happy & Sad & Surprised & Neutral & \textbf{Overall} \\
\midrule
FLUX & 1.4 & 1.3 & 1.4 & 0.0 & 0.0 & 0.0 & 0.2 & 0.6 \\
Proteus & 5.8 & 0.0 & 0.0 & 0.0 & 0.7 & 0.0 & 0.2 & 1.0 \\
SD3 & 5.7 & 1.7 & 20.7 & 0.1 & 2.4 & 0.2 & 0.5 & 4.5 \\
SANA & 2.5 & 0.2 & 0.4 & 0.0 & 0.0 & 0.0 & 0.0 & 0.4 \\
Hunyuan & 0.9 & 0.2 & 2.7 & 0.0 & 0.3 & 0.1 & 0.0 & 0.6 \\
Qwen & 3.3 & 0.0 & 5.6 & 0.0 & 0.5 & 0.0 & 0.3 & 1.4 \\
Kolors & 4.1 & 0.0 & 0.1 & 0.0 & 0.1 & 0.0 & 0.0 & 0.6 \\
Wan2.1 & 3.0 & 4.5 & 29.1 & 0.0 & 8.4 & 0.0 & 1.8 & 6.7 \\
\bottomrule
\end{tabular}
\end{table}

\newpage
\suppsection{Attribute Estimation}{supp:attr}

\begin{figure}[htbp]
    \centering
    \includegraphics[width=0.8\linewidth]{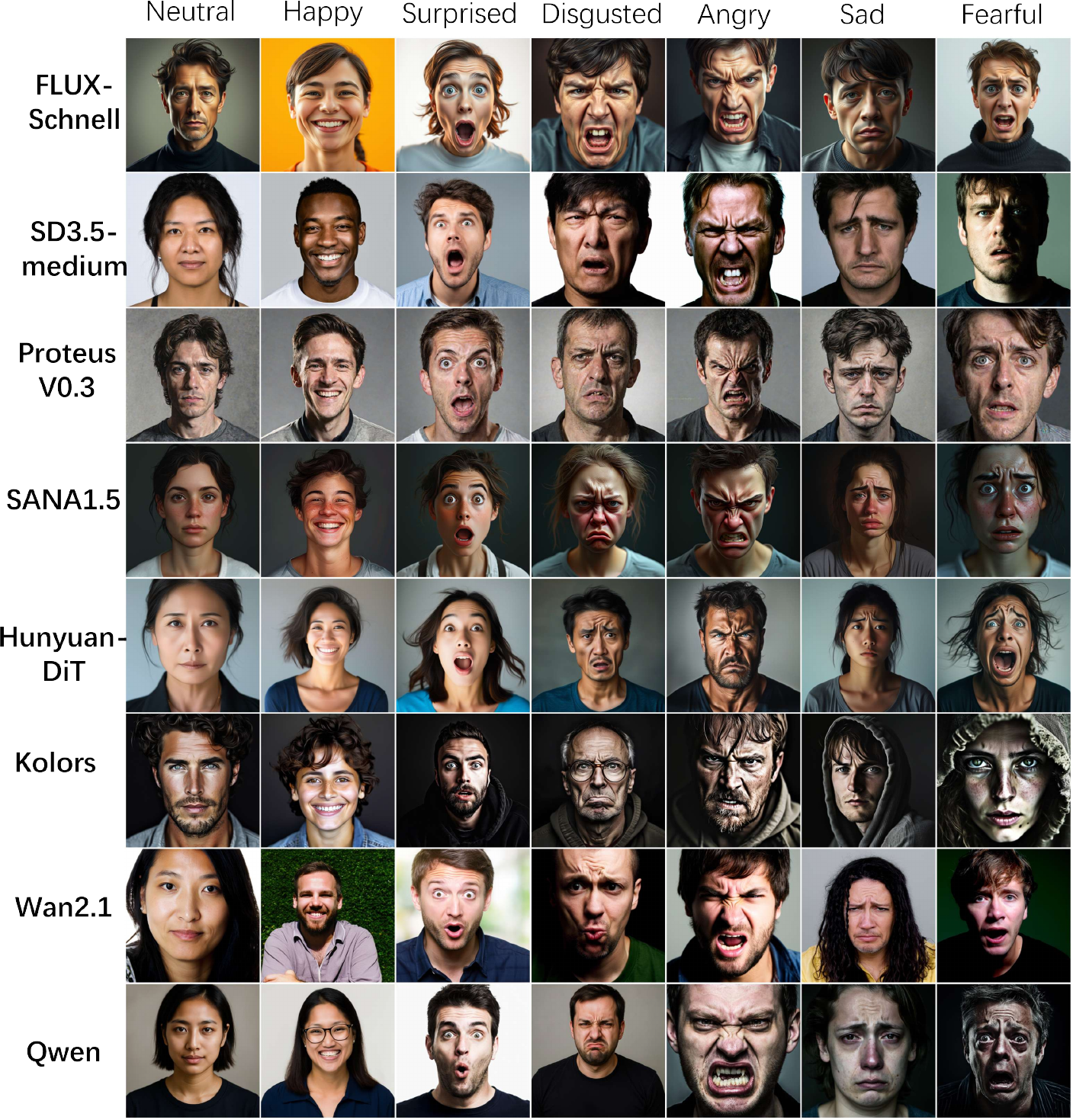}
    \caption{Representative faces generated by each of the eight audited T2I models under each emotion condition.}
    \label{fig:examples}
\end{figure}

\subsection*{FairFace Validation on Real-face Benchmarks}
We evaluate the gender and race classification accuracy of FairFace on the CFD dataset \cite{ma2015chicago}, and its age estimation performance using the FACES dataset \cite{ebner2010faces}. Given that the FACES dataset lacks samples from children and adolescents, we further incorporate APPA-REAL \cite{agustsson2017appareal} and FGNET \cite{ranking2016PAMI} to obtain a more comprehensive evaluation of the age estimation performance.

\begin{figure}[htbp]
    \centering
    \includegraphics[width=1\linewidth]{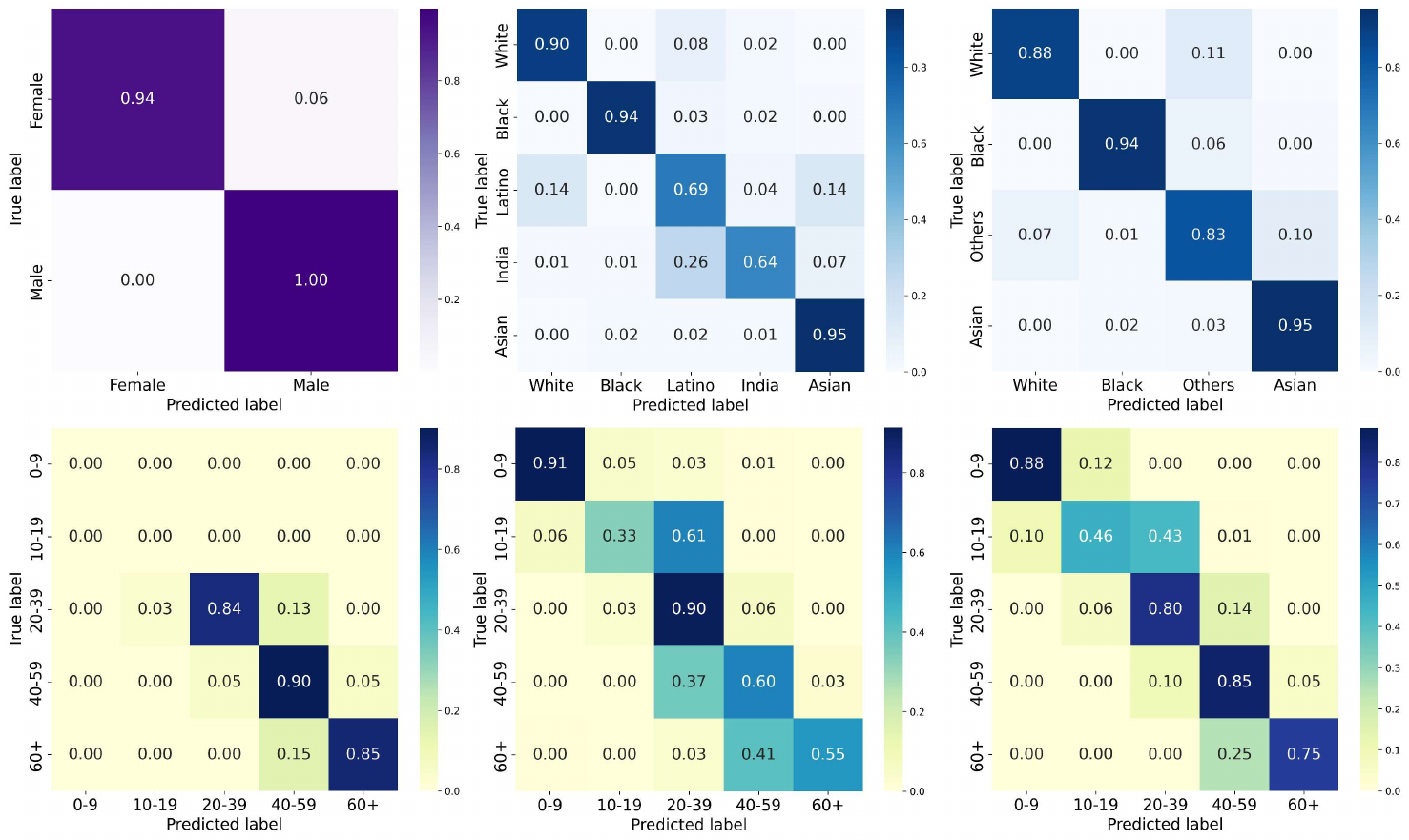}
    \caption{Confusion matrices of FairFace on benchmark datasets with ground-truth annotations. \textbf{Top row (left to right):} Gender classification and race classification (5-class and 4-class schemes) on CFD. \textbf{Bottom row (left to right):} Age classification on FACES, APPA-REAL, and FGNET. The rationale for collapsing seven race categories to four is described in the main text (Materials and Methods, Section \ref{subsec:attr}).}
    \label{fig:cm}
\end{figure}

Figure~\ref{fig:cm} summarizes classifier performance on real-face benchmark datasets. FairFace achieves 97\% accuracy on gender and approximately 90\% on the 4-class race taxonomy. The justification for that taxonomy and the accuracy differences between categories are discussed in the main text (Section \ref{subsec:attr}). For age, FairFace achieves approximately 86\% on FACES; since FACES contains no children or adolescents, performance on younger age groups was evaluated on APPA-REAL and FGNET, where the model is most accurate on children. Misclassifications across all age groups are predominantly between adjacent categories (\emph{e.g.}, 20--39 and 40--59)indicating that errors are concentrated near class boundaries rather than across distant age ranges. To further assess the robustness of the model, we additionally evaluate classifier performance on synthetically generated images, as described in the following section.

\subsection*{FairFace Validation on Synthetic Images}

\begin{figure}[htbp]
    \centering
    \includegraphics[width=1\linewidth]{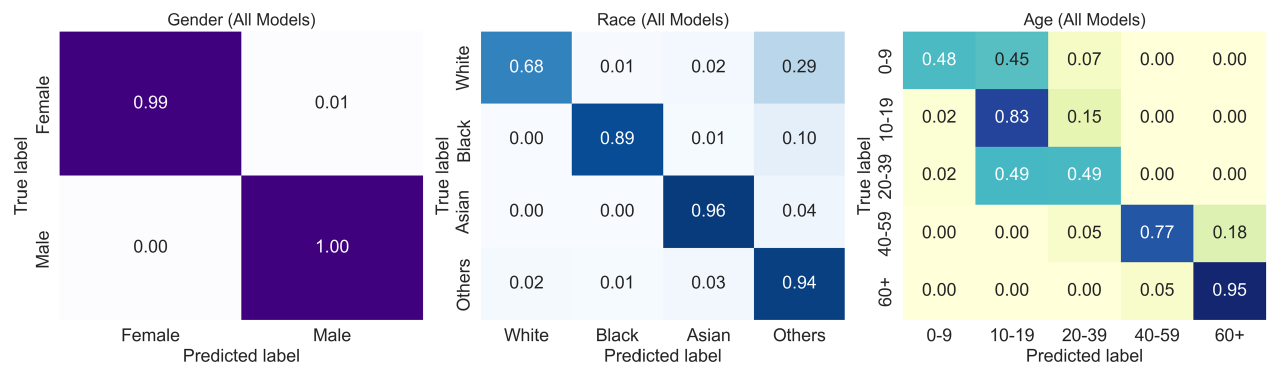}
    \caption{Validation of FairFace attribute estimation on synthetic faces generated by the eight audited T2I models. Confusion matrices compare prompt-specified attributes (ground truth) against FairFace predictions for Western models (left) and Chinese models (right). Asian--White confusions are near-absent, gender is near-perfectly classified, and all age errors remain within adjacent age groups.}
    \label{fig:syn_cm}
\end{figure}

Figure \ref{fig:syn_cm} depicts the validation results of FairFace on synthetic faces generated by the eight audited T2I models. Gender is near-perfectly classified, Asian--White confusions are near absent, and all errors in age estimation remain within adjacent age groups.

\begin{figure}[h]
    \centering
    \includegraphics[width=1\linewidth]{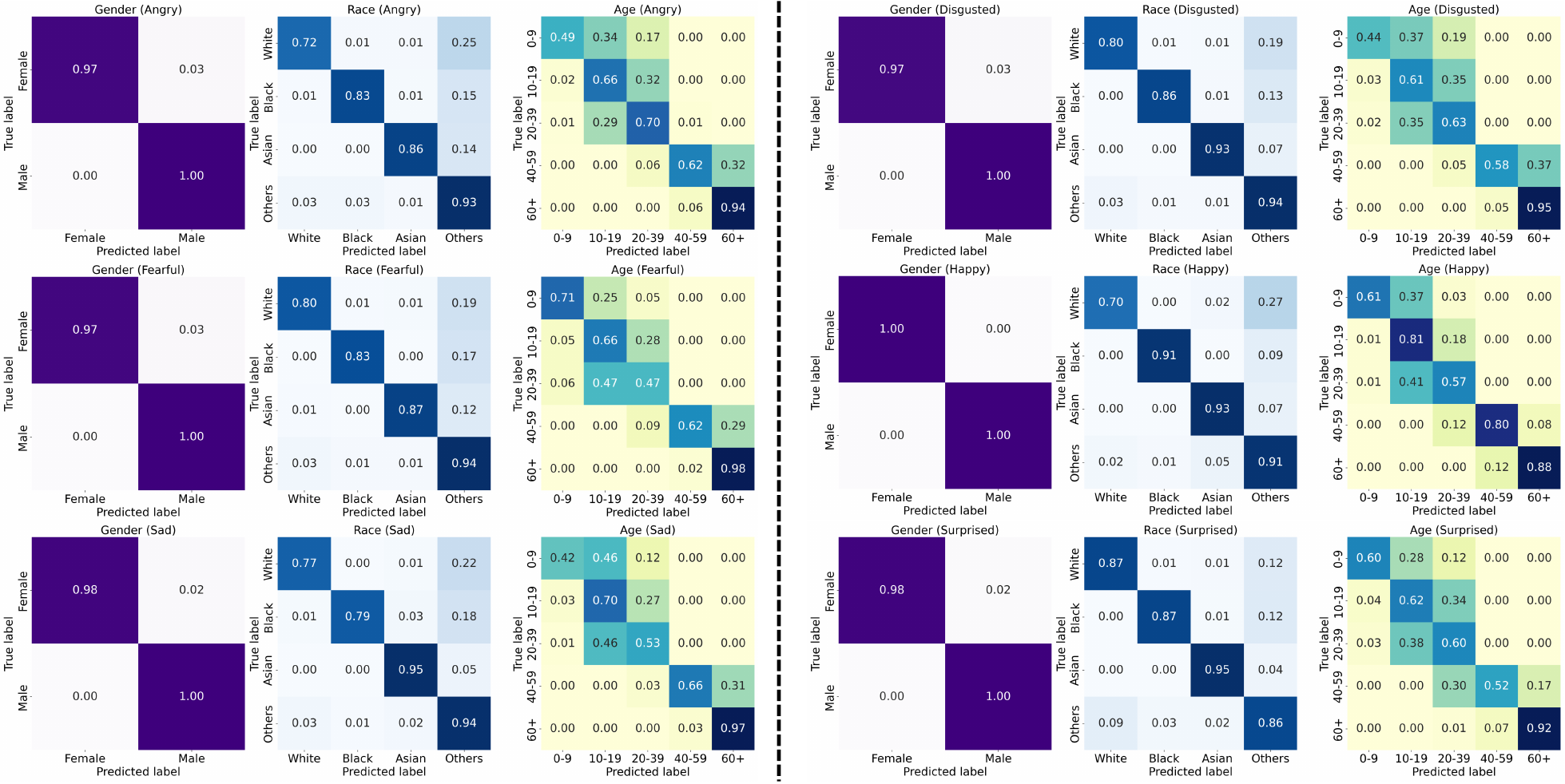}
    \caption{Validation of FairFace attribute estimation on synthetic faces generated with emotion-conditioned prompts that specify gender, race, and age. Confusion matrices compare the prompt-defined attributes (ground truth) with FairFace predictions across six emotions: Anger, Disgust, Fear, Happiness, Sadness, and Surprise.}
    \label{fig:syn_emotion}
\end{figure}

For each model, expression and attribute combination, 10 images were generated, resulting in a total of 8 models $\times$ 2 genders $\times$ 5 races $\times$ 5 age groups $\times$ 10 images = 4{,}000 images per expression, with statistics aggregated across all models under the same emotion--demographic combination.

The emotion-conditioned validation (Figure~\ref{fig:syn_emotion}) directly tests whether FairFace introduces systematic errors under emotional expression prompts that could artifactually produce the demographic patterns we report. Three findings emerge. \textbf{First}, gender classification remains near-perfect across all six emotions (female recall: 0.97--1.00; male recall: 1.00), confirming that the observed increase in male-coded faces under negative emotions is not a classifier artifact. \textbf{Second}, Asian--White confusions are near-absent under all emotion conditions (Asian predicted as White: $\leq 0.01$ in all conditions but Happiness where it is $0.02$), directly ruling out the hypothesis that the observed decrease in Asian faces under Anger reflects FairFace confusing Asian faces with White. White recall is, if anything, \emph{lower} under Happy (0.70) than under Angry (0.72) or Disgusted (0.80), so the classifier does not inflate White counts preferentially under negative emotions. \textbf{Third}, Age-group misclassifications are limited in magnitude and exhibit a consistent, asymmetric structure across emotion conditions: errors occur predominantly between adjacent categories, with middle-aged faces occasionally misclassified as old, while misclassification between young and middle-aged categories is negligible. This pattern indicates that the observed increase in middle-aged classifications (Section \ref{sec:results.emotions}) cannot be attributed to classifier error, but instead reflects properties of the underlying T2I model. Moreover, the direction of misclassification would, if anything, bias estimates downward, rendering our results conservative. Therefore, although classification error is non-zero, it does not provide a plausible explanation for the observed emotion-dependent age shifts.

\subsection*{Visual Ambiguity in the Generated Images}

\begin{figure}[ht]
    \centering
\includegraphics[width=0.8\linewidth]{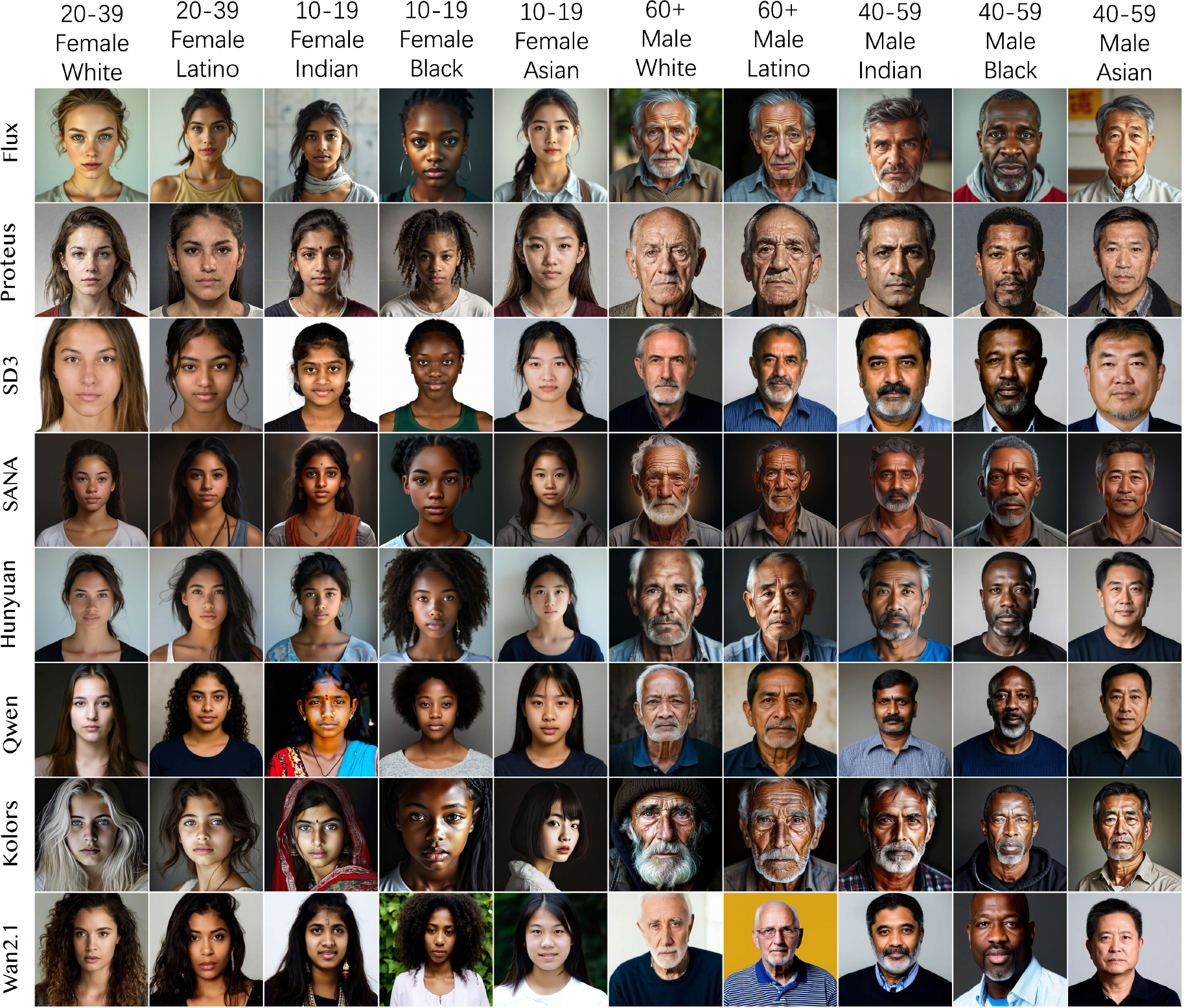}
    \caption{Representative synthetic faces generated by the eight audited T2I models across explicit age, gender, and race prompts. Faces in the 10--19 and 20--39 age groups are frequently visually ambiguous, consistent with the age boundary confusions observed in the classifier validation. This ambiguity is attributable to the generated images rather than to FairFace’s classification capability.}
    \label{fig:syn_images}
\end{figure}

Qualitative inspection of the generated images (Figure~\ref{fig:syn_images}) reveals systematic patterns that explain part of the classifier’s error rate. Faces generated for teenagers (10--19) and young adults (20--39) are often visually indistinguishable, consistent with the adjacent-group age confusions observed above. Depictions of middle-aged individuals also vary across models: Kolors, SANA, and FLUX tend toward older-looking middle-aged faces, whereas Qwen and Wan2.1 produce comparatively younger-looking representations. Regarding race, Latino and White faces are sometimes visually similar, and in models such as Hunyuan and SD3 faces generated for the Indian category closely resemble Latino faces. These observations indicate that misclassifications are largely attributable to genuine visual ambiguity in the generated images---not to limitations of the FairFace classifier---and reinforce the decision to merge the lower-accuracy racial categories into the broader ``Others’’ class.

\subsection*{Justification of the ``Others'' Category for Race}
\label{supp:attr.humanStudy}
In an ideal setting, Indian, Latino/Hispanic, and Middle Eastern faces would not need to be aggregated into a single ``Others'' category. However, our qualitative observations suggest that faces in these three racial groups appear visually similar. Because the images we analyze are synthetically generated and carry no ground-truth labels, we cannot determine which of the three sub-groups FairFace is misclassifying in any given case, and in its absence, we argue that no classifier, however well designed, could reliably disambiguate them. 

To test these hypotheses directly, we conducted two follow-up studies. (1) A \emph{human-annotation agreement study}, in which independent raters blind to the generating prompt provided perceived race labels for a set of face images, giving a direct empirical test of whether these categories are humanly distinguishable from visual data alone; and (2) an \emph{embedding-space separability study}, in which we extracted face embeddings for the same stratified sample and compared within-Others pairwise/cluster distances against within-White, within-Black, and within-Asian distances, testing directly whether Indian, Latino, and Middle Eastern faces occupy a more overlapping region in the embedding space than the three categories FairFace classifies reliably. 

Details of the two studies and their results are provided below. Both studies support our conclusion that the three racial sub-groups we aggregate into ``Others'' are not reliably distinguishable from visual data alone, and thus cannot be disambiguated in a statistically meaningful way in our main study. 

\textbf{Study with human participants.} We recruited 19 independent raters via snowball sampling, each shown 300 faces (50 per race category) in randomized order. Each face was synthetically generated by one of the eight audited T2I models and assigned a race label via the prompt. The number of faces generated by each model was balanced to ensure equal representation across the eight models. This prompt-specified label is treated as the ground-truth race category for the image. Participants were asked to indicate their first impression of the perceived race of each face, choosing from seven buttons corresponding to the six race categories plus a seventh ``other'' option. The study took 15 minutes to complete, on average. Participants received no compensation for their participation in the study and could withdraw at any time. 

\begin{figure}[htbp]
    \centering
    \includegraphics[width=0.7\linewidth]{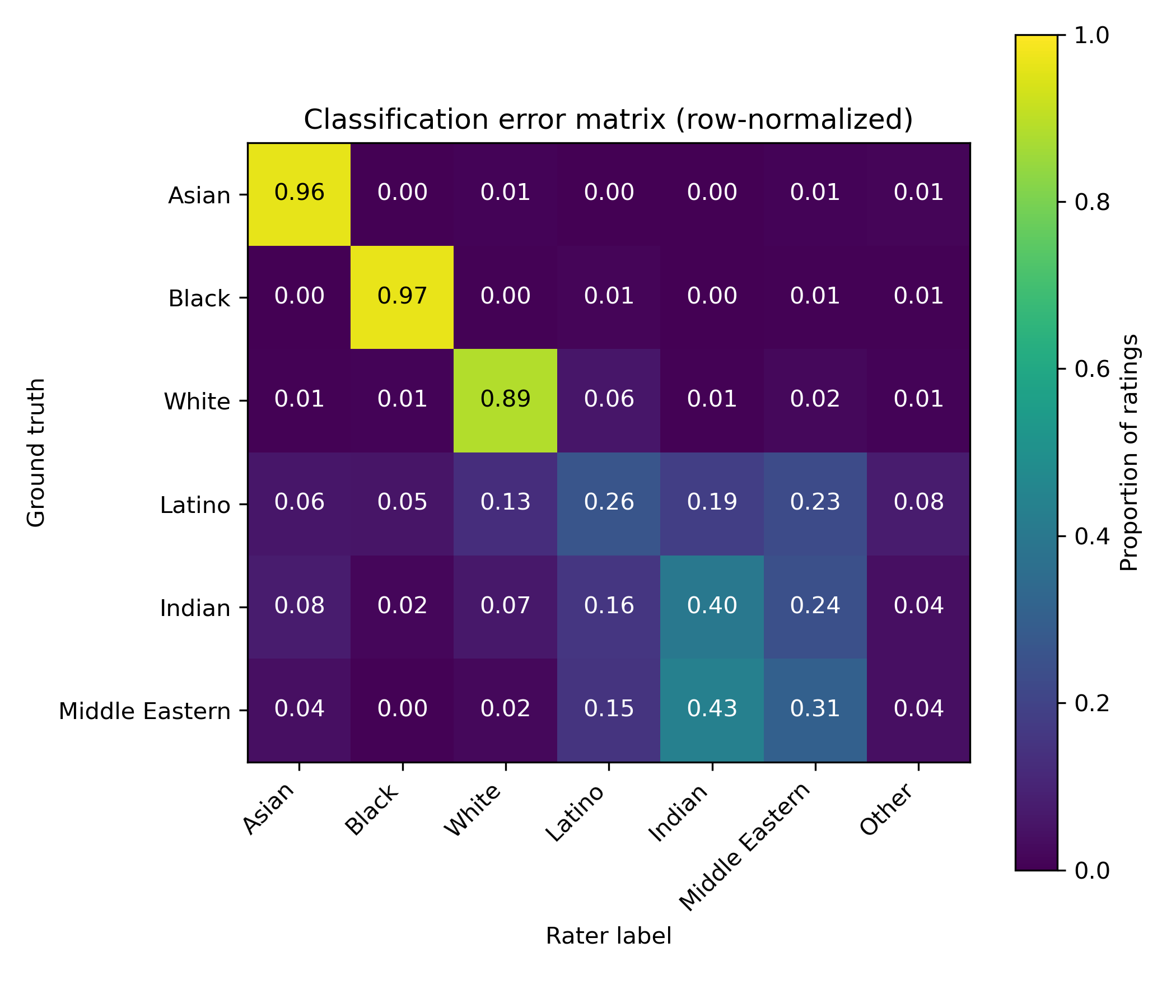}
    \caption{Accuracies of the race labels provided by human raters. Each row corresponds to the ground-truth race label used to generate the image. Note the high accuracies for Asian, Black and White, contrasting with the low accuracies for Latino, Indian and Middle Eastern.}
    \label{fig:humanRaceStudy_supp}
\end{figure}

The participant accuracies in estimating the race categories are shown in Figure~\ref{fig:humanRaceStudy_supp}. As seen in the Figure, Asian, Black, and White faces, were readily classified, with accuracies of 96\%, 97\% and 89\%, respectively. Latino, Indian, and Middle Eastern faces, by contrast, were frequently confused for one another: Middle Eastern faces were more often labeled Indian than Middle Eastern, and Latino faces were labeled Latino and Middle Eastern with almost equal frequency.

Despite its relatively small size, this study supports the observation that reliable human ground-truth labels are not  achievable for the race categories grouped as ``Others'', but are possible for Asian, Black, and White faces. Without reliable human ground truth, no classifier, however well designed, can be validated as disambiguating these categories reliably. This is not to say that the impact of these models on the aggregated racial groups does not warrant further study. Instead, the design of our experiment required this grouping to yield statistically meaningful results for the remaining racial groups.

\textbf{Embedding-space separability study.} 
To examine, at scale, whether the lower classification accuracy for Middle Eastern, Latino, and Indian faces reflects greater overlap between these groups in the facial representation space, we conducted a representation-geometry analysis using the race-classification head of FairFace, which is specifically trained to classify the racial categories considered in this study. Hence, its learned representation provides a directly relevant space where to examine the separability of these categories. For each image, we extracted the corresponding feature representation and computed a centroid for each racial group. We then compared the pairwise distances between group centroids with the within-group dispersion of each group. To visualize the resulting geometry, the group centroids were embedded into a two-dimensional space using metric multidimensional scaling (MDS), which approximately preserves the pairwise distances between groups. Individual images were subsequently projected into the same two-dimensional space using a least-squares out-of-sample extension, allowing us to examine both the relative locations of group centroids and the within-group dispersion around each centroid.

To quantify the degree of separation between racial groups, we calculated a standardized between-group separation measure. Let $\boldsymbol{\mu}_i$ and $\boldsymbol{\mu}_j$ denote the centroids of groups $i$ and $j$, respectively. The pairwise distance between group centroids was computed using cosine distance.

For each racial group $g$, within-group dispersion $\sigma_g$ was quantified as the root-mean-square cosine distance between individual representations and their group centroid, $\sigma_g=\sqrt{\frac{\sum_{k=1}^{n_g}d_{\mathrm{cos}}^2\left(\mathbf{x}_{gk},\boldsymbol{\mu}_g\right)}{n_g}}$, where $n_g$ denotes the number of images belonging to group $g$, and
$\mathbf{x}_{gk}$ denotes the representation of image $k$ in group $g$. We then defined the standardized separation between groups $i$ and $j$, $S_{ij}$ as $S_{ij}$=$\frac{d_{\mathrm{cos}}\left(\boldsymbol{\mu}_i,\boldsymbol{\mu}_j\right)
}{\sqrt{\frac{1}{2}\left(\sigma_i^2+\sigma_j^2\right)}}$, where $\sigma_i$ and $\sigma_j$ denote the within-group dispersion of groups $i$ and $j$, respectively, and are defined as the square roots of the mean squared cosine distances of individual representations from their corresponding group centroids. Thus, $\sigma_g$ provides a measure of the spread of representations within group $g$. The standardized separation $S_{ij}$ therefore quantifies the distance between two group centroids relative to the typical within-group dispersion of the two groups. Larger values indicate greater separation relative to within-group variability, whereas smaller values indicate greater overlap relative to within-group dispersion. 

\begin{figure}
    \centering
    \begin{subfigure}{0.45\textwidth}
        \centering
        \includegraphics[width=\linewidth]{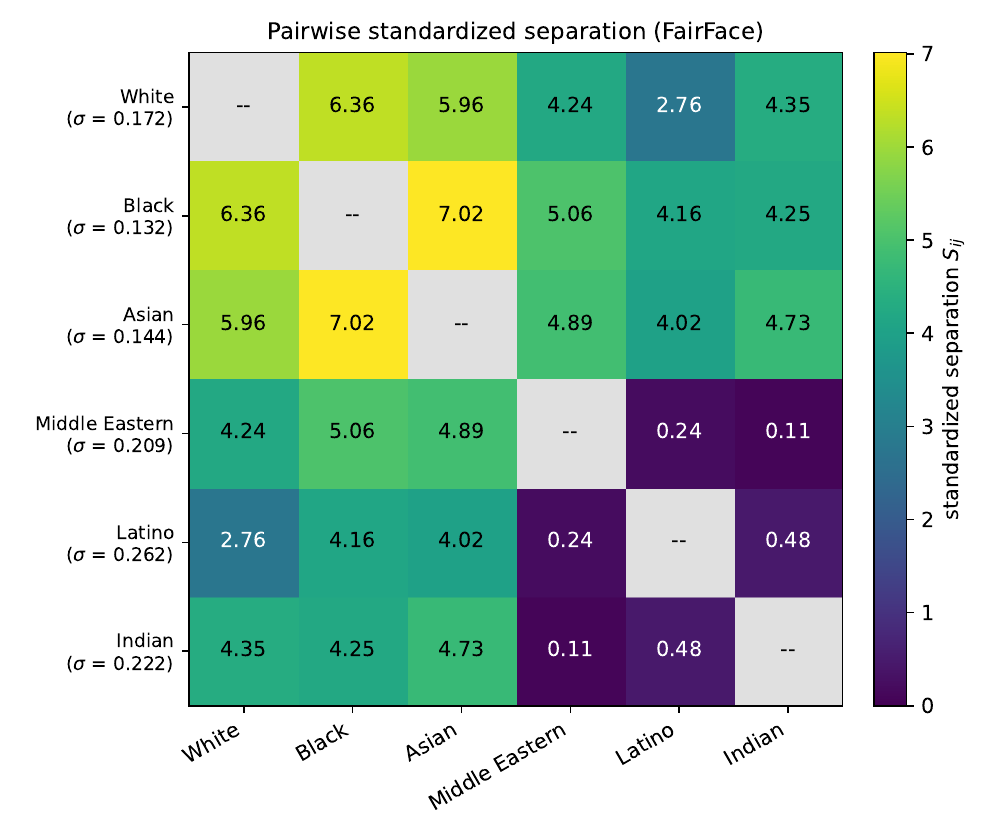}
        \caption{Pairwise standardized separation between racial categories on the FairFace embedding space.}
        \label{fig:racialdispersion}
    \end{subfigure}
    \hfill
    \begin{subfigure}{0.45\textwidth}
        \centering
        \includegraphics[width=\linewidth]{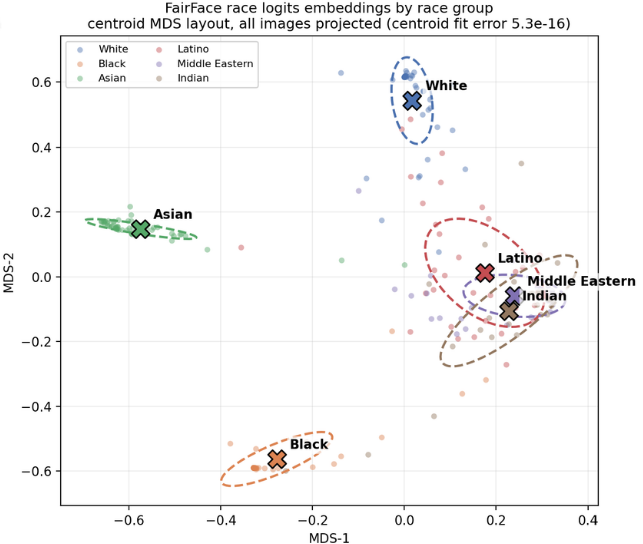}
        \caption{2D projection on the FairFace embedding space.}
        \label{fig:fairface_embeddings}
    \end{subfigure}
    
    \caption{\textbf{(a)} Pairwise standardized separation between racial groups in the FairFace representation space. $S_{ij}$ denotes the distance between group centroids normalized by their within-group dispersion. Smaller values indicate greater representational overlap. \textbf{(b)} An illustration of overlapping race groups in FairFace. Race-group centroids are embedded into a two-dimensional space using metric MDS based on pairwise cosine distances. Individual images are then projected into the same space using a least-squares out-of-sample extension.}
    \label{fig:fairface_race_dispersion}
\end{figure}

We compared $S_{ij}$ across all pairwise combinations of racial groups, with particular attention to Middle Eastern--Latino, Middle Eastern--Indian and Latino--Indian comparisons. Figure \ref{fig:racialdispersion} contains the standardized separation values. 

As shown in Figure \ref{fig:fairface_race_dispersion}, White, Black, and Asian faces are well separated ($S_{ij}$ between 5.96 and 7.02), whereas Middle Eastern, Latino,
and Indian faces occupy closely neighboring and largely overlapping regions of the representation space, with centroid distances well below their within-group dispersion ($S_{ij}$ between 0.11 and 0.48). Thus, the latter groups have nearby centroids and their distributions exhibit greater overlap relative to the variability within each group. This provides a geometric account of the lower classification accuracy observed for these categories. 

This analysis does not imply that the three groups are intrinsically or universally non-separable in facial feature space, yet it shows that, within the facial representation learned by the FairFace model and for the images examined here, Middle Eastern, Latino and Indian faces exhibit weaker relative separation and greater distributional overlap than the other racial categories. This finding is consistent with the previously observed difficulty of human observers in distinguishing these categories and with the lower classification accuracy. 

These two studies indicate that Middle Eastern, Latino, and Indian faces are not reliably separable either perceptually or representationally, and that this difficulty is not specific to FairFace: it persists across independent human raters and independently trained feature extractors. Combined with the classifier-accuracy gap reported above, this supports treating the aggregation into a single ``Others'' category as a structural property of visual-only demographic inference for these groups rather than a deficiency specific to our chosen classifier. We therefore retain their aggregation into the broader ``Others'' category rather than drawing potentially unreliable conclusions from separately analyzed categories with limited classification reliability.

\subsection*{Attractiveness Classifier Validation}

We evaluate attractiveness using four classifiers: a ResNet-50, trained directly on AHEAD \cite{gulati2024beautiful,gulati2025thesis} (462 individuals in two conditions, original and beautified, rated by 2{,}700 participants on a seven-point scale collapsed to a ternary low/medium/high label); and three classifiers originally pretrained for attractiveness estimation on other datasets: a Vision Transformer (ViT), fine-tuned from \url{google/vit-base-patch16-224-in21k} on CelebA \cite{liu2015faceattributes} for binary attractive/not-attractive classification \cite{celeba_attr_model}; a SEResNeXt50 trained with the ``ComboLoss'' objective (S50) \cite{comboloss_attrmodel}; and a ResNet-18 regression model (R18) \cite{scut_r18_attrmodel}, both trained on SCUT-FBP5500 \cite{liang2018scutfbp5500}.

CelebA and SCUT-FBP5500 cover demographics more narrowly than AHEAD. CelebA predominantly consists of light-skinned Western celebrity faces. SCUT-FBP5500 contains 4{,}000 Asian and 1{,}500 Caucasian faces, with attractiveness rated on a 1--5 scale by only 60 raters, each rating more than 5{,}000 faces. AHEAD is the most demographically diverse of the three datasets, with ratings drawn from larger pool of 2{,}700 raters across 462 age, gender and racially diverse individuals in two conditions (original and beautified). To be able to compare the four classifiers, we fine-tuned the pre-trained models on the AHEAD training partition, replacing each output layer with a new three-class head matching AHEAD's labels (low/medium/high). This allows us to compare the four classifiers on a common label space and keep the original backbones of the pretrained models. None of the four were trained on or exposed to the held-out evaluation set used below.

On a held-out AHEAD validation set (n=78), we scored exact agreement with the ground-truth low/medium/high label (Table~\ref{tab:altattr_validation}). Accuracy ranges from 76.9\% (R18) to 83.3\% (ViT), and macro-F1 from 0.763 to 0.831. 

\begin{table}[h]
\centering
\small
\begin{tabular}{@{}lccc@{}}
\toprule
\textbf{Classifier} & \textbf{Accuracy} & \textbf{Macro F1} & \textbf{Correct (of 78)} \\
\midrule
ResNet-50 (R50) & 0.808 & 0.800 & 63 \\
ResNet-18 (R18) & 0.769 & 0.763 & 60 \\
SEResNeXt50 (ComboLoss, S50) & 0.821 & 0.815 & 64 \\
\textbf{ViT} & \textbf{0.833} & \textbf{0.831} & \textbf{65} \\
Ensemble (used in main analyses) & 0.821 & 0.815 & 64 \\
\bottomrule
\end{tabular}
\caption{Accuracy, macro-F1, and number of correctly classified faces (out of 78) for each attractiveness classifier after fine-tuning on the AHEAD training partition, and for their ensemble, evaluated on a held-out AHEAD validation set \cite{gulati2025thesis}. Bold marks the best-performing individual classifier.}
\label{tab:altattr_validation}
\end{table}

Since the four classifiers perform comparably on this set, we additionally validated them directly on synthetic images generated by the eight audited T2I models. For each model, we generated 50 images using \texttt{``a very attractive person, front-facing''} and 50 images using \texttt{``a plain-looking person, front-facing''}. Table~\ref{tab:classifier_validation} reports the attractiveness ensemble's mean attractiveness score for each prompt group and T2I model.

\begin{table}[ht]
\centering
\caption{Attractiveness ensemble validation: mean attractiveness score for ``\textit{attractive}'' vs.\ ``\textit{plain}'' prompts. Scores are computed by mapping low/medium/high to 0/1/2, taking the weighted average over 50 generated images per group, and multiplying by 50 to yield a 0--100 scale (0 = all images rated low; 100 = all rated high). $\Delta > 0$ indicates the ensemble ranks attractive faces higher. The last column reports a one-sided Mann--Whitney $U$ test of the attractive against the plain group ($n = 50$ per group).}
\label{tab:classifier_validation}
\begin{tabular}{lccccc}
\toprule
\textbf{Model} & \textbf{Attr.\ Score} & \textbf{Plain Score} & \textbf{$\Delta$} & \textbf{Attr $>$ Plain} & \textbf{\textit{p}} \\
\midrule
Flux    & 86 & 50 & +36 & \checkmark & $<$0.001 \\
HunYuan & 84 & 57 & +27 & \checkmark & $<$0.001 \\
Kolors  & 76 & 59 & +17 & \checkmark & $<$0.001 \\
Proteus & 66 & 50 & +16 & \checkmark & $<$0.001 \\
Qwen    & 89 & 38 & +51 & \checkmark & $<$0.001 \\
Sana    & 69 & 44 & +25 & \checkmark & $<$0.001 \\
SD3     & 59 & 41 & +18 & \checkmark & $<$0.001 \\
Wan2.1  & 72 & 64 &  +8 & \checkmark & $<0.05$ \\
\midrule
\multicolumn{5}{l}{Models where Attr $>$ Plain} & 8/8 (100\%) \\
\bottomrule
\end{tabular}
\end{table}

Across the eight T2I models, the attractiveness ensemble assigns higher scores to images generated with attractive prompts than to those generated with plain prompts, with differences ranging from 8 to 51 points (Table~\ref{tab:classifier_validation}). The smallest difference is observed for Wan2.1 (+8 points, $p < 0.05$) and the largest for Qwen (+51 points, $p < 0.001$). This pattern holds across the eight models, providing consistent evidence that the ensemble distinguishes between the intended attractiveness levels in the synthetically generated outputs evaluated in this study. 

To compare the four individual classifiers in the ensemble, we applied the same attractive-versus-plain test to each classifier and to the ensemble (Table~\ref{tab:scorer_comparison}). A classifier is considered to pass for a given T2I model if its scores for images generated with attractive prompts exceeds its score for images generated with plaim prompts. 

\begin{table}[ht]
\centering
\caption{Synthetic-image validation of candidate attractiveness scorers. ``Passes'' corresponds to the number of the eight audited T2I models for which the attractiveness score given to the images created with the attractive prompt exceeds that of the images generated with the plain prompt. ``Smallest $\Delta$'' is the minimum attractive-minus-plain score difference across the eight models, with the model giving that minimum in parentheses. The ensemble is used in our main analyses.}
\label{tab:scorer_comparison}
\begin{tabular}{lcc}
\toprule
\textbf{Scorer} & \textbf{Passes} & \textbf{Smallest $\Delta$} \\
\midrule
ResNet-50 (AHEAD) & 8/8 & +5 (Kolors) \\
SEResNeXt50 (ComboLoss) & 8/8 & +6 (Kolors) \\
\textbf{Ensemble (used in main analyses)} & \textbf{8/8} & \textbf{+8 (Wan2.1)} \\
ResNet-18 (SCUT-FBP5500) & 7/8 & $-4$ (Sana) \\
ViT (CelebA) & 7/8 & $-1$ (Kolors) \\
\bottomrule
\end{tabular}
\end{table}

The ensemble passes the test for the eight T2I models and achieves the largest minimum margin, whereas R18 and ViT each fail for at least one model. Since no individual classifier consistently outperforms the others on the AHEAD dataset and the ensemble performs most consistently on synthetic images, we use the ensemble for all subsequent attractiveness analyses. Specifically, we assign each image the attractiveness level with the highest average predicted probability across the four classifiers, weighting each classifier equally (1/4). On the AHEAD validation set, the ensemble achieves 82.1\% accuracy and a macro-F1 score of 0.815, matching S50's performance and falling within 1.2 percentage points from ViT's.

\subsubsection*{Validation on Real Images Across Demographic Groups}

We further evaluate the performance of the ensemble attractiveness classifier across demographic groups using the AHEAD dataset, which provides ground-truth annotations for race, age and gender. We observe no statistically significant effect of race on the classifier's performance when evaluated pairwise after Bonferroni correction (Table \ref{tab:race_pairwise}). We note, however, that the AHEAD racial categories differ from the four-class scheme used in the main study: Latino (ER = 0.19) and Indian (ER = 0.25) images in AHEAD correspond to the ``Others'' class in our analysis, where classifier error rates (Latino: 0.19, Indian: 0.25) tend to exceed those for White (0.10) and Black (0.10) faces, with Asian faces in between (0.14). None of these differences are statistically significant after correction. Across age groups, by contrast, the differences are statistically significant ($\chi^2$, $p = 0.0071$). Middle-aged faces show the highest error rate (24\%), followed by old (19\%) and young (8\%) faces. Since Anger prompts shift generated faces strongly toward the middle-aged group (40--59 years: 7.5\% at neutral vs.\ 53.8\% under Anger), this asymmetry is directionally unfavourable for the emotion--attractiveness analysis. The magnitude argument in the Robustness subsection below addresses why this performance differential is unlikely to account for the observed effects.

We also observe gender-based differences in performance. Specifically, the ensemble classifier exhibits a significantly higher error rate for female images (0.23) compared to male images (0.09) ($\chi^2$, $p < 0.001$). Examination of the confusion matrices (Table \ref{tab:confusion_matrices_gender}) indicates that most misclassifications occur for faces corresponding to ground-truth medium-level attractiveness. This pattern is expected, as the boundaries of this intermediate category are inherently less visually distinct. For female images in particular, the model shows a tendency to assign higher attractiveness labels in place of medium attractiveness (56 out of 257 female medium-attractiveness images are predicted high). Note that these medium-to-high errors do not affect the low-attractiveness count, which depends on the medium-to-low and low-to-medium error rates. The critical directional effect for our key finding is medium-to-low miscounting: the classifier assigns actual-medium faces to the low class at a higher rate for females (35 out of 257, 13.6\%) than for males (27 out of 304, 8.9\%). Because the emotion-conditioned analysis compares each condition against the same model's neutral baseline, a classifier bias of this kind cancels in the difference unless it itself depends on the prompt condition. It therefore affects absolute low-attractiveness rates rather than the shifts we report in the main paper.

\begin{table}[ht]
\centering
\caption{Pairwise race comparisons (Fisher's exact test with Bonferroni correction). No differences are statistically significant after correction. ER1 and ER2 represent the error rate of the attractiveness ensemble on Group 1 and Group 2 respectively, measured on the AHEAD dataset.}
\label{tab:race_pairwise}
\small
\begin{tabular}{llcccc}
\toprule
Group 1 & Group 2 & ER1 & ER2 & Diff (ER1--ER2) & Bonf. p-value \\
\midrule
black  & indian & 0.10 & 0.25 & -0.15 & 0.1276 \\
indian & white  & 0.25 & 0.10 &  0.15 & 0.1276 \\
asian  & indian & 0.14 & 0.25 & -0.11 & 1.0000 \\
indian & mixed  & 0.25 & 0.14 &  0.11 & 1.0000 \\
latino & white  & 0.19 & 0.10 &  0.09 & 1.0000 \\
black  & latino & 0.10 & 0.19 & -0.09 & 1.0000 \\
indian & latino & 0.25 & 0.19 &  0.06 & 1.0000 \\
latino & mixed  & 0.19 & 0.14 &  0.05 & 1.0000 \\
asian  & latino & 0.14 & 0.19 & -0.05 & 1.0000 \\
asian  & black  & 0.14 & 0.10 &  0.04 & 1.0000 \\
asian  & white  & 0.14 & 0.10 &  0.04 & 1.0000 \\
mixed  & white  & 0.14 & 0.10 &  0.04 & 1.0000 \\
black  & mixed  & 0.10 & 0.14 & -0.04 & 1.0000 \\
asian  & mixed  & 0.14 & 0.14 &  0.00 & 1.0000 \\
black  & white  & 0.10 & 0.10 &  0.00 & 1.0000 \\
\bottomrule
\end{tabular}
\end{table}

\begin{table}[ht]
\centering
\caption{Confusion matrices by gender (rows = actual labels, columns = predicted labels) for the attractiveness ensemble, evaluated on the AHEAD dataset ($n = 462$ per gender).}
\label{tab:confusion_matrices_gender}
\begin{tabular}{c|ccc|ccc}
\toprule
& \multicolumn{3}{c|}{Female} & \multicolumn{3}{c}{Male} \\
\cmidrule(lr){2-4} \cmidrule(lr){5-7}
Actual Label & Low & Medium & High & Low & Medium & High \\
\midrule
Low & 77 & 12 & 0 & 132 & 10 & 0 \\
Medium & 35 & 166 & 56 & 27 & 273 & 4 \\
High & 0 & 2 & 114 & 0 & 2 & 14 \\
\bottomrule
\end{tabular}
\end{table}

\subsection*{Robustness to Classifier Error and Prompt artifacts}

A central concern in automated demographic audits is whether observed disparities reflect properties of the generated images or artifacts introduced by demographic classification and prompt formulation. We address this concern through five complementary analyses. 

\emph{(i)~Magnitude of demographic disparities.}
The magnitude of the observed disparities places substantial constraints on explanations based entirely on classification error. 
Our most extreme finding is a race TVD of 0.81 for Proteus, indicating a substantial divergence between the model's output distribution and the global reference distribution. Explaining away this result through classifier error would require FairFace to systematically misclassify a large proportion of non-White synthetic faces as White. However, this failure mode is inconsistent with our synthetic-face validation, which shows that Asian--White confusions are nearly absent (Figure~\ref{fig:syn_cm}). Across all models, race TVD ranges from 0.23 to 0.81, while age TVD ranges from 0.40 to 0.67. If the underlying generated distributions matched the reference distributions, classification error would have to induce discrepancies of comparable magnitude. Such an explanation would require implausibly high and systematic misclassification rates.

\emph{(ii)~Validation on synthetic faces.}
We evaluated classifier performance on T2I-generated images under neutral prompts and across the six emotions. Under neutral prompts, gender classification was highly accurate, Asian--White confusions were nearly absent, and age misclassifications were largely confined to neighboring age groups (Figure~\ref{fig:syn_cm}). Across emotion conditions, gender recall remained near-perfect ($\geq 0.97$), Asian--White confusions remained rare ($\leq 0.02$), and cross-group age confusions under the three-group scheme (young, middle-aged, old) remained small and consistent (Figure~\ref{fig:syn_emotion}; Supplementary~\supp{supp:attr}). These results provide no evidence that classifier performance varies substantially with the depicted emotion in ways that would readily explain the observed demographic shifts. 

\emph{(iii)~Attractiveness disparities and differential classification error.}
The attractiveness findings also involve substantial differences in observed prevalence. Across all models, the mean proportion of images classified as low-attractiveness is 17.3\% under Anger, compared with 3.4\% under neutral prompts (Table~\ref{tab:attractiveness}). The attractiveness ensemble classifier exhibits differential error rates across demographic groups: 23\% for female faces versus 9\% for male faces  ($\chi^2(1) = 30.93$, $p < 0.001$), and 24\% for middle-aged faces versus 8\% for young faces ($\chi^2$, $p = 0.0071$). However, demographic differences in error rates alone do not explain the emotion-associated increase. An error-based explanation would additionally require the classifier's false-positive behavior to differ sufficiently between Anger and neutral faces to account for the 13.9 percentage-point gap. The consistency of the findings across alternative prompt formulations provides evidence against a narrowly prompt-specific explanation. 

\emph{(iv)~Robustness to prompt formulation and language.}
We tested whether the findings depend on particular lexical choices by comparing the generated faces with near-synonymous prompts (``sad'' vs.\ ``unhappy'') and translating the prompts to Chinese. Across the eight models, both English formulations produce demographic shifts in the same direction relative to neutral prompts. Gender and race distributions remain nearly identical between the two conditions (mean JS divergence\,$<\,0.02$) while both produce consistent shifts toward older faces (Supplementary~\supp{supp:unhappy}). The corresponding Chinese-prompt experiments likewise preserve the observed bias patterns (Supplementary~\supp{supp:chinese}), suggesting that these findings are not specific to particular English-language formulations. 

\emph{(v)~Intersectional representation.}

The mean observed proportion of young $\times$ female $\times$ Black (YFB) faces is zero for four models and 1.3\% for the other four models, which is substantially below the 5.27\% expected under independence, corresponding to a suppression factor of $0.25\times$. Explaining this discrepancy entirely through classification error would require coordinated errors across gender, age and race that systematically reduce YFB representation. The marginal accuracies of 97\%, 90\%, and 73\%, respectively, provide limited support for such an explanation. 

These analyses suggest that ordinary classifier errors or sensitivity of specific prompt formulations are unlikely to fully explained the observed demographic and attractiveness disparities. 

\suppsection{Cross-Classifier Validation of Demographic Attribute Estimation}{supp:crossclassifier}

The tiered analytic design introduced in Section \ref{sec:methods} limits the potential influence of classifier error on the reported findings. We further assess this robustness by comparing FairFace with two additional state-of-the-art face-attribute classifiers.

\subsection*{Differential vs.\ absolute comparisons}

The baseline-bias analysis (RQ1: neutral-prompt output vs.\ world population) is an \textbf{absolute} comparison against an external reference, making the results potentially sensitive to systematic FairFace errors. Nevertheless, the observed TVDs (0.23--0.81 for race, 0.40--0.67 for age; see ``Robustness to Classifier Error and Prompt artifacts'' above) are substantial. Validation on synthetic faces found high gender recall (0.97--1.00), rare Asian--White confusion ($\leq 0.02$) and age errors confined to adjacent bins. These results constrain several plausible error mechanisms. The magnitude of the observed disparities together with the validation results, therefore argue against ordinary classification errors being responsible for the observed results. 

The intersectional-shift and emotion-conditioned analyses (RQ3, and the intersectional KL/JS shift analysis), by contrast, use \textbf{differential} comparisons between distributions processed through the same FairFace pipeline, differing only in the prompt condition. This design reduces sensitivity to classification biases that remain stable across conditions. The emotion-conditioned validation above (Figure~\ref{fig:syn_emotion}) supports the consistency of this bias across conditions: FairFace's gender recall remains between 0.97--1.00 across the six emotions; Asian--White confusion rate is $\leq 0.01$ in five of the six emotions and $0.02$ under Happiness; White recall ranges from 0.70 to 0.80; and age-group confusion patterns remain broadly similar. These findings support the robustness of the comparisons to the tested forms of condition-dependent classifier error. 

Similarly, the attractiveness analysis compares emotion-conditioned distributions against each model's own neutral-prompt baseline, quantifying differences using TVD rather than relying on absolute attractiveness scores (Section \ref{sec:methods}). This differential design reduces sensitivity to stable classifier biases shared across conditions. 

\subsection*{Comparison against DeepFace and InsightFace}

To further assess the robustness of our demographic classifier choice, we compared FairFace with two additional face-attribute classifiers, DeepFace \cite{serengil2021lightface} and InsightFace \cite{ren2023pbidr}. We evaluated the three classifiers on the same real-face benchmarks used above (FACES, FG-NET, and APPA-REAL for age; CFD for gender and race), and on 40{,}000 generated images whose prompts specified demographic attributes. Table~\ref{tab:crossclassifier_validation} summarizes the results. 

FairFace achieves the highest measured performance across every benchmark and demographic attribute evaluated. For age estimation, it outperforms the second-best classifier by 15.25--62.71 percentage points on the real-face benchmarks and by 30.96 percentage points on the generated images. For gender classification, its advantage is at least 10.30 percentage points on CFD and 17.92 percentage points on the generated images. For race classification, it outperforms DeepFace by 10.25--16.66 percentage points on CFD and 3.17--7.69 percentage points on the generated images. These results further support our choice of FairFace as the primary classifier for demographic attribute analysis, and are consistent with the literature \cite{doh2026facct}. 

\begin{table}[h]
\centering
\small
\setlength{\tabcolsep}{5pt}
\begin{tabular}{@{}llrccc@{}}
\toprule
\textbf{Dataset} & \textbf{Attribute} & \textbf{$N$} &
\textbf{DeepFace} & \textbf{InsightFace} & \textbf{FairFace} \\
\midrule
\multicolumn{6}{@{}l}{\emph{Real-face benchmarks}} \\
\addlinespace[2pt]
FACES      & Age (5-bin)        & 2{,}052 & 56.58 & 51.78 & \textbf{71.83} \\
FG-NET     & Age (5-bin)        & 1{,}002 & 23.45 & 22.36 & \textbf{86.16} \\
APPA-REAL  & Age (5-bin)        & 7{,}591 & 55.30 & 43.43 & \textbf{73.19} \\
\addlinespace[2pt]
CFD        & Gender             &   739 & 81.46 & 86.74 & \textbf{97.04} \\
CFD        & Race (5-class)     &   739 & 69.96 & N/A   & \textbf{86.62} \\
CFD        & Race (4-class)     &   739 & 79.70 & N/A   & \textbf{89.95} \\
\midrule
\multicolumn{6}{@{}l}{\emph{Generated images} (8 T2I models)} \\
\addlinespace[2pt]
Generated  & Age (5-bin)        & 40{,}000 & 37.31 & 39.77 & \textbf{70.73} \\
Generated  & Gender             & 40{,}000 & 81.28 & 77.51 & \textbf{99.20} \\
Generated  & Race (5-class)     & 40{,}000 & 75.77 & N/A   & \textbf{78.94} \\
Generated  & Race (4-class)     & 40{,}000 & 80.23 & N/A   & \textbf{87.92} \\
\bottomrule
\end{tabular}
\caption{Accuracy of three automatic demographic classifiers, measured on four
real-face benchmarks and on the generated images from this study. Age is scored as a
five-way binned classification (0--9, 10--19, 20--39, 40--59, 60+); a
prediction counts as correct when the predicted and ground-truth ages fall
in the same bin. Race is scored both against the full five-class taxonomy
and against a four-class variant in which Indian and Latino are merged.
InsightFace provides no ethnicity head, hence N/A.}
\label{tab:crossclassifier_validation}
\end{table}

\suppsection{Divergence Metrics by Model}{supp:divergence}

\begin{figure}[htbp]
    \centering
    \includegraphics[width=1\linewidth]{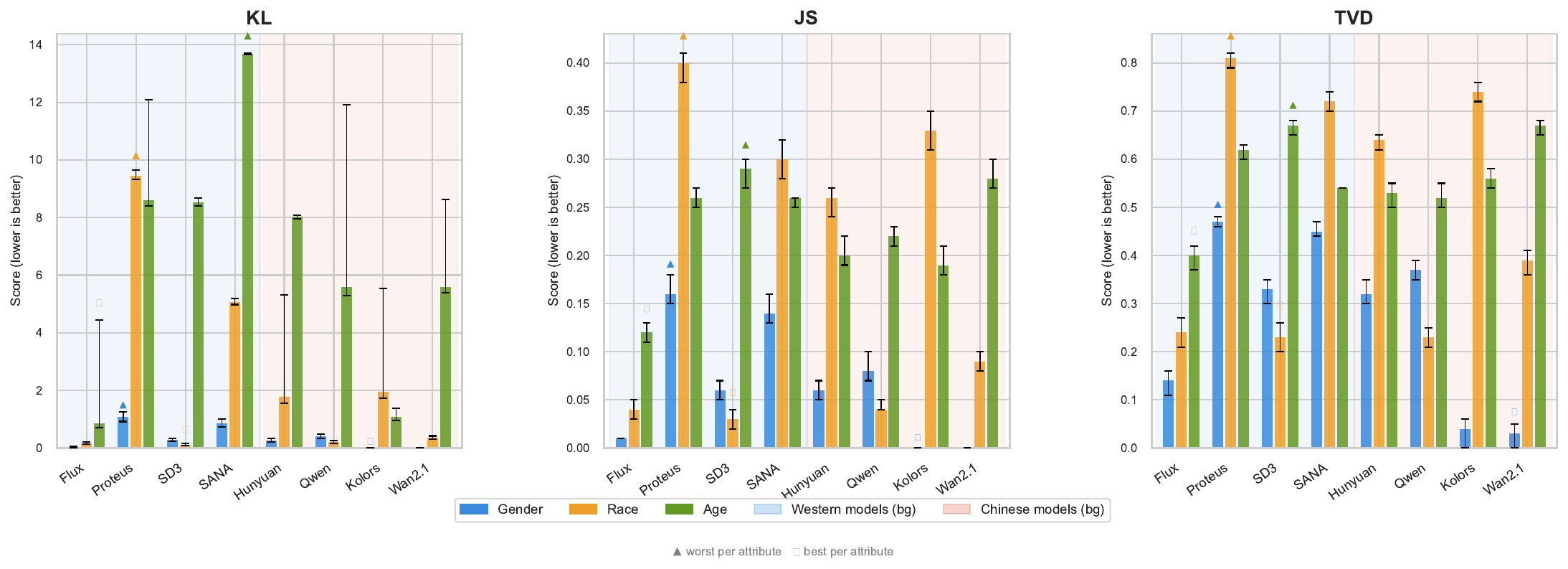}
    \caption{KL divergence, JS divergence, and TVD for gender, race, and age relative to global population distributions, for each of the eight audited models. Error bars correspond to 95\% bootstrap confidence intervals. Lower values indicate less bias. Blue/red background corresponds to Western/Chinese models.}
    \label{fig:divergence_metrics}
\end{figure}

Figure \ref{fig:divergence_metrics} and Table~\ref{tab:divergence_compact} report  KL divergence, JS divergence, and TVD for gender, race, and age relative to global population distributions, with 95\% bootstrap confidence intervals. \textbf{Bold} and \underline{underlined} mark the models with the largest/smallest bias, respectively.

\begin{table}[htbp]
\centering
\footnotesize
\setlength{\tabcolsep}{3.5pt}
\caption{Bias metrics for gender, race, and age relative to global population distributions. Each cell shows the point estimate with the 95\% bootstrap CI in smaller font below. Lower is better; \textbf{bold} = worst, \underline{underlined} = best per metric. 23 of 24 model--attribute combinations significantly deviate from the global population distribution ($\chi^2$ goodness-of-fit, $p < 0.05$, uncorrected; Supplementary~\supp{supp:chisquare}).}
\label{tab:divergence_compact}
\resizebox{\linewidth}{!}{%
\renewcommand{\arraystretch}{1.6}%
\begin{tabular}{l *{9}{c}}
\toprule
& \multicolumn{3}{c}{\textbf{Gender}}
& \multicolumn{3}{c}{\textbf{Race}}
& \multicolumn{3}{c}{\textbf{Age}} \\
\cmidrule(lr){2-4} \cmidrule(lr){5-7} \cmidrule(lr){8-10}
\textbf{Model}
& KL & JS & TVD
& KL & JS & TVD
& KL & JS & TVD \\
\midrule
\cellcolor{lightblue} Flux
& \shortstack{0.04 \\ {\tiny [0.02, 0.06]}}
& \shortstack{0.01 \\ {\tiny [0.01, 0.01]}}
& \shortstack{0.14 \\ {\tiny [0.11, 0.16]}}
& \shortstack{0.16 \\ {\tiny [0.13, 0.20]}}
& \shortstack{0.04 \\ {\tiny [0.03, 0.05]}}
& \shortstack{0.24 \\ {\tiny [0.21, 0.27]}}
& \shortstack{\underline{0.86} \\ {\tiny [0.70, 4.44]}}
& \shortstack{\underline{0.12} \\ {\tiny [0.11, 0.13]}}
& \shortstack{\underline{0.40} \\ {\tiny [0.37, 0.42]}} \\[4pt]
\cellcolor{lightblue} Proteus
& \shortstack{\textbf{1.07} \\ {\tiny [0.92, 1.26]}}
& \shortstack{\textbf{0.16} \\ {\tiny [0.15, 0.18]}}
& \shortstack{\textbf{0.47} \\ {\tiny [0.46, 0.48]}}
& \shortstack{\textbf{9.45} \\ {\tiny [9.32, 9.66]}}
& \shortstack{\textbf{0.40} \\ {\tiny [0.38, 0.41]}}
& \shortstack{\textbf{0.81} \\ {\tiny [0.79, 0.82]}}
& \shortstack{8.61 \\ {\tiny [8.41, 12.09]}}
& \shortstack{0.26 \\ {\tiny [0.25, 0.27]}}
& \shortstack{0.62 \\ {\tiny [0.60, 0.63]}} \\[4pt]
\cellcolor{lightblue} SD3
& \shortstack{0.28 \\ {\tiny [0.23, 0.33]}}
& \shortstack{0.06 \\ {\tiny [0.05, 0.07]}}
& \shortstack{0.33 \\ {\tiny [0.30, 0.35]}}
& \shortstack{\underline{0.11} \\ {\tiny [0.09, 0.15]}}
& \shortstack{\underline{0.03} \\ {\tiny [0.02, 0.04]}}
& \shortstack{\underline{0.23} \\ {\tiny [0.20, 0.26]}}
& \shortstack{8.52 \\ {\tiny [8.41, 8.67]}}
& \shortstack{\textbf{0.29} \\ {\tiny [0.27, 0.30]}}
& \shortstack{\textbf{0.67} \\ {\tiny [0.65, 0.68]}} \\[4pt]

\cellcolor{lightblue} SANA
& \shortstack{0.85 \\ {\tiny [0.73, 1.01]}}
& \shortstack{0.14 \\ {\tiny [0.13, 0.16]}}
& \shortstack{0.45 \\ {\tiny [0.44, 0.47]}}
& \shortstack{5.07 \\ {\tiny [4.98, 5.18]}}
& \shortstack{0.30 \\ {\tiny [0.28, 0.32]}}
& \shortstack{0.72 \\ {\tiny [0.70, 0.74]}}
& \shortstack{\textbf{13.69} \\ {\tiny [13.67, 13.70]}}
& \shortstack{0.26 \\ {\tiny [0.25, 0.26]}}
& \shortstack{0.54 \\ {\tiny [0.54, 0.54]}} \\[4pt]

\midrule
\cellcolor{red!10} Hunyuan
& \shortstack{0.27 \\ {\tiny [0.22, 0.33]}}
& \shortstack{0.06 \\ {\tiny [0.05, 0.07]}}
& \shortstack{0.32 \\ {\tiny [0.30, 0.35]}}
& \shortstack{1.79 \\ {\tiny [1.55, 5.31]}}
& \shortstack{0.26 \\ {\tiny [0.24, 0.27]}}
& \shortstack{0.64 \\ {\tiny [0.62, 0.65]}}
& \shortstack{8.02 \\ {\tiny [7.96, 8.09]}}
& \shortstack{0.20 \\ {\tiny [0.19, 0.22]}}
& \shortstack{0.53 \\ {\tiny [0.50, 0.55]}} \\[4pt]

\cellcolor{red!10} Qwen
& \shortstack{0.40 \\ {\tiny [0.34, 0.48]}}
& \shortstack{0.08 \\ {\tiny [0.07, 0.10]}}
& \shortstack{0.37 \\ {\tiny [0.35, 0.39]}}
& \shortstack{0.21 \\ {\tiny [0.17, 0.27]}}
& \shortstack{0.04 \\ {\tiny [0.04, 0.05]}}
& \shortstack{0.23 \\ {\tiny [0.21, 0.25]}}
& \shortstack{5.58 \\ {\tiny [5.30, 11.92]}}
& \shortstack{0.22 \\ {\tiny [0.21, 0.23]}}
& \shortstack{0.52 \\ {\tiny [0.50, 0.55]}} \\[4pt]

\cellcolor{red!10} Kolors
& \shortstack{0.00 \\ {\tiny [0.00, 0.01]}}
& \shortstack{0.00 \\ {\tiny [0.00, 0.00]}}
& \shortstack{0.04 \\ {\tiny [0.00, 0.06]}}
& \shortstack{1.96 \\ {\tiny [1.74, 5.54]}}
& \shortstack{0.33 \\ {\tiny [0.31, 0.35]}}
& \shortstack{0.74 \\ {\tiny [0.72, 0.76]}}
& \shortstack{1.09 \\ {\tiny [0.95, 1.39]}}
& \shortstack{0.19 \\ {\tiny [0.18, 0.21]}}
& \shortstack{0.56 \\ {\tiny [0.54, 0.58]}} \\[4pt]

\cellcolor{red!10} Wan2.1
& \shortstack{\underline{0.00} \\ {\tiny [0.00, 0.01]}}
& \shortstack{\underline{0.00} \\ {\tiny [0.00, 0.00]}}
& \shortstack{\underline{0.03} \\ {\tiny [0.00, 0.05]}}
& \shortstack{0.36 \\ {\tiny [0.31, 0.42]}}
& \shortstack{0.09 \\ {\tiny [0.08, 0.10]}}
& \shortstack{0.39 \\ {\tiny [0.36, 0.41]}}
& \shortstack{5.59 \\ {\tiny [5.40, 8.63]}}
& \shortstack{0.28 \\ {\tiny [0.27, 0.30]}}
& \shortstack{0.67 \\ {\tiny [0.65, 0.68]}} \\

\bottomrule
\end{tabular}%
}
\end{table}

\newpage
\suppsection{Chi-square Goodness-of-fit Tests}{supp:chisquare}

The divergence measures reported in the main text (KL divergence, JS divergence, and TVD) quantify the \emph{magnitude} of the gap between each model's output distribution and the global population reference, but they do not assess whether that gap could be attributed to sampling variability alone. To provide formal statistical confirmation, we apply a $\chi^2$ goodness-of-fit test for each model--attribute combination. For attribute $d$ with $K$ categories and observed counts $O_1, \ldots, O_K$ from $N = 1{,}000$ generated images, the expected count under the global reference is $E_i = N \cdot P_{\mathrm{world}}(d_i)$. The test statistic is:

\[
\chi^2 = \sum_{i=1}^{K} \frac{(O_i - E_i)^2}{E_i}, \quad df = K-1,
\]

\noindent with $K=2$ categories for gender ($df=1$), $K=5$ age groups ($df=4$), and $K=4$ race classes ($df=3$). Table~\ref{tab:chisq_combined} reports results for all 8 models and 3 attributes (24 tests total). Twenty-three of 24 combinations are statistically significant at $p < 0.05$. The exception is Wan2.1 for gender ($p = 0.11$), which also shows the smallest gender TVD of any model.

\begin{table}[htbp]
\centering
\small
\caption{$\chi^2$ goodness-of-fit tests for gender ($df=1$), age ($df=4$), and race ($df=3$) against global population reference distributions. Each cell reports the test statistic and $p$-value. Western models are shaded blue, Chinese models red. Values of $p < 10^{-300}$ underflow double precision and are reported as $<10^{-300}$. $\dagger$~not significant at $\alpha = 0.05$.}
\label{tab:chisq_combined}
\resizebox{\linewidth}{!}{%
\begin{tabular}{lcccccc}
\toprule
& \multicolumn{2}{c}{\textbf{Gender}} & \multicolumn{2}{c}{\textbf{Age}} & \multicolumn{2}{c}{\textbf{Race}} \\
\cmidrule(lr){2-3}\cmidrule(lr){4-5}\cmidrule(lr){6-7}
\textbf{Model} & $\chi^2$ & $p$ & $\chi^2$ & $p$ & $\chi^2$ & $p$ \\
\midrule
\cellcolor{lightblue}Flux    & 75.19   & $4.3\times10^{-18}$  & 821.97  & $1.3\times10^{-176}$ & 430.14  & $6.6\times10^{-93}$  \\
\cellcolor{lightblue}Proteus & 872.75  & $8.2\times10^{-192}$ & 1847.49 & $<10^{-300}$         & 5073.77 & $<10^{-300}$         \\
\cellcolor{lightblue}SD3     & 422.77  & $6.1\times10^{-94}$  & 2123.97 & $<10^{-300}$         & 247.07  & $2.8\times10^{-53}$  \\
\cellcolor{lightblue}Sana    & 824.14  & $3.1\times10^{-181}$ & 1808.75 & $<10^{-300}$         & 3987.94 & $<10^{-300}$         \\
\cellcolor{red!10}HunYuan    & 422.26  & $7.9\times10^{-94}$  & 1403.35 & $1.3\times10^{-302}$ & 1888.24 & $<10^{-300}$         \\
\cellcolor{red!10}Qwen       & 547.91  & $3.6\times10^{-121}$ & 1399.30 & $9.8\times10^{-302}$ & 246.27  & $4.2\times10^{-53}$  \\
\cellcolor{red!10}Kolors     & 4.93    & $0.026$              & 1520.43 & $<10^{-300}$         & 4231.77 & $<10^{-300}$         \\
\cellcolor{red!10}Wan2.1     & 2.52    & $0.11^\dagger$       & 2123.70 & $<10^{-300}$         & 1039.78 & $4.2\times10^{-225}$ \\
\bottomrule
\end{tabular}}
\end{table}

\newpage
\suppsection{Intersectional Analysis with Neutral Expression}{supp:intersectional}

\begin{table}[htbp]
\centering
\caption{Least Represented Intersectional Cells (Mean $P(\mathbf{d})$ Across 8 Models)}
\label{tab:least_represented}
\resizebox{\textwidth}{!}{%
\begin{tabular}{lcccccccccc}
\toprule
\textbf{Cell} & \textbf{Mean} & \textbf{FLUX} & \textbf{HunYuan} & \textbf{Kolors} & \textbf{Proteus} & \textbf{Qwen} & \textbf{SANA} & \textbf{SD3} & \textbf{Wan2.1} \\
\midrule
Old $\times$ Female $\times$ Black   & 0.0000 & 0.0000 & 0.0000 & 0.0000 & 0.0000 & 0.0000 & 0.0000 & 0.0000 & 0.0000 \\
Old $\times$ Male $\times$ Black     & 0.0001 & 0.0010 & 0.0000 & 0.0000 & 0.0000 & 0.0000 & 0.0000 & 0.0000 & 0.0000 \\
Old $\times$ Female $\times$ Asian   & 0.0001 & 0.0000 & 0.0000 & 0.0010 & 0.0000 & 0.0000 & 0.0000 & 0.0000 & 0.0000 \\
Old $\times$ Female $\times$ White   & 0.0003 & 0.0000 & 0.0000 & 0.0020 & 0.0000 & 0.0000 & 0.0000 & 0.0000 & 0.0000 \\
Middle-aged $\times$ Female $\times$ Black & 0.0004 & 0.0010 & 0.0000 & 0.0000 & 0.0000 & 0.0020 & 0.0000 & 0.0000 & 0.0000 \\
Old $\times$ Female $\times$ Others  & 0.0006 & 0.0010 & 0.0000 & 0.0040 & 0.0000 & 0.0000 & 0.0000 & 0.0000 & 0.0000 \\
Old $\times$ Male $\times$ Others    & 0.0008 & 0.0040 & 0.0000 & 0.0020 & 0.0000 & 0.0000 & 0.0000 & 0.0000 & 0.0000 \\
Old $\times$ Male $\times$ Asian     & 0.0009 & 0.0070 & 0.0000 & 0.0000 & 0.0000 & 0.0000 & 0.0000 & 0.0000 & 0.0000 \\
\bottomrule
\end{tabular}%
}
\end{table}

The eight cells in Table~\ref{tab:least_represented} show a specific age/gender/race asymmetry in the least-represented intersections. Age: seven of the eight cells involve \emph{old} faces and the remaining cell involves \emph{middle-aged} faces; \emph{young} faces do not appear among the eight least-represented intersections. Gender: five cells involve \emph{female} faces and three involve \emph{male} faces. Race: Black appears in three cells, Asian and Others in two
each, and White in one (Old $\times$ Female $\times$ White, 0.0003). White is thus the least frequent racial category in this specific list, contrasting with the systematic over-representation of White faces at the marginal level (Section \ref{sec:results}). These results show that the extreme tail of the intersectional distribution is dominated by older age groups, with both genders and multiple racial categories represented. This pattern complements the young $\times$ female $\times$ Black case discussed in the main text
(Section \ref{sec:results}), and illustrates that severe intersectional under-representation occurs at both ends of the age spectrum.

\begin{table}[H]
\centering
\caption{Near-Zero Intersectional Cells per Model ($P(\mathbf{d}) < 0.01$, out of 24 cells)}
\label{tab:near_zero}
\begin{tabular}{lcc}
\toprule
\textbf{Model} & \textbf{Near-Zero Cells} & \textbf{Proportion} \\
\midrule
FLUX      & 11 & 45.8\% \\
Qwen      & 13 & 54.2\% \\
Wan2.1    & 15 & 62.5\% \\
SD3       & 17 & 70.8\% \\
HunYuan   & 18 & 75.0\% \\
Kolors    & 18 & 75.0\% \\
Proteus   & 20 & 83.3\% \\
SANA      & 20 & 83.3\% \\
\midrule
\textbf{Mean} & \textbf{16.5 $\pm$ 3.3} & \textbf{69\%} \\
\bottomrule
\end{tabular}
\end{table}

Table~\ref{tab:top5_regional} reports the top-5 intersectional cells for each regional aggregate distribution.

\begin{table}[H]
\centering
\caption{Top-5 intersectional cells for aggregate Western and Chinese distributions.}
\label{tab:top5_regional}
\begin{tabular}{clcclc}
\toprule
\multicolumn{3}{c}{\textbf{Western Models}} & \multicolumn{3}{c}{\textbf{Chinese Models}} \\
\cmidrule(lr){1-3} \cmidrule(lr){4-6}
\textbf{Rank} & \textbf{Cell} & \textbf{$P$} & \textbf{Rank} & \textbf{Cell} & \textbf{$P$} \\
\midrule
1 & young-male-white   & 26.5\% & 1 & young-female-asian & 23.2\% \\
2 & young-female-white & 26.2\% & 2 & young-male-white   & 22.1\% \\
3 & young-male-asian   & 15.3\% & 3 & young-female-white & 14.5\% \\
4 & young-male-others  &  7.9\% & 4 & young-male-others  & 13.3\% \\
5 & young-female-asian &  6.7\% & 5 & young-male-black   &  7.5\% \\
\bottomrule
\end{tabular}
\end{table}

\begin{table}[H]
\centering
\caption{Observed probability of young $\times$ female $\times$ Black per model.}
\label{tab:yfb}
\begin{tabular}{lcc}
\toprule
\textbf{Model} & \textbf{Observed $P$} & \textbf{Suppression vs.\ $P_{\exp}$} \\
\midrule
Hunyuan  & 0.000 (0.00\%) & --- \\
Kolors   & 0.000 (0.00\%) & --- \\
Qwen     & 0.017 (1.70\%) & $0.32\times$ \\
Wan2.1   & 0.064 (6.40\%) & $1.22\times$ \\
FLUX     & 0.019 (1.90\%) & $0.36\times$ \\
Proteus  & 0.000 (0.00\%) & --- \\
SANA     & 0.000 (0.00\%) & --- \\
SD3      & 0.006 (0.60\%) & $0.11\times$ \\
\midrule
\textbf{Mean} & \textbf{0.013 (1.30\%)} & $\mathbf{0.25\times}$ \\
\bottomrule
\end{tabular}
\end{table}

\begin{table}[H]
\centering
\caption{Pairwise JS divergences within Western and Chinese model groups.}
\label{tab:js_within}
\begin{tabular}{llcllc}
\toprule
\multicolumn{3}{c}{\textbf{Western Models}} & \multicolumn{3}{c}{\textbf{Chinese Models}} \\
\cmidrule(lr){1-3} \cmidrule(lr){4-6}
\textbf{Model 1} & \textbf{Model 2} & \textbf{$D_{\mathrm{JS}}$} & 
\textbf{Model 1} & \textbf{Model 2} & \textbf{$D_{\mathrm{JS}}$} \\
\midrule
FLUX    & Proteus & 0.3747 & Hunyuan & Kolors  & 0.5972 \\
FLUX    & SANA    & 0.3259 & Hunyuan & Qwen    & 0.4673 \\
FLUX    & SD3     & 0.1224 & Hunyuan & Wan2.1  & 0.4306 \\
Proteus & SANA    & 0.5356 & Kolors  & Qwen    & 0.3723 \\
Proteus & SD3     & 0.4859 & Kolors  & Wan2.1  & 0.1378 \\
SANA    & SD3     & 0.4874 & Qwen    & Wan2.1  & 0.2035 \\
\midrule
\multicolumn{2}{l}{\textbf{Mean}} & \textbf{0.3887} & 
\multicolumn{2}{l}{\textbf{Mean}} & \textbf{0.3681} \\
\bottomrule
\end{tabular}
\end{table}

\suppsection{Permutation Test for Western vs.\ Chinese Homogenization}{supp:permutation}

To test whether Western and Chinese text-to-image models exhibit systematically different degrees of demographic bias, we apply an exact permutation test. With only $n = 4$ models per group, parametric tests (e.g., $t$-test) are unreliable due to insufficient degrees of freedom. An exact permutation test is preferred because it makes no distributional assumptions and is exact rather than approximate.

For each demographic attribute (gender, race, age), we compute the per-model TVD from the global reference distribution under the neutral prompt, yielding 8 TVD scores: 4 for Western models (\textsc{Flux}, \textsc{Proteus}, \textsc{SD3}, \textsc{SANA}) and 4 for Chinese models (\textsc{Hunyuan}, \textsc{Qwen}, \textsc{Kolors}, \textsc{Wan2.1}). The observed test statistic is the absolute difference between the two group means:
\begin{equation}
    T_{\text{obs}} = \left| \bar{X}_{\text{Western}} - \bar{X}_{\text{Chinese}} \right|
\end{equation}

Under the null hypothesis that regional group membership has no effect on TVD, all $\binom{8}{4} = 70$ ways of partitioning the 8 models into two groups of 4 are equally likely. We enumerate all 70 partitions exhaustively and compute $T$ for each, yielding an exact reference distribution. The p-value is the proportion of partitions whose test statistic is at least as extreme as the observed value:
\begin{equation}
    p = \frac{\left|\left\{\, \text{partition} : T \geq T_{\text{obs}} \,\right\}\right|}{70}
\end{equation}

Because the enumeration is exhaustive, the test is exact and deterministic, requiring no random seed or Monte Carlo approximation. The minimum achievable two-sided p-value with this design is $2/70 \approx 0.029$.

The results are as follows. For gender, $T_{\text{obs}} = 0.157$ and $p = 0.200$; for race, $T_{\text{obs}} = 0.000$ and $p = 1.000$; for age, $T_{\text{obs}} = 0.013$ and $p = 0.886$. None of the three attributes reaches significance, confirming that the degree of demographic bias is statistically indistinguishable between Western and Chinese models.

\newpage
\suppsection{Unhappy vs.\ Sad}{supp:unhappy}

Although ``sad'' and ``unhappy'' are used interchangeably in everyday language, it is not obvious that generative models treat them as equivalent affective prompts. This section tests whether the two words induce the same demographic output distributions. We compare the marginal distributions over gender, race, and age produced by each of the eight models under the two prompts, and extend the analysis to the joint demographic space to identify intersectional combinations that exhibit the largest probability shifts between them.

\begin{figure}[H]
\centering
\begin{minipage}[c]{0.49\linewidth}
\centering
{\footnotesize
\setlength{\tabcolsep}{3pt}
\renewcommand{\arraystretch}{1.0}
\begin{tabular}{l
    S[table-format=1.3]
    S[table-format=1.3]
    S[table-format=1.3]
    S[table-format=1.3]}
\toprule
Model & {JS (Age)} & {JS (Gender)} & {JS (Race)} & {JS (Joint)} \\
\midrule
\cellcolor{lightblue}FLUX & \textbf{0.326} & 0.013 & 0.009 & \textbf{0.348} \\
\cellcolor{lightblue}Proteus      & 0.191 & 0.006 & 0.009 & 0.207 \\
\cellcolor{lightblue}SANA         & 0.083 & \textbf{0.029} & \underline{0.001} & 0.104 \\
\cellcolor{lightblue}SD3          & 0.091 & 0.015 & 0.008 & 0.137 \\
\midrule
\cellcolor{red!10}Hunyuan      & 0.015 & 0.005 & 0.012 & 0.067 \\
\cellcolor{red!10}Kolors       & 0.012 & \underline{0.001} & 0.004 & \underline{0.024} \\
\cellcolor{red!10}Qwen         & \underline{0.004} & 0.020 & 0.016 & 0.043 \\
\cellcolor{red!10}Wan2.1       & 0.023 & \underline{0.001} & \textbf{0.027} & 0.064 \\
\midrule
\textbf{Mean} & 0.093 & 0.011 & 0.011 & 0.124 \\
\bottomrule
\end{tabular}
}
\end{minipage}
\hfill
\begin{minipage}[c]{0.49\linewidth}
\centering
\includegraphics[width=\linewidth]{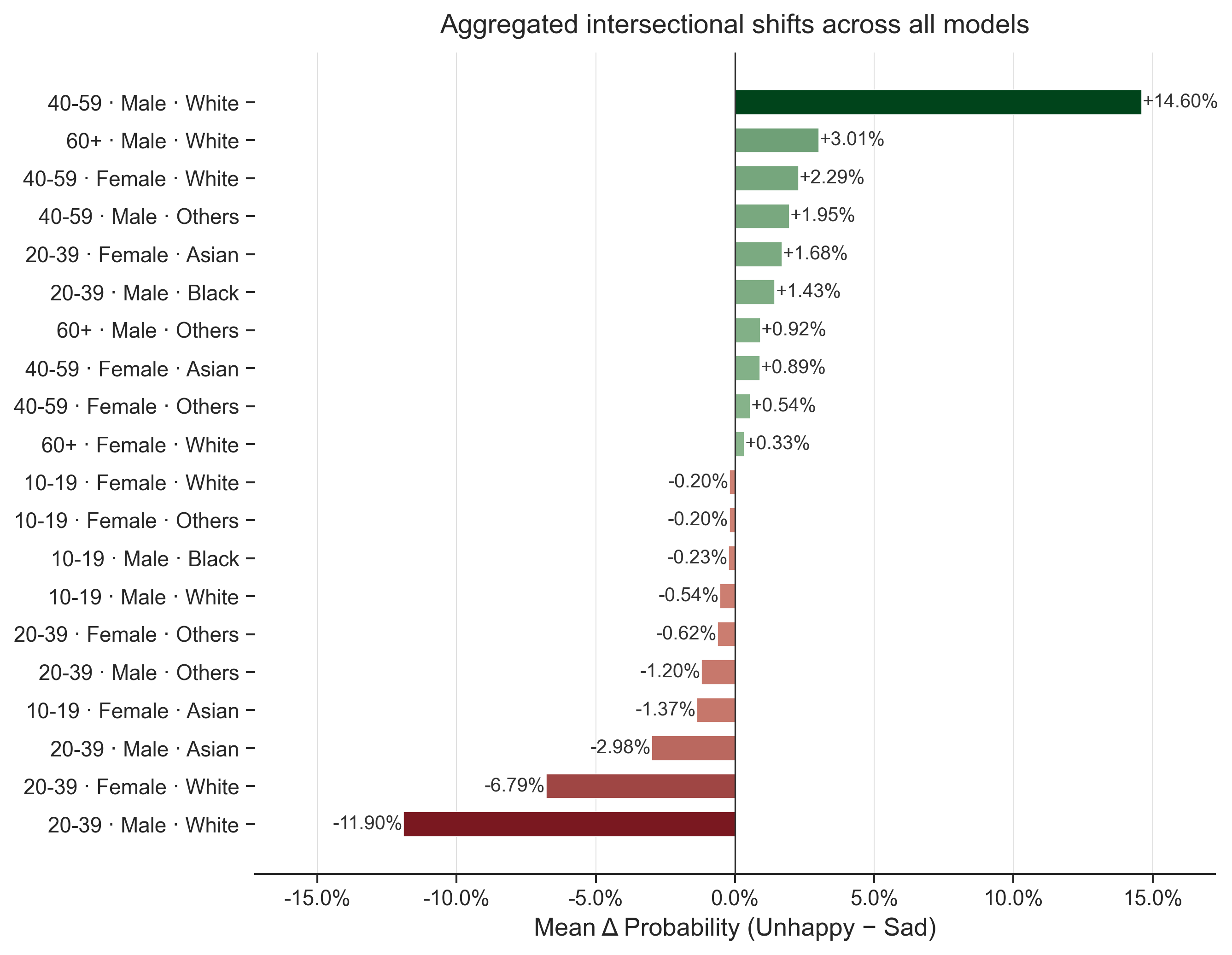}
\end{minipage}
\vspace{0.6em}
\caption{
\textbf{Left:} Jensen--Shannon divergence between the distributions of faces created with sad vs unhappy prompts across demographic attributes. The worst result is highlighted in bold and the best result is underlined.
\textbf{Right:} Top demographic intersections with the largest probability shifts between unhappy and sad prompts (unhappy -- sad), aggregated across models.
}
\label{fig:js_and_intersection_shift}
\end{figure}

Figure~\ref{fig:js_and_intersection_shift} shows that ``sad'' and ``unhappy'' produce near-identical gender and race distributions across the eight models (mean JS\,$<\,0.02$ for both attributes) and shift age distributions in the same direction---toward older faces---relative to the neutral baseline. This directional consistency corroborates the main findings rather than qualifying them: ``unhappy'' independently replicates the valence-primary demographic pattern observed for ``sad.'' The mean joint JS across models is 0.124 (range: 0.024--0.348), with larger values for some Western models (FLUX: 0.348, Proteus: 0.207) reflecting differences in the magnitude of the age shift rather than its direction. Chinese models are consistently close (joint JS: 0.024--0.067).

At the intersectional level (right panel), non-negligible shifts do emerge: transitioning from ``sad’’ to ``unhappy’’ systematically shifts probability mass away from younger faces toward middle-aged and older male combinations. This mirrors the broader pattern observed across negatively valenced emotions in the main results, and suggests that ``unhappy’’ is encoded slightly more negatively than ``sad’’ in the models’ semantic spaces. This difference is subtle and does not alter the directional conclusions of the study: both prompts produce qualitatively equivalent demographic output distributions, confirming that the findings are not sensitive to this specific lexical choice.

\suppsection{Chinese Prompts vs.\ English Prompts}{supp:chinese}

\begin{figure}[H]
    \centering
    \includegraphics[width=1\linewidth]{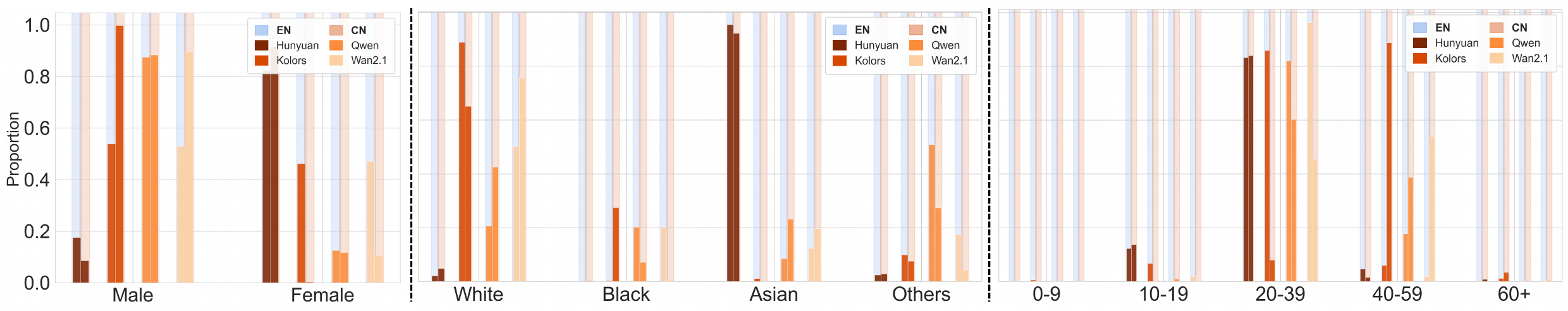}
    \caption{Demographic attribute distributions (gender, race, age) of the four Chinese T2I models under English (EN) and Chinese (ZH) neutral prompts. Despite using the models' native language, White overrepresentation persists across all models. The most striking difference is observed for Kolors, which generates $\approx$50\% female faces under English prompts but $\approx$0\% under Chinese prompts---a near-complete erasure of female representation that is absent in the other three Chinese models.}
    \label{fig:cn_en}
\end{figure}

The four Chinese T2I models overrepresent White faces when prompted in English (main text, Section \ref{subsec:rq2}). To assess whether this reflects a prompt-language artifact, we replaced each English neutral prompt with its direct Chinese translation, keeping all other generation settings unchanged. Western models were excluded as they do not accept Chinese-language input. We computed gender, race, and age distributions for each model under both conditions and report the comparison in Figure~\ref{fig:cn_en}.

Across the four models, switching to Chinese prompts produces observable but generally modest distributional changes. The overrepresentation of White faces persists in all models, confirming that this bias is embedded in the generative prior rather than driven by the English-language interface. For Hunyuan, Qwen, and Wan2.1, Chinese prompts slightly increase the proportion of Asian faces, but the change is not large enough to bring race distributions close to the Chinese population profile. Age and gender distributions remain broadly stable across prompt languages for these three models.

The exception is \textbf{Kolors}, which exhibits a dramatic and qualitatively distinct response to Chinese-language prompts: the proportion of female faces collapses from approximately 50\% under English prompts to approximately 0\% under Chinese prompts, a near-complete erasure of female representation. This effect is absent in Hunyuan, Qwen, and Wan2.1, ruling out a general property of Chinese-language T2I generation. It instead suggests that Kolors' Chinese-language encoder maps neutral person-descriptors to a representation space in which the default person is unambiguously male, a bias that is invisible under English-only evaluation and that would go undetected without multilingual auditing. This finding directly motivates the governance recommendation in the main text that cross-linguistic bias audits be established as a minimum standard for globally deployed models.

\newpage
\suppsection{Emotion-induced Biases}{supp:emotion}

\begin{table}[htbp]
\centering
\caption{Gender proportion by emotion, pooled across 8 models ($N = 8{,}000$ per condition). $\Delta$ denotes the change in the proportion of images of men relative to the same proportion under the neutral prompt. Chi-square tests compare each emotion condition against neutral.}
\label{tab:gender_emotion}
\begin{tabular}{lcccccc}
\toprule
\textbf{Emotion} & \textbf{Male (\%)} & \textbf{Female (\%)} & $\boldsymbol{\Delta}$ \textbf{Male (\%)} & $\boldsymbol{\chi^2}$ & \textbf{\textit{p}-value} \\
\midrule
Neutral    & 57.55 & 42.45 & —      & —       & — \\
Angry      & \textbf{95.43} &  4.57 & +37.88 & 3188.52 & $<$0.001*** \\
Disgusted  & \textbf{90.75} &  9.25 & +33.20 & 2298.46 & $<$0.001*** \\
Fearful    & \textbf{73.32} & 26.68 & +15.77 &  439.42 & $<$0.001*** \\
Happy      & 57.65 & 42.35 & +0.10  &    0.01 & 0.911 \\
Sad        & \textbf{73.86} & 26.14 & +16.31 &  471.64 & $<$0.001*** \\
Surprised  & \textbf{79.51} & 20.49 & +21.96 &  893.64 & $<$0.001*** \\
\bottomrule
\multicolumn{6}{l}{\small $^{***}p < 0.001$} \\
\end{tabular}
\end{table}

\begin{table}[htbp]
\centering
\caption{Race proportion by emotion condition, pooled across 8 models ($N = 8{,}000$ per condition). $\Delta$ denotes the change in each race group proportion relative to neutral. Chi-square tests compare each emotion condition against neutral.}
\label{tab:race_emotion}
\resizebox{\textwidth}{!}{
\begin{tabular}{lcccccccccc}
\toprule
\textbf{Emotion} & \textbf{White} & \textbf{Black} & \textbf{Asian} & \textbf{Others} & $\boldsymbol{\Delta}$\textbf{White} & $\boldsymbol{\Delta}$\textbf{Black} & $\boldsymbol{\Delta}$\textbf{Asian} & $\boldsymbol{\Delta}$\textbf{Others} & $\boldsymbol{\chi^2}$ & \textbf{\textit{p}-value} \\
\midrule
Neutral & 49.04 & 8.01 & 26.27 & 16.68 & — & — & — & — & — & — \\
Angry & \textbf{63.08} & 13.11 & 2.46 & 21.35 & +14.04 & +5.10 & -23.81 & +4.67 & 1863.61 & $<$0.001*** \\
Disgusted & \textbf{81.44} & 1.35 & 2.11 & 15.10 & +32.40 & -6.66 & -24.16 & -1.58 & 2674.50 & $<$0.001*** \\
Fearful & \textbf{81.77} & 3.94 & 5.92 & 8.36 & +32.74 & -4.08 & -20.35 & -8.31 & 2016.26 & $<$0.001*** \\
Happy & \textbf{42.70} & 19.18 & 17.99 & 20.14 & -6.34 & +11.16 & -8.29 & +3.46 & 551.86 & $<$0.001*** \\
Sad & \textbf{72.66} & 2.70 & 10.57 & 14.06 & +23.62 & -5.31 & -15.70 & -2.61 & 1130.54 & $<$0.001*** \\
Surprised & \textbf{83.94} & 3.28 & 6.04 & 6.75 & +34.90 & -4.74 & -20.24 & -9.93 & 2242.25 & $<$0.001*** \\
\bottomrule
\multicolumn{11}{l}{\small $^{***}p < 0.001$, $^{**}p < 0.01$, $^{*}p < 0.05$} \\
\end{tabular}
}
\end{table}

\begin{table}[htbp]
\centering
\caption{Age proportion by emotion, pooled across 8 models ($N = 8{,}000$ per condition). $\Delta$ denotes the change in the proportion of faces belonging to each age group relative to neutral. Chi-square tests compare each emotion condition against neutral.}
\label{tab:age_emotion}
\resizebox{\textwidth}{!}{
\begin{tabular}{lcccccccccccc}
\toprule
\textbf{Emotion} & \textbf{0-9} & \textbf{10-19} & \textbf{20-39} & \textbf{40-59} & \textbf{60+} & $\boldsymbol{\Delta}$\textbf{0-9} & $\boldsymbol{\Delta}$\textbf{10-19} & $\boldsymbol{\Delta}$\textbf{20-39} & $\boldsymbol{\Delta}$\textbf{40-59} & $\boldsymbol{\Delta}$\textbf{60+} & $\boldsymbol{\chi^2}$ & \textbf{\textit{p}-value} \\
\midrule
Neutral & 0.10 & 11.55 & 80.30 & 7.54 & 0.51 & — & — & — & — & — & — & — \\
Angry & 0.00 & 0.30 & 40.35 & \textbf{53.83} & 5.53 & -0.10 & -11.25 & -39.95 & +46.29 & +5.01 & 5046.90 & $<$0.001*** \\
Disgusted & 0.00 & 0.38 & \textbf{56.29} & 40.80 & 2.54 & -0.10 & -11.18 & -24.01 & +33.26 & +2.02 & 3122.16 & $<$0.001*** \\
Fearful & 0.10 & 1.07 & \textbf{75.31} & 21.04 & 2.48 & +0.00 & -10.48 & -4.99 & +13.50 & +1.96 & 1321.45 & $<$0.001*** \\
Happy & 0.09 & 7.59 & \textbf{86.39} & 5.53 & 0.41 & -0.01 & -3.96 & +6.09 & -2.01 & -0.10 & 109.16 & $<$0.001*** \\
Sad & 0.00 & 2.14 & \textbf{77.00} & 19.85 & 1.01 & -0.10 & -9.41 & -3.30 & +12.31 & +0.50 & 987.29 & $<$0.001*** \\
Surprised & 0.04 & 2.95 & \textbf{96.60} & 0.41 & 0.00 & -0.06 & -8.60 & +16.30 & -7.12 & -0.51 & 1082.33 & $<$0.001*** \\
\bottomrule
\multicolumn{13}{l}{\small $^{***}p < 0.001$, $^{**}p < 0.01$, $^{*}p < 0.05$} \\
\end{tabular}}
\end{table}

\begin{table}[htbp]
\centering
\caption{Proportion of faces classified with low/medium/high attractiveness by emotion, pooled across 8 models ($N = 8{,}000$ per condition). $\Delta$ denotes the change in each attractiveness level proportion relative to neutral. Chi-square tests compare each emotion condition against neutral.}
\label{tab:attractiveness_emotion}
\begin{tabular}{lcccccccc}
\toprule
\textbf{Emotion} & \textbf{Low} & \textbf{Medium} & \textbf{High} & $\boldsymbol{\Delta}$\textbf{Low} & $\boldsymbol{\Delta}$\textbf{Medium} & $\boldsymbol{\Delta}$\textbf{High} & $\boldsymbol{\chi^2}$ & \textbf{\textit{p}-value} \\
\midrule
Neutral & 3.39 & 80.58 & 16.04 & — & — & — & — & — \\
Angry & 17.27 & \textbf{77.98} & 4.75 & +13.89 & -2.60 & -11.29 & 5196.95 & $<$0.001*** \\
Disgusted & \textbf{45.99} & 52.33 & 1.69 & +42.60 & -28.25 & -14.35 & 44677.37 & $<$0.001*** \\
Fearful & 27.22 & \textbf{70.24} & 2.54 & +23.84 & -10.34 & -13.50 & 14434.59 & $<$0.001*** \\
Happy & 4.30 & \textbf{83.51} & 12.19 & +0.91 & +2.94 & -3.85 & 102.17 & $<$0.001*** \\
Sad & 20.62 & \textbf{74.98} & 4.40 & +17.24 & -5.60 & -11.64 & 7723.84 & $<$0.001*** \\
Surprised & 20.36 & \textbf{76.88} & 2.76 & +16.98 & -3.70 & -13.27 & 7697.69 & $<$0.001*** \\
\bottomrule
\multicolumn{9}{l}{\small $^{***}p < 0.001$, $^{**}p < 0.01$, $^{*}p < 0.05$} \\
\end{tabular}
\end{table}

This section provides the statistical test details and supporting tables for the emotion-conditioned bias analyses.

\paragraph{Statistical testing.}
For each model $m$, attribute $d$, and emotion $e$, we compare the count distribution under emotion $e$ (1,000 images) against the neutral-prompt distribution $e_0$ (1,000 images) using a 2-sample chi-squared test on the $K \times 2$ contingency table:
\[ \chi^2 = \sum_{i=1}^{K} \frac{(O_i(e) - E_i(e))^2}{E_i(e)}, \quad E_i(e) = 1{,}000 \times \hat{P}(i \mid e_0), \quad df = K-1. \]
This yields $8\ \text{models} \times 4\ \text{attributes} \times 6\ \text{emotions} = 192$ tests in total. We apply Bonferroni correction with significance threshold $\alpha^* = 0.05 / 192 \approx 2.6 \times 10^{-4}$. Of the 192 combinations, 180 (93.8\%) are statistically significant after correction, confirming that emotion-induced demographic shifts are a pervasive rather than model- or attribute-specific phenomenon.

\begin{figure}[htbp]
    \centering
    \includegraphics[width=0.7\linewidth]{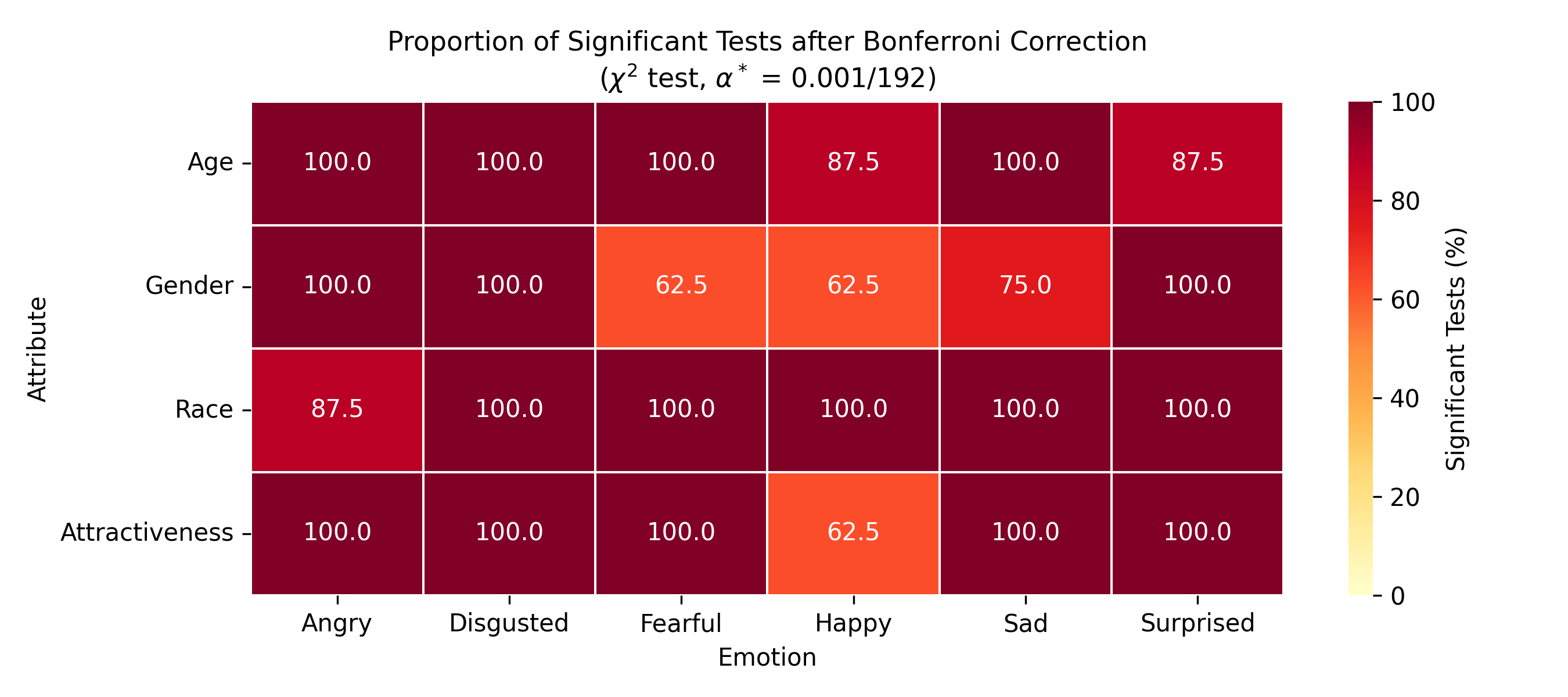}
    \caption{Heatmap of the proportion of significant distributional shifts after Bonferroni correction for each attribute $\times$ emotion combination.}
    \label{fig:attr_chisq}
\end{figure}

Figure~\ref{fig:attr_chisq} shows the percentage of significant distribution shifts after Bonferroni correction, allowing identification of which emotion--attribute combinations drive the largest shifts. Disgust yields a significant distribution shift across all attributes. Race and attractiveness levels are the most impacted attributes across emotions and gender is the least impacted attribute.

Table~\ref{tab:claim_a} details the racial composition shift under anger pooled across all eight models.

Table~\ref{tab:age_emotion} shows the corresponding age-distribution shifts under anger and disgust, the two emotions producing the largest age divergences.

\begin{table}[H]
\centering
\caption{Effect of anger on racial representation, pooled across 8 models ($N=8{,}000$ per condition). White faces increase by 14 percentage points while Asian faces decrease by 24 percentage points.}
\label{tab:claim_a}
\begin{tabular}{lcccc}
\toprule
\textbf{Condition} & \textbf{White} & \textbf{Black} & \textbf{Asian} & \textbf{Others} \\
\midrule
Neutral & 3,923 (49.04\%) & 641 (8.01\%)    & 2,102 (26.28\%) & 1,334 (16.68\%) \\
Angry   & 5,046 (63.08\%) & 1,049 (13.11\%) &   197 (2.46\%)  & 1,708 (21.35\%) \\
\midrule
$\Delta$ & \textbf{+14.04pp} & +5.10pp & $-$23.82pp & +4.67pp \\
\bottomrule
\end{tabular}
\smallskip\\
\footnotesize Angry vs.\ Neutral: $\chi^2 = 1863.61$, $df = 3$, $p < 10^{-300}$.
\end{table}

\begin{table}[htbp]
\centering
\caption{Age distributions under angry and disgusted conditions vs.\ neutral baseline, pooled across 8 models ($N=8{,}000$ per condition). Both emotions shift the distribution strongly away from young (20--39) toward middle-aged (40--59) faces.}
\label{tab:age_emotion2}
\begin{tabular}{lrrrrr}
\toprule
\textbf{Condition} & \textbf{0--9} & \textbf{10--19} & \textbf{20--39} & \textbf{40--59} & \textbf{60+} \\
\midrule
Neutral   & 8 (0.10\%)  & 924 (11.55\%)  & 6,424 (80.30\%) & 603 (7.54\%)    & 41 (0.51\%)  \\
Angry     & 0 (0.00\%)  & 24 (0.30\%)    & 3,228 (40.35\%) & 4,306 (53.83\%) & 442 (5.53\%) \\
$\Delta$  & $-$0.10pp   & $-$11.25pp     & $-$39.95pp      & \textbf{+46.29pp} & +5.01pp    \\
\midrule
Neutral   & 8 (0.10\%)  & 924 (11.55\%)  & 6,424 (80.30\%) & 603 (7.54\%)    & 41 (0.51\%)  \\
Disgusted & 0 (0.00\%)  & 30 (0.38\%)    & 4,503 (56.29\%) & 3,264 (40.80\%) & 203 (2.54\%) \\
$\Delta$  & $-$0.10pp   & $-$11.18pp     & $-$24.01pp      & \textbf{+33.26pp} & +2.02pp    \\
\bottomrule
\end{tabular}

\begin{minipage}{\linewidth}
\smallskip
\raggedright
\footnotesize
Angry vs.\ Neutral: $\chi^2 = 5046.90$, $df = 4$, $p < 0.001$.\quad
Disgusted vs.\ Neutral: $\chi^2 = 3122.16$, $df = 4$, $p < 0.001$.
\end{minipage}
\end{table}

\paragraph{Attractiveness shifts.}
Tables~\ref{tab:attractiveness} and~\ref{tab:delta_lowatt} provide the full attractiveness results. Table~\ref{tab:attractiveness} reports the mean proportion of low-, medium-, and high-attractiveness faces per emotion condition, averaged across the eight models. Table~\ref{tab:delta_lowatt} reports the increase in low-attractiveness proportion ($\Delta P_{\text{low-att}}$) relative to the neutral baseline, broken down by demographic group and emotion. Happiness is the only emotion with small changes across groups, ranging from $-6.6$ points (Middle-aged) to $+2.3$ points (non-White). The four negatively-valenced emotions increase the low-attractiveness share in every group, with increases ranging from $+32.3$ (Middle-aged) to $+47.6$ (Female) points under Disgust and $+2.6$ (Middle-Aged) to $+14.9$ (non-White) points under Anger.

\begin{table}[H]
\centering
\caption{Mean proportion (\%) of low-, medium-, and high-attractiveness faces by emotion condition, averaged across 8 models. The six emotion conditions reduce the high-attractiveness share relative to the neutral baseline ($16.0\%$): by $11.2$ to $14.3$ points for the negative emotions, by $13.2$ points for Surprise and by $3.8$ points for Happiness ($12.2\%$), the smallest reduction. The low-attractiveness share rises by $42.6$ points under Disgust, $13.9$ under Anger, and $0.9$ under Happiness, relative to the neutral baseline ($3.4\%$).}
\label{tab:attractiveness}
\begin{tabular}{lccccccc}
\toprule
\textbf{Attractiveness} & \textbf{Neutral} & \textbf{Angry} & \textbf{Disgusted} & \textbf{Fearful} & \textbf{Happy} & \textbf{Sad} & \textbf{Surprised} \\
\midrule
Low    &  3.4 & 17.3 & 46.0 & 27.2 &  4.3 & 20.6 & 20.4 \\
Medium & 80.6 & 78.0 & 52.3 & 70.2 & 83.5 & 75.0 & 76.9 \\
High   & 16.0 &  4.8 &  1.7 &  2.5 & 12.2 &  4.4 &  2.8 \\
\bottomrule
\end{tabular}
\end{table}

\begin{table}[H]
\centering
\caption{Increase in low-attractiveness proportion ($\Delta P_{\text{low-att}}$, percentage points) relative to the neutral baseline, by demographic group and emotion. Neutral baselines: Male 4.0\%, Female 4.0\%; White 2.8\%, non-White 2.8\%; Young 2.4\%, Middle-aged 18.1\%. Under Happiness, changes range from $-6.6$ points (Middle-aged) to $+2.3$ points (non-White). Within each dimension and emotion the mean is taken over the same set of models for every group, so that groups remain comparable; a model contributes only when all groups of that dimension have at least 10 faces for it under both conditions. The number of contributing models is given in parentheses. Cells marked --- have too few faces in any model (the 60+ group is omitted entirely for the same reason: it accounts for only 0.51\% of neutral faces). Bold negative values indicate a reduction in low-attractiveness under that condition. }
\label{tab:delta_lowatt}
\begin{tabular}{llcccccc}
\toprule
Dimension & Group & Angry & Disgusted & Fearful & Happy & Sad & Surprised \\
\midrule
\multirow{2}{*}{Gender} & Male & +4.8 & +38.5 & +25.5 & \textbf{-0.5} & +15.5 & +16.4 \\
 & Female & +6.6 & +47.6 & +27.2 & +2.0 & +14.6 & +17.4 \\
 & \textit{(models)} & \textit{(4)} & \textit{(7)} & \textit{(8)} & \textit{(8)} & \textit{(8)} & \textit{(8)} \\
\midrule
\multirow{2}{*}{Race} & White & +14.7 & +43.2 & +20.9 & +1.9 & +16.2 & +9.0 \\
 & non-White & +14.9 & +41.6 & +27.2 & +2.3 & +25.9 & +10.9 \\
 & \textit{(models)} & \textit{(8)} & \textit{(8)} & \textit{(7)} & \textit{(8)} & \textit{(8)} & \textit{(6)} \\
\midrule
\multirow{2}{*}{Age} & Young & +8.5 & +32.4 & +17.6 & \textbf{-0.4} & +11.1 & --- \\
 & Middle-aged & +2.6 & +32.3 & +12.1 & \textbf{-6.6} & +5.8 & --- \\
 & \textit{(models)} & \textit{(7)} & \textit{(7)} & \textit{(7)} & \textit{(7)} & \textit{(7)} & \textit{---} \\
\bottomrule
\end{tabular}
\end{table}

\end{document}